\documentclass[aps,prd,twocolumn,superscriptaddress,showpacs,nofootinbib]{revtex4-1}

\usepackage[utf8]{inputenc}

\usepackage{amsmath}
\usepackage{amssymb}
\usepackage{graphicx} 
\usepackage{dcolumn}  
\usepackage{bm}       
\usepackage{placeins}
\usepackage{braket} 
\usepackage{relsize} 
\usepackage{hyperref}
\usepackage[capitalise]{cleveref} 
\usepackage{commath} 
\usepackage{tikz}
\tikzset{every picture/.style={line width=0.3mm}}
\usetikzlibrary{decorations.markings}
\usetikzlibrary{calc}
\definecolor{pred}{RGB}{238,28,37}
\definecolor{pblue}{RGB}{48,49,146}
\definecolor{pgreen}{RGB}{00,163,80}
\def\centerarc[#1](#2)(#3:#4:#5){ \draw[#1] ($(#2)+({#5*cos(#3)},{#5*sin(#3)})$) arc (#3:#4:#5);}
\makeatletter
\newsavebox{\@brx}
\newcommand{\llangle}[1][]{\savebox{\@brx}{\(\m@th{#1\langle}\)}
  \mathopen{\copy\@brx\kern-0.5\wd\@brx\usebox{\@brx}}}
\newcommand{\rrangle}[1][]{\savebox{\@brx}{\(\m@th{#1\rangle}\)}
  \mathclose{\copy\@brx\kern-0.5\wd\@brx\usebox{\@brx}}}
\makeatother

\DeclareMathOperator{\Tr}{Tr}
\DeclareMathOperator\sign{sgn}
\renewcommand\Re{\mathrm{Re}}
\renewcommand\Im{\mathrm{Im}}

\newcommand{\Vhbar}{}
\newcommand{\m}{\omega_0}

\newcommand{\smallfrac}[2]{\mbox{\small ${\displaystyle \frac{#1}{#2}}$}}

\newcommand{\im}{\mathrm{i}}
\newcommand{\e}{\mathrm{e}}

\numberwithin{equation}{section}
\renewcommand{\theequation}{\arabic{section}.\arabic{equation}}

\DeclareMathOperator*{\SumInt}{%
\mathchoice%
  {\ooalign{$\displaystyle\sum$\cr\hidewidth$\displaystyle\int$\hidewidth\cr}}
  {\ooalign{\raisebox{.14\height}{\scalebox{.7}{$\textstyle\sum$}}\cr\hidewidth$\textstyle\int$\hidewidth\cr}}
  {\ooalign{\raisebox{.2\height}{\scalebox{.6}{$\scriptstyle\sum$}}\cr$\scriptstyle\int$\cr}}
  {\ooalign{\raisebox{.2\height}{\scalebox{.6}{$\scriptstyle\sum$}}\cr$\scriptstyle\int$\cr}}
}

\begin{document}

\newcommand{\JLU}{Institut f\"ur Theoretische Physik,
  Justus-Liebig-Universit\"at, 
  35392 Giessen, Germany}   
\newcommand{\HFHF}{Helmholtz Research Academy Hesse for FAIR (HFHF), Campus Giessen, 35392 Giessen, Germany}

\title{Real-time methods for spectral functions}

\author{Johannes V. Roth}
\affiliation{\JLU}

\author{Dominik Schweitzer}
\affiliation{\JLU}

\author{Leon J. Sieke}
\affiliation{\JLU}

\author{Lorenz von Smekal}
\affiliation{\JLU}
\affiliation{\HFHF}

\date{May 2022}

\begin{abstract}
In this paper we develop and compare different real-time methods to calculate spectral functions. These are classical-statistical simulations, the Gaussian state approximation (GSA), and the functional renormalization group (FRG) formulated on the Keldysh closed-time path. Our  test-bed system is the quartic anharmonic oscillator, a single self-interacting bosonic degree of freedom, coupled to an external heat bath providing dissipation analogous to the Caldeira-Leggett model. As our benchmark we use the spectral function from exact diagonalization with constant Ohmic damping. To extend the GSA for the open system, we solve the corresponding Heisenberg-Langevin equations in the Gaussian approximation.  
For the real-time FRG, we introduce a novel general prescription to construct causal regulators based on introducing scale-dependent fictitious heat baths. 
Our results explicitly demonstrate how the discrete transition lines of the quantum system gradually build up the broad continuous structures in the classical spectral function as temperature increases. At sufficiently high temperatures, classical, GSA and exact-diagonalization results all coincide. The real-time FRG is able to reproduce the effective thermal mass, but overestimates broadening and only qualitatively describes higher excitations, at the present order of our combined vertex and loop expansion. As temperature is lowered, the GSA follows the ensemble average of the exact solution better than the classical spectral function. In the low-temperature strong-coupling regime, the qualitative features of the exact result are best captured by our real-time FRG calculation, with quantitative improvements to be expected at higher truncation orders. 
\end{abstract}

\maketitle

\section{Introduction}

Non-perturbative studies in thermal field theory are most commonly done within the imaginary time formalism \cite{Matsubara_1955}, where time is analytically continued to replace $t \to -\im \tau$ with the Euclidean time variable  $\tau \in [0, \beta]$, where $\beta=1/T$ is the inverse temperature of the equilibrium system. Along the imaginary-time axis one introduces (anti-)periodic boundary conditions for the fields.
This is the basis for the powerful path-integral representation of the partition function (see standard texts, e.g.~\cite{Kapusta:2006pm,Wipf:2013vp}). 
If one is interested in real-time quantities such as response functions or transport coefficients, however, this analytic continuation has to be undone. Because the Euclidean time interval is compact at finite temperature, the information is incomplete without further input, and the underlying Wick rotation cannot simply be inverted by back-substituting $ \tau \to \im t $. But even at vanishing temperature, this typically leads to ill-posed inverse numerical problems.    

Methods working directly in real-time have therefore become increasingly popular over the last decades, despite their generally higher computational requirements. They avoid the analytic-continuation problem and allow studying off-equilibrium systems. The price is that standard Monte-Carlo simulations for ab-initio studies as in Euclidean spacetime are not possible, since importance sampling fails due to the sign problem.
Other computational methods are needed for real-time computations.

Here we are particularly interested in real-time methods for spectral functions, which generally contain the complete spectrum of quasi-particle, multi-particle and collective excitations of a system contributing to a given correlation function, often including transport coefficients in particular low-frequency limits as relevant e.g.~for hydrodynamic descriptions. Spectral functions can be formally defined as expectation values of the unequal-time commutators of the corresponding fields. After Fourier transformation, they represent the density of states containing all possible excitations in that channel. Under certain circumstances functional equations for originally Euclidean correlation functions such as  Dyson-Schwinger \cite{Strauss:2012dg,Fischer:2020xnb,Horak:2020eng,Horak:2021pfr} or FRG flow equations \cite{Kamikado:2013sia,Tripolt:2013jra,Tripolt:2014wra,Tripolt:2021jtp,Jung:2021ipc}  can be analytically continued back to the real-frequency domain before they are being solved. For phenomenological scattering theory and resolvent based real-time approaches to calculate spectral functions, see e.g.~Refs.~\cite{Guhr:1997ve,Samanta:2020pez,Lo:2019who}. 
The flexibility to include dissipative or diffusive dynamics and the different dynamic critical behavior have so far been beyond these approaches, however.

In this work, we therefore focus on genuine real-time methods as more general and versatile alternatives. In particular, we employ three different non-perturbative real-time methods for calculating spectral functions in a simplified test system consisting of a single  self-interacting bosonic degree of freedom, the quartic anharmonic oscillator, which can also be considered as the $\lambda \phi^4$ theory of self-interacting real scalar fields in $(0+1)$ dimensions, coupled to an external heat bath which is modeled as an ensemble of harmonic oscillators after Caldeira and Leggett~\cite{CALDEIRA1983587,Kamenev:2011,Weiss_2012,Ingold_2002}. 
This rather simple system has the advantage that its spectral function can be calculated \emph{exactly} using a standard discretization of the Schr\"odinger equation, which allows to qualitatively discuss and quantitatively compare in detail the different approaches and approximation schemes.

The different approaches we compare here are \emph{classical-statistical simulations}~\cite{Aarts:2001yx,Berges:2009jz,Schlichting:2019tbr,Schweitzer:2020noq,Schweitzer:2021iqk}, which describe the purely classical time evolution according to Langevin-type equations of motion, the \emph{Gaussian state approximation} (GSA) introduced in Refs.~\cite{Buividovich:2017kfk,Buividovich:2018scl}, where one additionally considers the time evolution of two-point functions, and the \emph{functional renormalization group} (FRG)~\cite{Wetterich:1992yh,Berges:2000ew,Gies:2006wv,Polonyi:2001se,Pawlowski:2005xe,Schaefer:2006sr,Kopietz:2010zz,Braun:2011pp,Friman:2011zz} which we use here on the closed time path (CTP, also called the Schwinger-Keldysh contour)~\cite{Huelsmann:2020xcy,Tan:2021zid,Berges:2012ty,Mesterhazy:2015uja,Duclut:2016jct,Canet:2006xu}, where we build on the previous study of Ref.~\cite{Huelsmann:2020xcy} and take a closer look at the necessary causal structure of the regulators
\cite{Duclut:2016jct}.

This paper is organized as follows.
In \cref{quart-anh-osc} we start by briefly summarizing the discretization scheme and formulas needed for an exact calculation of the spectral function.
The following three \cref{gauss-state-approx,class-sf,rt-frg} describe in detail the three real-time calculation methods for the spectral function, i.e.~the  classical-statistical simulations in \cref{class-sf}, the Gaussian state approximation in \cref{gauss-state-approx}, and the real-time FRG in \cref{rt-frg}.
Our results from the different methods are presented, compared and discussed in detail in \cref{results}, and our conclusions are given with a brief outlook on possible further   studies in \cref{conclusion}. Several appendices are added with further technical details and derivations especially for the Heisenberg-Langevin equations of motion in the GSA, and our regulators and truncation scheme for the real-time FRG flows.

\section{Quartic anharmonic oscillator}
\label{quart-anh-osc}

We consider a single quartic anharmonic oscillator of unit mass, defined by the Hamiltonian
\begin{align}
    \label{hamiltonian-anh-osc}
    \hat H = \frac{\hat{p}^2}{2} + \frac{\m^2}{2} \hat{x}^2 + \frac{\lambda}{4!}\hat{x}^4,
\end{align}
in thermal contact with an external heat bath in equilibrium.
The heat bath is modeled as an ensemble of harmonic oscillators, which is generally known as the Caldeira-Leggett model in the literature~\cite{CALDEIRA1983587,Kamenev:2011,Weiss_2012,Ingold_2002}, and which we will explain in more detail in Section~\ref{gauss-state-approx}.

The anharmonic oscillator can be interpreted as a self-interacting single-component real scalar field theory in $(0+1)$~dimensions. It serves here as a benchmark system for comparing different methods for calculating spectral functions, since its spectrum can be numerically determined with essentially arbitrary precision using a discretization of the Schr\"odinger equation on a lattice.
Therefore, we will refer to the Schr\"odinger discretization method as the \emph{exact-diagonalization} solution.

The spectral function is defined as the real distribution given by the thermal expectation value of the commutator of two Heisenberg (field) operators taken at unequal times as follows \cite{Parisi:1988nd,Aarts_2001},
\begin{equation}
    \label{spectralFunction}
    \rho(t-t') = \im \langle [\hat{x}(t),\hat{x}(t')] \rangle_\beta,
\end{equation}
where the average is taken over the canonical ensemble $e^{-\beta \hat{H}}/Z$ at temperature $T=1/\beta $ with the partition function $Z = \Tr e^{-\beta \hat{H}}$.
To obtain a real distribution also in the frequency domain from the real and odd $\rho(-t) = - \rho(t)$, a factor of $2\pi\im $ is commonly absorbed in the definition of its Fourier transform,
\begin{equation}
    \rho (\omega) \equiv \frac{1}{2\pi\im} \int \dif t \, \rho(t) \, \e^{\im \omega t} , \label{unconvFT}
\end{equation}
which is then positive, $\rho(\omega)\ge 0 $, for $ \omega>0$, also odd $\rho(-\omega) = -\rho(\omega)$, and normalized according to
\begin{equation}
    \int_{-\infty}^\infty \!\dif\omega \,\omega \rho(\omega) = 1 . 
\end{equation}
Without dissipation, this spectral function may be expressed as a sum over energy-eigenstates (see for example Chapter 6.2 of Ref.~\cite{Kapusta:2006pm}),
\begin{align}
    \rho(\omega) &= \frac{1}{Z} \sum_{m,n} \e^{-\beta E_n} \, \Big( \delta(\omega - E_m + E_n) \nonumber \\
    &\hskip 1.2cm  -\delta(\omega + E_m - E_n)\Big) \, |\langle n| \hat{x} |m\rangle|^2 . \label{sf-in-eigenstates}
\end{align}
For the non-interacting theory ($\lambda = 0$), i.e.~the harmonic oscillator with frequency $\m$, this reduces to
\begin{equation}
    \rho_0(\omega)  = \sign(\omega) \, \delta(\omega^2 -\m^2)  = \,  \frac{1}{\pi} \, \Im \, G^R_0(\omega).
\end{equation}
When the free oscillator is coupled to an Ohmic heat bath, one describes an \emph{open} quantum system and the retarded Green function $G_0^R$
acquires an additional damping term with constant $\gamma >0 $, corresponding to ${G^R_{0,\gamma} }^{\!\!-1} = -(\omega^2 -\m^2 + \im \gamma \omega ) $, and the spectral function becomes
\begin{align}
    \rho_{0,\gamma}(\omega) &= \, \frac{1}{\pi} \, \Im \, G^R_{0,\gamma}(\omega) \\
    &= \frac{1}{\pi} \, \frac{\gamma\omega}{(\omega^2-\m^2)^2 +\gamma^2\omega^2} \nonumber\\
    &= \int_{-\infty}^\infty \dif \omega' \omega' \rho_0(\omega' )  \,  \frac{1}{\pi} \, \frac{\gamma\omega}{(\omega^2- {\omega'}^2)^2 +\gamma^2\omega^2}
    . \nonumber
\end{align}
This of course describes the collisional broadening of the free spectral function $\rho_0(\omega) $  due to the Ohmic heat bath, together with the frequency shift of the damped harmonic oscillator from the poles in $G^R_{0,\gamma}(\omega)$ at
\begin{equation}
\omega = \pm \, \sqrt{\omega_0^2-\gamma^2/4} -\im\gamma/2 . 
\end{equation}
The interaction with the environment in the open system introduces the new parameter $\gamma$ which essentially quantifies how strongly the single quartic oscillator \eqref{hamiltonian-anh-osc} couples to the degrees of freedom in the external heat bath.
The damping $\gamma$ then causes the spectral function to be broadened even in the harmonic case (with $\lambda = 0$), because  the system particle can decay into heat-bath excitations, corresponding to a finite lifetime $\sim 1/\gamma$.
Hence, the thermal expectation value in \eqref{spectralFunction} is understood over a reduced density operator where interactions with the environment have been traced out.
In the limit $\gamma \to 0^+$ we recover the usual $\varepsilon$-prescription of the retarded/advanced propagators, and effectively consider a system in infinitesimal contact with an external heat bath, described by the canonical density operator $\hat{\rho} = e^{-\beta \hat{H}}/Z$. 

Applying the same collisional broadening to the spectral function of the anharmonic oscillator in (\ref{sf-in-eigenstates}), one analogously obtains
\begin{align}
    \rho_\gamma(\omega) &= \frac{1}{Z} \sum_{m,n} \e^{-\beta E_n} \, 
    |\langle n| \hat{x} |m\rangle|^2 \,  2\Delta E_{mn}
    \nonumber \\
    &\hskip 1.4cm  \times  \, \frac{1}{\pi}\, \frac{\gamma\omega}{(\omega^2 - \Delta E_{mn}^2)^2 + \gamma^2 \omega^2} , \label{sfgamma-in-eigenstates}
\end{align}
with  $\Delta E_{mn} =E_m-E_n$. The  only assumption here is that the width $\gamma$ is not affected by the anharmonicity, in particular, that it does not acquire a frequency dependence for $\lambda \not= 0$.\footnote{Although this is reasonable for small $\gamma $
in the Ohmic bath, it will not hold in more realistic situations with ultraviolet cutoff $\omega_\text{D} $ as in the Drude model for the bath,
where memory effects will necessarily occur on time scales shorter than $\omega_\text{D}^{-1} $, inducing a frequency dependent damping $\gamma (\omega) $ on scales $\omega \sim \omega_\text{D}$ \cite{Ingold_2002}.}
Otherwise this is an exact expression which we will use for our benchmark calculations.

\begin{figure}[t]
\hspace{-.1cm}\includegraphics[width=0.95\linewidth]{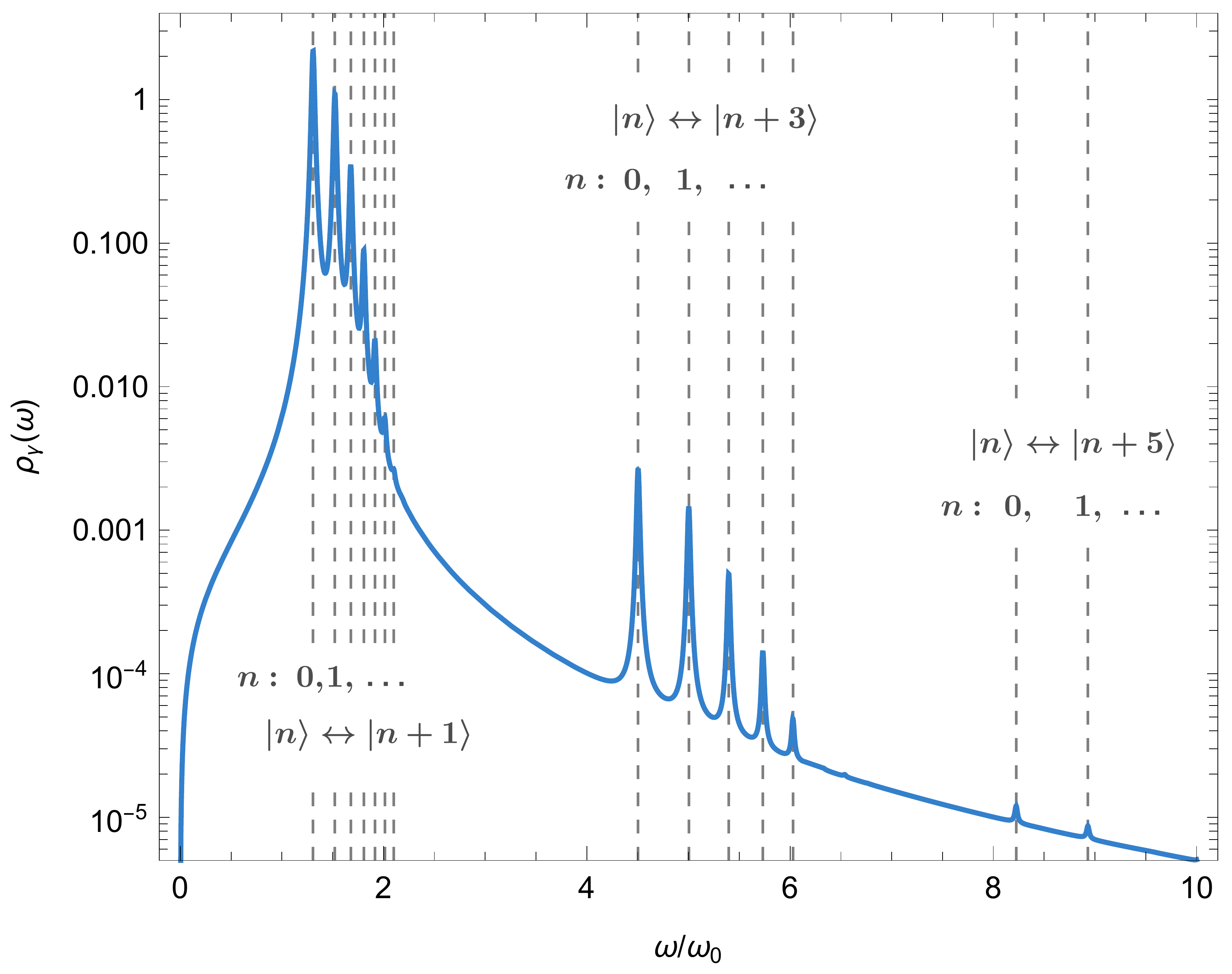}
\caption{Exemplary spectral function (in units of $\m^{-2}$) from exact diagonalization with damping according to Eq.~(\ref{sfgamma-in-eigenstates}), for $T/\m = 1$, $\lambda/\m^3 = 4$ and $\gamma/\m = 0.03$. The dashed vertical lines mark the discrete energy differences of the individual transitions as obtained from the exact diagonalization.\label{QanOscSF}}
\end{figure}

To compute the spectral function of the anharmonic oscillator with this constant broadening
we discretize the coordinate $x$ on a sufficiently large interval and solve the eigenvalue problem for the corresponding finite Hamiltonian matrix, obtained from (\ref{hamiltonian-anh-osc}) in coordinate space, by exact diagonalization. We then verify that the interval is large enough to cover the support of all relevant eigenfunctions at a given temperature and that the discretization is fine enough to obtain precise results sufficiently high up in the spectrum. For the parameters $\lambda$, $\gamma$ and $T$ used in Sec.~\ref{results} typical interval sizes are  $x \in [-20, 20]$ (in units of $1/\sqrt{\m}$) 
with $\sim 3000 $ grid points. An example is shown in Figure~\ref{QanOscSF} where we plot the spectral function of the quartic anharmonic oscillator  at a rather large coupling of $\lambda/\m^3 = 4 $ for a temperature  $T/\m = 1 $, and with a comparatively small damping of $\gamma/\m = 0.03 $, from Eq.~\eqref{sfgamma-in-eigenstates}. In particular, the small damping allows resolving the individual transitions: (a) between adjacent energy levels $\ket n $ and $\ket{n+1} $ in the main peak which are split up because they are no-longer equidistant when $\lambda\not= 0$, (b)  transitions across three levels between $\ket n $ and $\ket{n+3} $ in the second sequence of peaks at higher frequencies which are also due to the sizable  $\lambda >0  $, and (c) analogous transitions across five levels $\ket n $ and $\ket{n+5} $, here for frequencies above $8 \m$. The finite temperature manifests itself in the contributions from the individual transitions with $n\ge 1$  in each sequence which vanish for $T\to 0$ where only the corresponding  ground-state transitions $\ket 0 \leftrightarrow \ket 1$, $\ket 3$, $\ket 5$, \dots, survive.
In the results section below we will implicitly assume all dimensionful quantities to be quoted in the appropriate units of $\m $.

\section{Classical Spectral functions}
\label{class-sf}

In the classical-statistical limit \cite{Aarts:2001yx,Berges:2009jz,Schlichting:2019tbr,Schweitzer:2020noq,Schweitzer:2021iqk}, the full Heisenberg equations of motion are truncated by approximating all quantum mechanical expectation values of products of operators by the corresponding products of their expectation values, i.e.~$\langle \hat{O}_1 \dots \hat{O}_n\rangle \rightarrow \langle\hat O_1\rangle \dots \langle \hat O_n \rangle$, which can be formalized via the real-time path integral formulation for classical-statistical systems \cite{Martin:1973zz,Kamenev:2011,Hertz:2016vpy}.
One then arrives at the well-known Langevin equations of motion describing the purely classical dissipative dynamics  \cite{Kamenev:2011,Weiss_2012},
\begin{align}
     \dod{}{t} X &= P, \label{class-eom-x} \\
     \dod{}{t} P &= - \m^2 X - \frac{\lambda}{6} X^3 -\gamma P + \xi(t) ,
     \label{class-eom-p}
\end{align}
where $X(t) $ and $P(t)$ are the expectation values of $\hat x(t) $ and $\hat p(t)$ in coherent states, and the stochastic fluctuating force $\xi$  is given by a Gaussian white noise with zero mean and variance $2\gamma T$,
\begin{align}
    \langle\xi(t)\rangle_\beta &= 0, \\
    \langle\xi(t) \xi(t')\rangle_\beta &= 2\gamma T\, \delta(t-t').
\end{align}
As mentioned in the previous section, the \emph{spectral function} $\rho$ can be defined as the thermal expectation value of the commutator of two operators, cf.~\eqref{spectralFunction} for our oscillator here. It is related to the corresponding time-ordered Green's function $G^\mathbb T$ via 
\begin{equation}
    G^\mathbb{T}(t,t') = F(t,t')-\frac{\im}{2}\rho(t,t') [\Theta(t-t')-\Theta(t'-t)],
    \label{eq:decomposition}
\end{equation}
where $F(t,t^\prime)$ is the \emph{statistical two-point function} defined as the expectation value of the corresponding anti-commutator \cite{Parisi:1988nd,Aarts_2001}, i.e.~here,
\begin{equation}
    F(t,t') = \frac{1}{2}\langle\{\hat x(t),\hat x(t')\}\rangle_\beta.
\end{equation}
Since these equilibrium two-point functions all depend only on the time difference $t-t'$, their Fourier transforms  $F(\omega)$, $\rho(\omega)$ depend on a single frequency $\omega $, where for $\rho(\omega )$ we use the definition in Eq.~(\ref{unconvFT}), but for the even function $F(-\omega )=F(\omega )$ the conventional form so that
\begin{equation}
F(t-t') = \int \frac{\dif\omega}{2\pi} \, F(\omega) \, \e^{-\im\omega(t-t')} 
. \label{convFT}
\end{equation}
In thermal equilibrium one furthermore applies the periodicity or \emph{Kubo-Martin-Schwinger} (KMS) condition,
\begin{equation}
    G^\mathbb{T}(t,t') = G^\mathbb{T}(t', t+\im\beta),
\end{equation}
in the decomposition of Eq.~(\ref{eq:decomposition}), in order to derive the \emph{fluctuation-dissipation relation} (FDR), e.g.~see \cite{Aarts_1998},
\begin{align}
    F(\omega) &=  \left(2 n_\text{B}(\omega)+{1}\right) \, \pi \rho(\omega)
    \label{eq:FDT} \nonumber\\
    &= \coth\left(\frac{\omega}{2T}\right) \, \pi \rho(\omega) ,
\end{align}
where we have used the special convention for the definition of $\rho(\omega)$ in (\ref{unconvFT}), and $ n_\text{B}(\omega) =1/(\mathrm{e}^{\beta \omega}-1 ) $.
In the classical limit $T\gg \omega$ we approximate $\coth(\omega/2T) \approx 2T/\omega$.
The classical FDR from (\ref{eq:FDT}) then relates the corresponding classical two-point functions,
\begin{equation}
    F_\text{c}(\omega) = \frac{T}{\omega} \, 2\pi \rho_\text{c}(\omega) . \label{class-rho-FDR}
\end{equation}
In the time domain, undoing the Fourier transform, this reads
\begin{equation}
    \rho_\text{c}(t-t') = -\frac{1}{T} \,\partial_t F_\text{c}(t-t'). 
    \label{eq:class_spectral_func}
\end{equation}
Furthermore, because the statistical two-point function is in the classical limit given by the purely thermal correlation function 
\begin{equation}
    F_\text{c}(t-t') = \langle X(t) X(t')\rangle_\beta - \langle X(t)\rangle_\beta \langle X(t')\rangle_\beta,
\end{equation}
the spectral function (\ref{eq:class_spectral_func}) can be written as \cite{Schweitzer:2020noq}
\begin{align}
    \rho_\text{c}(t-t')
    &= -\frac{1}{2T} \, \big\langle P(t) X(t') - X(t) P(t')\big\rangle_\beta,
    \label{eq:simple_spectr_func}
\end{align}
where  $P = \dot X $ is the \emph{conjugate momentum} which has zero mean in the thermal ensemble,  $\braket{P(t)}_\beta  =0$. Because of time-reversal invariance of the thermal expectation values, the two terms in  (\ref{eq:simple_spectr_func}) are the same, and the explicit anti-symmetrization in this definition of  $\rho_\text{c}(-t) = - \rho_\text{c}(t)$ can be introduced without loss.
Evaluating Eq.~(\ref{eq:simple_spectr_func}) 
provides a straightforward way of calculating the spectral function in the classical-statistical limit \cite{Aarts:2001yx,Berges:2009jz,Schlichting:2019tbr,Schweitzer:2020noq,Schweitzer:2021iqk}.

\section{Gaussian-state approximation}
\label{gauss-state-approx}

\subsection{Closed system}
Before considering the coupling to an environment, we first briefly discuss the Gaussian state approximation (GSA) for a closed system. The GSA is obtained by truncating the full Heisenberg equations of motion
\begin{equation}
    \dod{}{t} \hat{O} = \mathrm{i} \left[\hat{H}, \hat{O}\right]
\end{equation}
for the canonically conjugate Heisenberg operators $\hat{x}(t) $ and $\hat{p}(t) $:
\begin{subequations}
\begin{align}
    \dod{}{t} \hat{x} &= \hat{p}, \label{eq:partial_t x} \\
    \dod{}{t} \hat{p} &= -\m^2\hat{x} - \frac{\lambda}{6}\hat{x}^3
    \equiv -V'(\hat x). \label{eq:partial_t p}
\end{align}
\end{subequations}
The equations of motion for the expectation values can be obtained by averaging equations (\ref{eq:partial_t x}) and (\ref{eq:partial_t p}) over some density operator $\hat\rho$ describing the mixed initial state of the ensemble.
These equations then contain expectation values of the form $\langle\hat{x}^2(t)\rangle$ and $\langle\hat{x}^3(t) \rangle$, whose evolution equations in turn include expectation values of even higher-order combinations of $\hat{x}$ and $\hat{p}$. 
This leads to an infinite hierarchy of equations that cannot be solved analytically or numerically without further approximations. Moreover, to deal with expectation values of products of $\hat{x}$ and $\hat{p}$ we follow Ref.~\cite{Buividovich:2018scl} and introduce the Wigner transform of the density matrix in position eigenstates,
\begin{equation}
    w(x,p) = \int \dif y \, \e^{-\im py} \, \langle x+y/2|\hat\rho |x-y/2 \rangle , \label{wignerTrafo}
\end{equation}
which allows expressing the expectation values of symmetrized products of $\hat{x}$ and $\hat{p}$ in the form of classical phase space integrals, such as e.g.
\begin{align}
    \frac{1}{2} \, \langle \hat x \hat p + \hat p \hat x\rangle = \int \frac{\dif x\dif p}{2\pi} \, x\, p \, w(x,p) . \label{expWig}
\end{align} 
To truncate the infinite set of equations given by (\ref{eq:partial_t x}) and (\ref{eq:partial_t p}), the density matrix is itself approximated by a Gaussian, and can therefore be characterized by a Gaussian Wigner function likewise \cite{Buividovich:2017kfk},
\begin{align}
    w(x,p) &=     \label{eq:wignerfunc}\\
&\hskip -.4cm \mathcal{N}\exp
    \left\{
    \hspace{-0.5ex}
    -\frac{1}{2}
    \begin{pmatrix}
        x - X \\
        p - P
    \end{pmatrix}^{\hspace{-0.5ex}T}
    \hspace{-0.5ex}
    \begin{pmatrix}
        \sigma_{xx} & \sigma_{xp} \\
        \sigma_{xp} & \sigma_{pp}
    \end{pmatrix}^{\hspace{-0.5ex}-1}
    \hspace{-1ex}
    \begin{pmatrix}
        x - X \\
        p - P
    \end{pmatrix}
    \right\}. \nonumber
\end{align}
Here, the parameters $X \equiv \langle\hat{x}\rangle, P \equiv \langle\hat p \rangle$ describe the center of the Gaussian wave packet in coordinate and momentum space. As such they are not necessarily the expectation values in coherent states yet, here.
The symmetrized connected expectation values 
\[
\sigma_{ab} \equiv \llangle \hat a \hat b\rrangle \equiv \langle \hat a \hat b + \hat b \hat a \rangle/2 - \langle \hat a \rangle \langle \hat b \rangle
\]
characterize the dispersions of the wave packet, and $\mathcal{N}$ is a normalization factor.

Equations (\ref{eq:partial_t x}) and (\ref{eq:partial_t p}) are then averaged over the Gaussian state with the Wigner function (\ref{eq:wignerfunc}). 
Applying Wick's theorem as needed, one then obtains
\begin{subequations}
\begin{align}
    \dod{}{t} X &= P, \label{eq:eom1}\\
    \dod{}{t} P &= -\m^2 X -\frac{\lambda}{6}\left(X^3 + 3X\sigma_{xx}\right). \label{eq:eom2}
\end{align}
\end{subequations}
To evolve the dispersions $\sigma_{xx}$, $\sigma_{xp}$, and $\sigma_{pp}$, the Heisenberg equations for the corresponding symmetrized operator products are employed
\begin{subequations}
\begin{align}
    \dod{}{t} \hat{x}^2
    &= \hat{x} \hat{p} + \hat{p} \hat{x},
    \label{eq:dtxx}\\
    \dod{}{t} \frac{\hat{x}\hat{p} + \hat{p}\hat{x}}{2}
    &= \hat{p}^2 -\m^2\hat{x}^2 - \frac{\lambda}{6}\hat{x}^4,
    \label{eq:dtxp}\\
    \dod{}{t} \hat{p}^2
    &= -\m^2\left(\hat{p}\hat{x} + \hat{x}\hat{p}\right) - \frac{\lambda}{6}\left(\hat{p}\hat{x}^3 + \hat{x}^3\hat{p}\right). 
    \label{eq:dtpp}
\end{align}
\end{subequations}
Averaging equations (\ref{eq:dtxx}) -- (\ref{eq:dtpp}) over the Gaussian state with the Wigner function (\ref{eq:wignerfunc}), applying Wick's theorem, and subtracting the disconnected contributions, the remaining equations of motion are obtained as
\begin{subequations}
\begin{align}
    \dod{}{t} \sigma_{xx}
    &= 2 \sigma_{xp}, \label{eq:eom3}\\
    \dod{}{t} \sigma_{xp}
    &=\sigma_{pp} -\m^2\sigma_{xx} - \frac{\lambda}{2}\sigma_{xx}\left(X^2 + \sigma_{xx}\right) \nonumber\\
    &=\sigma_{pp} - \sigma_{xx}\, \mathcal{C}\bigl(X,\sigma_{xx}\bigr), \label{eq:eom4}\\
    \dod{}{t} \sigma_{pp}
    &=-2\m^2\sigma_{xp} -\lambda\sigma_{xp}\bigl(X^2 + \sigma_{xx}\bigr) \nonumber\\
    &=-2\sigma_{xp}\, \mathcal{C}\bigl(X,\sigma_{xx}\bigr), \label{eq:eom5}
\end{align}
\end{subequations}
where we have introduced the curvature of the potential
\begin{equation}
    \mathcal{C}\bigl(X,\sigma_{xx}\bigr) = \m^2 + \frac{\lambda}{2}\left(X^2 + \sigma_{xx}\right).
\end{equation}
The equations of motion (\ref{eq:eom1}), (\ref{eq:eom2}) and (\ref{eq:eom3}) -- (\ref{eq:eom5}) can be integrated numerically using a symplectic leapfrog algorithm, as described in Appendix \ref{sec:leapfrog_algorithm}, to obtain a complete description of the Gaussian state at any time.

We conclude this section with some general remarks on the formal structure of the GSA which will be particularly relevant for the systematic construction of a thermal equilibrium state in Section~\ref{ho-thermal-equi-state} below.
\begin{itemize}
    \item
    In general, we call a (possibly mixed) state $\hat{\rho}$ \emph{Gaussian}, if its Wigner transform \eqref{wignerTrafo} has the form of a Gaussian probability distribution \eqref{eq:wignerfunc} for some  $X,P,\sigma_{xx},\sigma_{xp},\sigma_{pp}$, with obvious generalization to an arbitrary number of degrees of freedom, where the $\sigma $'s are replaced by the covariance matrix $\Sigma$.
    \item
    Note that 
    a Gaussian state $\hat{\rho}_G$ defined in this way is not necessarily a pure state. This can be seen most easily by calculating the von Neumann entropy~\cite{Buividovich:2018scl}. One observes that the purity of a Gaussian state is related to the determinant $\det \Sigma$ of the covariance matrix $\Sigma$. This determinant is a product of pairs of symplectic eigenvalues $f_k$, one per bosonic degree of freedom (dof). With $N_\mathrm{dof} $ of them, Heisenberg's uncertainty relation then implies that 
 \[ \det\Sigma = \prod_k^{N_\mathrm{dof}} f_k^2 \ge \Big(\frac{1}{4}\Big)^{N_\mathrm{dof}}\! .\]
     On the other hand, the von Neumann entropy  
     vanishes and the Gaussian state is pure, if and only if  $f_k = 1/2$ for all dof's. For a single degree of freedom as above, for example, we have 
     \[
     f= \sqrt{\sigma_{xx}\sigma_{pp} - \sigma^2_{xp}} \] 
     and restricting to pure Gaussian states therefore defines a non-linear subset $\mathcal{G} \subset \mathcal{H} = L^2(\mathbb{R})$ of the full Hilbert space (of square-integrable functions) which can be parametrized by a 4-dimensional manifold with $X,P \in (-\infty,\infty)$, $\sigma_{xx}$, $\sigma_{pp} \in (0,\infty)$ and constrained by $\sigma_{xx}\sigma_{pp} \geq 1/4$. To ensure that the von Neumann entropy vanishes, the off-diagonal variance is then fixed up to a sign by 
     $\sigma_{xp} = \pm\sqrt{\sigma_{xx}\sigma_{pp} - 1/4}$. 
  \item
    To describe more general Gaussian states~$\hat{\rho}_G$, we again follow Ref.~\cite{Buividovich:2018scl} and define the set of mixed Gaussian states in terms of those density operators that can be written as mixtures of the pure Gaussian states in $\mathcal G$,
    \begin{align}
        \hat{\rho}_G = \SumInt_{\psi \in \mathcal{G}} p(\psi) |\psi\rangle \langle\psi| , \label{mixedStateGSA}
    \end{align}
    with probabilities $p(\psi)$ that are Gaussian likewise. 
\end{itemize}

\subsection{Caldeira-Leggett model}
\label{Caldeira-Leggett-Sec}

In order to study the dynamical properties of thermal equilibrium states, we introduce a coupling between the system, our anharmonic oscillator we will also refer to as the {\em particle}, and the environment consisting of an ensemble of harmonic oscillators, which models a fixed-temperature heat bath.
Such a model in the canonical operator formalism as well as in the functional path integral formulation (after Feynman and Vernon) has been discussed frequently in the literature, see for example Refs.~\cite{Grabert:1988yt,Ford:1988zz,Hakim:1985zz,Weiderpass_2020,Lampo_2016,Stauber_2006,Massignan_2015,Boyanovsky_2017_Jul,Boyanovsky_2017_Dec,Ingold_2002,Weiss_2012}.
Often Born and Markov approximations are employed, leading to master equations which are easy to solve but not generally accurate.
Exact solutions also have been obtained analytically~\cite{Boyanovsky_2017_Dec}.
However, this is unfortunately not the case for the anharmonic oscillator.

The total Hamiltonian under consideration then consists of those of the system $S$, the heat bath $B$ together with their interactions $I$, and reads~\cite{Ingold_2002,Weiss_2012}
\begin{subequations}
\begin{align}
    \hat H &= \hat H_S + \hat H_B + \hat H_I, \\
    \hat H_S &= \frac{\hat p^2}{2} + \frac{\m^2}{2}\hat x^2 + \frac{\lambda}{4!}\hat x^4, \label{system-hamiltonian}\\
    \hat H_B &= \sum_s \Big(  \frac{ \hat \pi_s^2}{2} + \frac{\omega_s^2}{2}\hat \varphi_s^2 \Big), \label{bath-hamiltonian}\\
    \hat H_I &= -\hat x\sum_s g_s \hat \varphi_s + \hat x^2\sum_s \frac{g_s^2}{2\omega_s^2}, \label{interaction-hamiltonian}
\end{align}
\end{subequations}
where $\hat \varphi_s, \hat \pi_s$ denote the coordinate and the conjugate momentum of the heat-bath oscillator with index $s$, $\omega_s$ is its eigenfrequency, and $g_s$ the coupling constant of its linear coupling to the coordinate $x$.
The quadratic term in $\hat H_I$ serves to exactly compensate the bath-induced (negative) shift of the oscillator frequency squared,
\begin{align}
    \Delta\m^2 = \sum_s \frac{g_s^2}{\omega_s^2}, \label{bathMassShift}
\end{align}
that would otherwise arise. This guarantees that $\m^2$ is the physically measured natural frequency of the non-interacting system oscillator with damping. 
Completing the square, we can then write the interaction-plus-bath part as
\begin{align}
    \hat H_B + \hat H_I = \sum_s \bigg( \frac{\hat \pi_s^2}{2} + \frac{\omega_s^2}{2} \left( \hat \varphi_s - \frac{g_s}{\omega_s^2} \hat x \right)^2 \bigg).
\end{align}

\subsubsection{Heisenberg-Langevin equations of motion}
\label{section:equations of motion}

Introducing a spectral function to describe the ensemble of bath modes by 
\begin{align}
    J(\omega) = \pi \sum_s \frac{g_s^2}{\omega_s} \left( \delta(\omega - \omega_s) - \delta(\omega + \omega_s) \right) , \label{bath-spectral-density}
\end{align}
 which corresponds to a positive-definite spectral density, one can eliminate the heat bath from the full set of  Heisenberg equations to derive the quantum equations of motion of the Caldeira-Leggett model for a general heat bath described by $J(\omega)$, see e.g. Refs.~\cite{Ingold_2002,Weiss_2012},
\begin{subequations}
\begin{align}
    \dod{}{t} \hat{x}(t) &= \hat{p}(t), \label{qeom-x} \\
    \dod{}{t} \hat{p}(t) &= - \int_0^t \dif t'\,\gamma(t-t') \hat{p}(t') - V'(\hat{x}(t)) + \hat{\xi}(t), \label{qeom-p}
\end{align}
\end{subequations}
with an operator-valued fluctuating force
\begin{align}
    \hat{\xi}(t) &= \sum_s g_s \bigg[ \Big( \hat{\varphi}_s(0) - \frac{g_s}{\omega_s^2} \hat{x}(0) \Big) \cos( \omega_s t ) \nonumber\\
    &\hskip 4cm + \frac{ \hat{\pi}_s(0) }{ \omega_s } \sin(\omega_s t) \bigg] \nonumber \\
    &\equiv \hat{\eta}(t) - \gamma(t) \hat{x}(0). \label{xi-op}
\end{align}
Here $ \hat{\eta}(t) $ is defined such that it only acts in the bath's Hilbert space,
and we have furthermore introduced the damping kernel
\begin{align}
    \gamma( t ) = 2 \int_0^\infty \frac{\dif \omega}{2\pi} \frac{ J(\omega) }{ \omega } \cos(\omega t) . \label{gamma-ohmic}
\end{align}
For the simplest case of an Ohmic bath with damping constant $ \gamma $ and a sharp cutoff at $ \omega=\Lambda $, we have
\begin{equation}
    J_\Lambda(\omega) = 2\gamma \omega\, \Theta( \Lambda-|\omega| ), \label{ohmic-spectral-density}
\end{equation} and thus
\begin{align}
    \gamma_\text{sharp}(t) = \frac{2 \gamma\Lambda}{\pi} \, \frac{ \sin( \Lambda t ) }{ \Lambda t }
    \xrightarrow{ \Lambda\to\infty } 2\gamma \delta(t). \label{gamma-sharp}
\end{align}
Note that the `transient term' $ -\gamma(t) \hat{x}(0) $ in Eq.~(\ref{xi-op}) corresponds to a sudden initial shift in the thermal distribution when the particle is connected to the bath~\cite{Weiss_2012} at $t=0$. It can therefore safely be omitted in all calculations as long as we are interested in times $ \Lambda t \gg 1 $, or analogously for $\omega_\text{D} t \gg 1$ with $\gamma(t) = \gamma \omega_\text D \exp\{-\omega_\text{D} t\} $ in the Drude model for the bath \cite{Ingold_2002}.

Equations (\ref{qeom-x}), (\ref{qeom-p}) together are commonly referred to as the Heisenberg-Langevin equations (HLEs) in the literature \cite{Gardiner_2000,Boyanovsky_2017_Dec}. They generalize the classical Langevin equations (\ref{class-eom-x}), (\ref{class-eom-p}) which describe the dissipative dynamics of the expectation values of position and momentum. The corresponding HLEs, on the other hand, encode the full dynamics of the quantum-mechanical operators in the Heisenberg picture \cite{Gardiner_2000}. A common approximation is to replace the operators in the HLEs by their expectation values and the quantum noise by a classical colored-noise source, which results in the so-called \emph{quasiclassical} Langevin equations \cite{Weiss_2012}. While these can provide a reasonable description of nearly harmonic systems \cite{Eckern_1990,Koch_1981}, other important general features of the HLEs such as the Heisenberg uncertainty principle for the operators $\hat x$ and $\hat p$ are lost. Finally, the classical Langevin equations are obtained in the Markovian limit in which all memory effects are disregarded and the noise becomes local in time (white noise limit).

As a first step towards a solution of the Heisenberg-Langevin equations in the GSA, we assume that at the initial time $t=0$ the bath is in equilibrium, while the system particle is described by some density matrix $\hat{\rho}_S$ with Gaussian Wigner function  as defined in Eq.~\eqref{eq:wignerfunc}, i.e.~we assume that the \emph{initial} state is given by
\begin{align}
    \hat{\rho}_0 \equiv \hat{\rho}(t_0) = \hat{\rho}_S \otimes \hat{\rho}_B .
\end{align}
We defer the precise specification of the equilibrium density matrix $\hat{\rho}_B$ of the heat bath to Section \ref{ho-thermal-equi-state} where we discuss some further subtleties associated with its GSA description.

Our goal here is to formulate a Langevin-type equation in the Gaussian state formalism, which preserves more of the features of the HLE than the quasiclassical or classical approach, thus having extended applicability.
First of all, we consider the most general Gaussian Wigner function $W$ that describes the entire system of oscillator and heat bath,
\begin{equation}
    W( \vec{\zeta},t ) = \mathcal{N}\exp\left\{ -\frac{1}{2} ( \vec{\zeta}-\vec{Z}(t) )^T \Sigma^{-1}(t) ( \vec{\zeta}-\vec{Z}(t) ) \right\},
    \label{general-gauss-state}
\end{equation}
where the vector $ \vec{\zeta} = (x, p, ..., \varphi_s, \pi_s, ... ) \in \Gamma $ denotes a point in the full phase space $\Gamma$ of the system, $ \vec{Z}(t) = ( X(t),P(t), ..., \Phi_s(t),\Pi_s(t), ... ) $ are the expectation values of the corresponding Heisenberg operators, and
\begin{equation}
    \Sigma =
    \left(\begin{array}{cc|cccc}
        \sigma_{xx} & \sigma_{xp}                   & ...    & \sigma_{x \varphi_s} & \sigma_{x \pi_s} & ... \\
        \sigma_{xp} & \sigma_{pp}                   & ...    & \sigma_{p \varphi_s} & \sigma_{p \pi_s} & ... \\ \hline
        \vdots               & \vdots               & \ddots & \\
        \sigma_{\varphi_s x} & \sigma_{\varphi_s p} &        & \sigma_{\varphi_s \varphi_s} & \sigma_{\varphi_s \pi_s}\\
        \sigma_{\pi_s x}     & \sigma_{\pi_s p}     &        & \sigma_{\varphi_s \pi_s}     & \sigma_{\pi_s \pi_s}\\
        \vdots               & \vdots               &        &                              &                          & \ddots
    \end{array}\right)
\end{equation}
represents the corresponding covariance matrix. Note that $\Sigma$ will in general also contain cross-correlations such as $ \sigma_{x\varphi_s} $ between the oscillator particle and the heat bath, which encode quantum entanglement.

In order to translate~(\ref{qeom-x}),~(\ref{qeom-p}) into corresponding equations of motion within the GSA~\cite{Buividovich:2017kfk}, in a systematic and thermodynamically consistent way, we adopt the following procedure:

\renewcommand{\labelenumi}{(\roman{enumi})}

\begin{enumerate}
\item 
Average the Heisenberg equations of motion (\ref{qeom-x}), (\ref{qeom-p}) together with (\ref{xi-op}) using the general Gaussian state~(\ref{general-gauss-state}) and write down the resulting equations for all dynamic quantities contained in $\vec{Z}$ and $\Sigma$. Evaluate correlation functions using Wick's theorem.

\item
Integrate out the heat-bath degrees of freedom by solving the equations of motion obtained from step~(i) for the bath oscillator coordinates $ \Phi_s, \Pi_s $ and the system-bath cross-correlations $ \sigma_{x \varphi_s}, \sigma_{p \varphi_s}, \sigma_{x \pi_s}, \sigma_{p \pi_s} $, under the assumption that the full solution $ X(t), P(t), \sigma_{xx}(t), \sigma_{xp}(t), \sigma_{pp}(t) $ is already known.
Use initial conditions of the form
\begin{align}
\hspace*{.9cm}    \Sigma(0) &=     \label{initial-sigma}\\[6pt]
    &\hskip -.4cm \left(\begin{array}{cc|cccc}
        \sigma_{xx}(0) & \sigma_{xp}(0)        & ...    & 0 & 0 & ... \\
        \sigma_{xp}(0) & \sigma_{pp}(0)        & ...    & 0 & 0 & ... \\ \hline
        \vdots         & \vdots                  & \ddots & \\
        0 & 0 &          &\sigma_{\varphi_s\varphi_s}(0) & 0          \\
        0 & 0 &          & 0          & \sigma_{\pi_s\pi_s}(0)  \\
        \vdots           & \vdots               &        &                              &                          & \ddots
   \end{array}\right),\nonumber
\end{align}
to obtain the 2-point correlators, but leave the initial phase-space variables  $ \Phi_s(0), \Pi_s(0) $ of the heat bath arbitrary.
Insert these formal solutions into the five remaining equations of motion for the Gaussian particle.

\item
Notice that the remaining five equations of motion for the particle only depend on the initial conditions of the bath, $ \Phi_s(0), \Pi_s(0), \Sigma_\text{bath}(0) $, where $\Sigma_\text{bath}(0)$ (the lower right corner of $\Sigma(0)$ in (\ref{initial-sigma})) describes all the 2-point correlators of the bath oscillators' phase-space variables.
Therefore, to introduce thermal fluctuations, all we need to do is
distribute the initial expectation values of the bath oscillators according to a thermal distribution (to be specified below), e.g.~via a fluctuating force term.
\end{enumerate}
Since the quantum average over the Gaussian Wigner function \eqref{general-gauss-state} in step (i) according to the prescription in \eqref{expWig} is entirely different from the thermal average over the initial conditions in step (iii), we denote the former by $\langle \cdots \rangle$ and the latter by $\langle \cdots \rangle_\beta$ with an additional index~$\beta =1/T$.

Before we can investigate the solution of the equations of motion from steps (i) and (ii), we first have to take a closer look at how to represent the `thermal initial state of the bath' which enters in step (iii) in the Gaussian approximation, i.e.~how are $ \Phi_s(0)$, $\Pi_s(0)$ and  $\Sigma_\text{bath}(0) $  distributed in a thermal-equilibrium state at a given temperature $T$.
This is surprisingly subtle, however, as discussed in the next subsection.

\subsubsection{Gaussian thermal equilibrium state}
\label{ho-thermal-equi-state}

Since the bath is described as an ensemble of harmonic oscillators in Gaussian mixed states, we still have to specify what precisely we mean by a `thermal equilibrium state' for a single bath oscillator in the harmonic case with  $ \lambda=0 $, where the Gaussian approximation becomes exact.
Constructing a thermal ensemble of Gaussian states that models a quantum canonical state at temperature $T$, along the lines described for mixed Gaussian states at the end of Section~\ref{gauss-state-approx}, is not entirely trivial. We have to express the full quantum-mechanical mixed thermal state of a harmonic oscillator, described by the density operator (with spectral representation in the energy-eigenstates $|n\rangle$ of the harmonic oscillator),
\begin{equation}
\hat{\rho}_\mathrm{\tiny HO} = e^{-\beta \hat{H}}/Z = Z^{-1} \sum_n \,  \e^{-\beta\m (n+1/2)} \ket n\bra n ,
\label{canonical-density-HO}
\end{equation} 
that acts on the full Hilbert space~$\mathcal{H}$,
in terms of a $\hat{\rho}_G$ acting on $\mathcal{G} \subset \mathcal{H}$ 
to describe a \emph{Gaussian} mixed state, by a density operator of the form \eqref{mixedStateGSA}.
In general, a mixed thermal state describing a canonical ensemble is not of this form, so this involves an approximation.
For the harmonic oscillator, however, it can be easily verified from the definition in \eqref{wignerTrafo} that the density operator~$\hat{\rho}_\mathrm{HO}$ in Eq.~\eqref{canonical-density-HO} does have a Gaussian Wigner function, which is given by
\begin{align}
    w_\mathrm{\tiny HO}(x,p)  =& \smallfrac{2}{F(\m)}\, \e^{-\frac{p^2 + \scriptsize \m^2 x^2}{\m F(\m)}} \, ,
    \label{WF-HO}\\
    &\mbox{where}
    \;\; F(\omega ) = \coth{\frac{\beta\omega}{2}} \nonumber
\end{align}
is its thermal distribution function. To express
the canonical equilibrium ensemble at temperature $T$, represented by the mixed-state density operator $\hat{\rho}_\mathrm{\tiny HO} $ in Eq.~(\ref{canonical-density-HO}), as a mixed Gaussian state of the form \eqref{mixedStateGSA},
we now use our classical phase-space variables $X = \langle\hat x\rangle_\alpha $ and $P =\langle \hat p\rangle_\alpha  $, here restricted again to the expectation values of $\hat x$ and $\hat p$ in coherent states
\begin{equation}
    |\alpha\rangle = e^{-|\alpha|^2/2} \sum_{n=0}^\infty \frac{\alpha^n}{\sqrt{n!}}\, |n\rangle . \label{defCohState}
\end{equation}
These states are characterized by the complex variable $\alpha $ whose real and imaginary parts are given by $X$ and $P$, 
\begin{equation}
    \alpha=\frac{1}{\sqrt{2\m}}\Big( \m  X + {\im} P \Big) . \label{coh-state-id}
\end{equation}
We therefore denote these coherent states simply by the two real phase-space variables in the following, i.e. 
\begin{equation}
    |X,P\rangle \equiv |\alpha\rangle .
\end{equation}
The coherent states of the harmonic oscillator have mini\-mal uncertainty with
\begin{subequations}
\begin{align}
    \sigma^0_{xx} &\equiv \langle \hat x^2\rangle_\alpha -\langle \hat x \rangle^2_\alpha = \frac{1}{2\m} , \\
    \sigma^0_{pp} &\equiv \langle \hat p^2\rangle_\alpha -\langle \hat p \rangle^2_\alpha = \frac{\m}{2} , \\
    \sigma^0_{xp} &\equiv \frac{1}{2} \langle \hat x\hat p + \hat p \hat x\rangle_\alpha -\langle \hat x \rangle_\alpha \langle\hat p \rangle_\alpha = 0 .
\end{align}
\end{subequations}
On the other hand, for the mixed equilibrium state in the canonical ensemble from (\ref{canonical-density-HO}) at temperature $T=1/\beta$, with $\langle \hat O \rangle_\beta = \Tr \hat O \hat \rho_\text{HO}  $, one has $\langle\hat x\rangle_\beta = 0 $ and $\langle \hat p\rangle_\beta =0  $, and the full variances $\sigma^{(\beta)}$ in the thermal state are readily computed (or read off from \eqref{WF-HO}) as,
\begin{subequations}
\begin{align}
    \sigma_{xx}^{(\beta)} &= \langle \hat x^2\rangle_\beta  = \frac{1}{\m} \Big( n_\text B(\m) +\frac{1}{2}  \Big) , \label{thermal-xx}\\
    \sigma_{pp}^{(\beta)} &= \langle \hat p^2\rangle_\beta = \m \Big( n_\text B(\m) +\frac{1}{2}  \Big)  , \label{thermal-pp}\\
    \sigma_{xp}^{(\beta)} &= \frac{1}{2} \langle \hat x\hat p + \hat p \hat x\rangle_\beta = 0 ,
    \label{thermal-xp}
\end{align}
\end{subequations}
where $n_\text B(\m) = 1/(\exp(\beta \m)-1)$ is the Bose-Einstein distribution.
We therefore see here explicitly that
the full thermal widths $\sigma_{xx}^{(\beta)}$ and $\sigma_{pp}^{(\beta)} $ of the oscillator's phase-space variables can be split into purely thermal or `classical' parts $\sigma_{xx}^c$, $\sigma_{pp}^c$ plus the purely `quantum' parts $\sigma_{xx}^0 $, $\sigma_{pp}^0$ from minimal uncertainty, as noted in Ref.~\cite{Buividovich:2018scl}, i.e.
\begin{equation}
    \sigma_{xx}^{(\beta)} = \sigma_{xx}^c + \sigma_{xx}^0, \quad
    \sigma_{pp}^{(\beta)} = \sigma_{pp}^c + \sigma_{pp}^0 , \label{sigmaSplit}
\end{equation}
with 
\begin{align}
    \sigma_{xx}^c = n_\text B(\m)/\m,
    \quad
    \sigma_{pp}^c = \m\, n_\text B(\m) .
    \label{classical-thermal=widths}
\end{align}
The minimal-uncertainty variances are already included in each coherent pure state. To define 
a mixed Gaussian state $\hat{\rho}_G$ with the variances of the thermal equilibrium ensemble, we therefore only include the classical thermal widths of (\ref{classical-thermal=widths}) in the incoherent sum~\cite{Buividovich:2018scl}, defining
\begin{align}
    \hat{\rho}_G &= \label{rho_ho_gsa}\\
& \tilde{\mathcal{N}} \int \dif X\dif P\,\exp\left\{ -\frac{X^2}{2\sigma_{xx}^c} -\frac{P^2}{2\sigma_{pp}^c} \right\} |X,P\rangle \langle X,P| , \nonumber 
\end{align}
with normalization factor $\tilde{\mathcal{N}}$, ensuring that $\Tr\,\hat{\rho}_G=1$, and which is clearly of the form \eqref{mixedStateGSA}.
The index $G$ here emphasizes that such a mixed Gaussian state is  in general \emph{not} equal to the mixed thermal quantum state $\hat{\rho}_Q$ in the canonical ensemble.\footnote{While every coherent state is Gaussian, the converse is not true. 
There are pure states that are Gaussian, by the definition in Eq.~\eqref{eq:wignerfunc}, which do \emph{not} correspond to any coherent state \eqref{defCohState} and are therefore not contained in the incoherent sum \eqref{rho_ho_gsa}.}
For the harmonic oscillator in Eq.~(\ref{canonical-density-HO}), however, we have $\hat\rho_\mathrm{\tiny HO} = \hat\rho_G$.

One may directly verify that such a thermal state is indeed a stationary solution of the harmonic Gaussian equations of motion (\ref{eq:eom1}), (\ref{eq:eom2}) and~(\ref{eq:eom3}) -- (\ref{eq:eom5}). In general, it maps Gaussian states to Gaussian states in the Hilbert space, as required. 
This finishes the discussion of the Gaussian thermal state of a single harmonic oscillator, and we now continue to model the entire system consisting of our \emph{anharmonic} oscillator coupled to an ensemble of harmonic oscillators, in mixed Gaussian states with thermal variances as described here, reintroducing the heat-bath index $s$.

\subsubsection{Heisenberg-Langevin equations in the GSA}

We now turn to the form of the Heisenberg-Langevin equations of motion \eqref{qeom-x}, \eqref{qeom-p} evaluated in the Gaussian approximation.
Steps (i) and (ii) from the quantum equations of motion (\ref{qeom-x}), (\ref{qeom-p}) with the Ohmic bath~(\ref{ohmic-spectral-density}),  on time scales $\Lambda t \gg 1$ for a sufficiently large cutoff $\Lambda $, cf.~\eqref{gamma-sharp}, lead to
\begin{subequations}
\begin{align}
    \dod{}{t} X &= P,  \label{gauss-eom-x} \\
    \dod{}{t} P &= -\left(\m^2 + \frac{\lambda}{2} \sigma_{xx}\right) X -\frac{\lambda}{6}X^3 -\gamma P + \xi(t),
    \label{gauss-eom-p}
\end{align}
\end{subequations}
for the center~$(X,P)$.
The derivation is the same as the one for the classical Langevin equations of motion, except for the application of Wick's theorem to the 3-point correlator $\langle \hat{x}^3(t)\rangle$.
The fluctuating force term $\xi(t)$ is given by the expectation value of the quantum stochastic force $\hat{\xi}(t)$ from \eqref{xi-op}, where the transient initial shift vanishes if the expectation value of $\hat x(0)$ does,
\begin{align}
    \xi(t) &\equiv \langle \hat{\xi}(t) \rangle   =  \langle \hat{\eta}(t) \rangle - \gamma(t)  \langle \hat{x}(0) \rangle  \label{xi-op-averaged} \\ 
    &= 
    \sum_s g_s \left[ \Phi_s(0)  \cos( \omega_s t ) + \frac{ \Pi_s(0) }{ \omega_s } \sin(\omega_s t) \right] . \nonumber
\end{align}
It thus only depends on the initial conditions of the bath oscillators' phase-space variables $ \Phi_s(0), \Pi_s(0) $. Although no-longer operator valued and hence classical, this GSA noise $\xi(t) $ is colored in general, however, as we will discuss in Subsection~\ref{gauss-colored-noise} below.

For the Gaussian widths, the analogous averaging of step~(i) leads to
\begin{subequations}
\begin{align}
    \dod{}{t} \sigma_{xx} &= 2 \sigma_{xp} ,
    \label{sigma-xx-fluc}
    \\
    \dod{}{t} \sigma_{xp} &= \sigma_{pp} - \sigma_{xx} \mathcal{C}(X,\sigma_{xx}) - \gamma \sigma_{xp}     \label{sigma-xp-fluc} \\
    &\hskip 1cm +\llangle \hat x(t)\hat \eta(t) \rrangle - 2\gamma \delta(t) 
   \sigma_{xx}(0)   ,
    \nonumber \\
    \dod{}{t} \sigma_{pp} &= - 2 \sigma_{xp} \mathcal{C}(X,\sigma_{xx}) - 2\gamma  \sigma_{pp} 
    \label{sigma-pp-fluc} \\
    &\hskip 1cm + 2 \llangle \hat p(t)\hat \eta(t) \rrangle -  4\gamma\delta(t) \sigma_{xp}(0) . 
\nonumber 
\end{align}
\end{subequations}
Here we have already assumed the Ohmic heat bath $J_\Lambda(\omega)$  in the limit $\Lambda\to\infty$ where the memory integrals collapse, cf.~\eqref{gamma-sharp}. The initial delta-functions can then safely be neglected. 

On the other hand, the irreducible correlators of the oscillator particle's position and momentum  with  the fluctuating force operator, $\llangle \hat x(t)\hat \eta(t) \rrangle $ and $ \llangle \hat p(t)\hat \eta(t) \rrangle $, both contain a logarithmically divergent contribution $d(\Lambda) \sim\ln(\Lambda/\mathcal C_0 )$ where $\mathcal C_0 $ is the $t=0$ initial value of the time-dependent curvature of the potential,
\begin{equation}
 \mathcal C(t) \equiv \mathcal C(X,\sigma_{xx}) =  \m^2 + \frac{\lambda}{2} (X^2(t) + \sigma_{xx}(t)).   \label{curve}
\end{equation}
We will show in Appendix~\ref{gauss-heat-bath-eoms-derivation} that this divergence can be absorbed by a formally infinite but time-independent shift of the particle's momentum width $\sigma_{pp} \to \sigma_{pp} -  d(\Lambda)  $ in such a way that it cancels from both equations \eqref{sigma-xp-fluc} and \eqref{sigma-pp-fluc} which then describe the time dependence of the ultraviolet-finite part of $\sigma_{pp}$, together with finite $\sigma_{xp}$ and $\sigma_{xx}$ at all times. Note that the divergence of $\sigma_{pp} $ with $\Lambda \to \infty$ is an unavoidable effect of the unrealistic assumption of an Ohmic bath without ultraviolet (UV) cutoff. It can be interpreted as corresponding to  the bath
continuously `measuring' the position of the particle with arbitrarily high `resolution' without UV cutoff for $\Lambda\to\infty $ \cite{Hakim:1985zz,Weiss_2012}.

Finally, for the evaluation of the irreducible correlators  $\llangle \hat x(t)\hat \eta(t) \rrangle $ and $ \llangle \hat p(t)\hat \eta(t) \rrangle $ according to rules (ii) and (iii), we need to make an additional adiabatic approximation as explained explicitly also in Appendix~\ref{gauss-heat-bath-eoms-derivation}.  In this adiabatic approximation we assume that we can average the curvature $\mathcal C$ of the potential in \eqref{curve} over time scales that are large compared to the relaxation time of the heat bath. The heat-bath oscillators are then considered as the fast degrees of freedom that can adjust to slow changes in the curvature $\mathcal C (t)$.  This adiabatic approximation 
then yields for the Gaussian widths,
\begin{subequations}
\begin{align}
    \dod{}{t} \sigma_{xx} &= 2 \sigma_{xp}, \label{eom-xx-ad} \\
    \label{eom-xp-ad}
    \dod{}{t} \sigma_{xp} &= \sigma_{pp} -  \mathcal{C}(t)\sigma_{xx}  - \gamma \sigma_{xp}   \\
    &\hskip 1.8cm + \mathcal C(t) F\big(\mathcal C(t)\big)  -\Delta_K\big(\mathcal C(t)\big) ,  \nonumber \\
    \dod{}{t} \sigma_{pp} &= - 2 \mathcal{C}(t) \sigma_{xp}  - 2 \gamma \sigma_{pp}  +2\gamma \Delta_K\big(\mathcal C(t)\big) , \label{eom-pp-ad} 
\end{align}
\end{subequations}
where for the Ohmic bath $J_\Lambda(\omega) $ with $\Lambda\to\infty $ the  fluctuating force $ \llangle \hat x(t)\hat \eta(t) \rrangle $ after ultraviolet subtraction yields $\mathcal C F - \Delta_K $. 
The first contribution is obtained from  
\begin{align}
F(\mathcal C) = \frac{1}{2\omega_\mathcal C } \bigg(\frac{1}{2}+ \frac{1}{\pi}   \arctan\Big(\frac{\omega_\mathcal C^2 -\gamma^2/4}{\gamma\omega_\mathcal C}\Big)\bigg) 
\end{align}
upon inserting $\mathcal C(t)$ together with an equally slowly varying frequency 
\begin{equation}
\omega_\mathcal C(t) \equiv \sqrt{\mathcal C(t)  -\gamma^2/4} > 0 ,
\end{equation}
assuming weak damping. The second contribution to the fluctuating force is the ultraviolet subtracted one, given by 
\begin{align}
    \Delta_K\big(\mathcal C(t)\big) &=  \big(\mathcal C(t) -\gamma^2/2\big)  \, F\big(\mathcal C(t)\big) \\
    &\hskip .4cm -  \big(\mathcal C_0 -\gamma^2/2\big)  \, F\big(\mathcal C_0\big) - \frac{\gamma}{2\pi} \, \ln\frac{\mathcal C(t)}{\mathcal C_0} \, , \nonumber
\end{align}
which determines the relevant (ultraviolet-finite) part of the fluctuating force  $ \llangle \hat p(t)\hat \eta(t) \rrangle $ on the ultraviolet subtracted $\sigma_{pp}$ as well, and which vanishes when $\mathcal C \equiv \mathcal C_0 $ is used at all times in the static limit, see Appendix~\ref{gauss-heat-bath-eoms-derivation}.

Together with Eqs.~\eqref{gauss-eom-x} and \eqref{gauss-eom-p}, these equations for the evolution of the Gaussian widths, from Eqs.~\eqref{eom-xx-ad} -- \eqref{eom-pp-ad}, constitute the full set of equations of motion for our particle in the anharmonic potential, and in contact with an external heat bath, i.e.~the Heisenberg-Langevin equations within the GSA in our adiabatic approximation. 

The static approximation is obtained from  Eqs.~\eqref{eom-xx-ad} -- \eqref{eom-pp-ad} by simply using the time-independent $\mathcal C = \mathcal C_0$, for which we have $\Delta_K(\mathcal C_0) = 0$. In this case, the equations for 
 the widths, cf.~\eqref{eom-xx-appendix} -- \eqref{eom-pp-appendix}, can be solved independently of those for the coordinates $X$, $P$, see Appendix~\ref{gauss-heat-bath-eoms-derivation}.
 The asymptotic behavior of the solution, cf.~Eq.~\eqref{statsol-app},  uniquely fixes
 \begin{equation}
    \sigma_{xx}(t) \to  F(\mathcal C_0), \;\; \mbox{for} \;\; t\to\infty .
    \label{statsol}
 \end{equation}
Therefore, with the static solution, there are only two effects remaining of the GSA in comparison with the classical time-evolution. These are (a) a time-dependent shift of the oscillator frequency in Eq.~\eqref{gauss-eom-p}, 
\begin{equation}
    \m^2 \to \m^2 + \frac{\lambda}{2} \, F(\mathcal C_0),  
\end{equation}
with the stationary value \eqref{statsol} of $\sigma_{xx}(t) $ for sufficiently late times,
and (b) a modified {\em colored noise} $\xi(t)$, which we will specify in Subsection~\ref{gauss-colored-noise} below.
At very high temperatures, the frequency shift is negligible, and the noise becomes {\em white} again, such that the correct classical limit is guaranteed to be recovered in this static approximation.

In the other direction, to go beyond the adiabatic approximation, one could in principle include  the feedback of the time dependence of the  curvature $\mathcal C(t)$ in \eqref{curve} on the off-diagonal variances between system particle and heat-bath oscillators via post-adiabatic corrections in the spirit of time-dependent perturbation theory, in the future, as briefly outlined in Appendix \ref{gauss-heat-bath-eoms-derivation} as well. For the results presented below, we have either used the static approximation with constant $\mathcal C_0$, cf.~Eqs.~\eqref{eom-xx-appendix} -- \eqref{eom-pp-appendix}, or the adiabatic approximation in Eqs.~\eqref{eom-xx-ad} -- \eqref{eom-pp-ad}, for comparison.

\subsubsection{Initial Conditions}

In the adiabatic approximation the thermal equilibrium value $\mathcal{C}_0$ is an inital condition that has to be know beforehand. It can be obtained by maximizing the von Neumann entropy at fixed energy.

For a mixed Gaussian state of our single bosonic degree of freedom the von Neumann entropy ${S=-\mathrm{Tr}(\hat{\rho}\ln\hat{\rho})}$ can be written in terms of the symplectic eigenvalue 
\[
f = \sqrt{\sigma_{xx}\sigma_{pp} - \sigma_{xp}^2} \]
of the correlation matrix
\begin{equation}
    \Sigma = \begin{pmatrix} \sigma_{xx} & \sigma_{xp} \\ \sigma_{xp} & \sigma_{pp} \end{pmatrix}
    \label{eq:correlation_matrix} ,
\end{equation}
which is related to the pair of eigenvalues $\lambda_\pm = \pm \im f $ of $\Sigma \Omega$, where $\Omega $
is the symplectic matrix for the canonically conjugate variables $x$ and $p$.
Thermal equilibrium implies $\sigma_{xp}=0$ and we thus have $f=\sqrt{\sigma_{xx}\sigma_{pp}}$. 

The von Neumann entropy can then be written as
\begin{equation}
    S = \left(f+\tfrac{1}{2}\right)\ln\left(f+\tfrac{1}{2}\right) - \left(f-\tfrac{1}{2}\right)\ln\left(f-\tfrac{1}{2}\right).
    \label{eq:entropy}
\end{equation}
Because we also have $X=0$ and $P=0$ in thermal equilibrium, the expectation value of energy in the Gaussian state reduces to
\begin{equation}
E=\frac{1}{2}\sigma_{pp}
+ \frac{\omega_0^2}{2}\sigma_{xx}
+ \frac{\lambda}{8}\sigma_{xx}^2.
\end{equation}
We can thus express $\sigma_{pp}$ in terms of $E$ and $\sigma_{xx}$ and write
\begin{equation}
    f^2 = 2E\sigma_{xx} -\omega_0^2\sigma_{xx}^2-\frac{\lambda}{4}\sigma_{xx}^3.
\end{equation}
Because the entropy (\ref{eq:entropy}) increases  monotonically with $f$, it reaches its maximum when $f$ does, which is the case when $\partial f^2 / \partial \sigma_{xx}=0$. This yields
\begin{align}
    E &= \omega_0^2 \sigma_{xx} +\frac{3\lambda}{8}\sigma_{xx}^2, \\
    f^2 &= \left(\omega_0^2 + \frac{\lambda}{2}\sigma_{xx}\right)\sigma_{xx}^2.
\end{align}
The temperature $T$ is now introduced using 
\begin{equation}
    T = \frac{\partial E}{\partial S} = \frac{\partial E}{\partial f}\left(\frac{\partial S}{\partial f}\right)^{-1}
    = \frac{\partial E}{\partial \sigma_{xx}}\left(\frac{\partial f}{\partial \sigma_{xx}}\right)^{-1}\left(\frac{\partial S}{\partial f}\right)^{-1} \!\!.
    \nonumber
\end{equation}
Working out the  partial derivatives with respect to $f$ and $\sigma_{xx}$, we thus  obtain
\begin{align}
 T   &= 
    \sqrt{ \omega_0^2+\frac{\lambda}{2}\sigma_{xx} } 
    \, \left(
    \ln \frac{\sigma_{xx}\sqrt{\omega_0^2 +\frac{\lambda}{2}\sigma_{xx}}+\frac{1}{2}}{\sigma_{xx}\sqrt{\omega_0^2 +\frac{\lambda}{2}\sigma_{xx}}-\frac{1}{2}}\right)^{-1}
      \nonumber \\
    &=
    \sqrt{ \mathcal{C}_0 } \, \left(
    \ln \frac{4\left(\mathcal{C}_0 - \omega_0^2\right)\sqrt{\mathcal{C}_0}+ \lambda }{4\left(\mathcal{C}_0- \omega_0^2\right)\sqrt{\mathcal{C}_0}- \lambda }\right)^{-1}, 
   \label{eq:C0_of_beta} 
\end{align}
with $\mathcal{C}_0 = \m^2+\frac{\lambda}{2}\sigma_{xx}$
in the interacting case for $\lambda \neq 0$.
Before we start a simulation at a given the temperature, we can therefore calculate $\mathcal{C}_0$ numerically via \eqref{eq:C0_of_beta}. 
In the static approximation this is then fixed, and so is $\sigma_{xx}$ 
in the HLEs \eqref{gauss-eom-x} and \eqref{gauss-eom-p} for $X$ and $P$.

The underlying initial conditions for the widths $\sigma_{\varphi_s\varphi_s}$ and $ \sigma_{\pi_s\pi_s} $  of the heat-bath oscillators in Eq.~\eqref{initial-sigma}  correspond to the thermal harmonic-oscillator variances Eqs.~\eqref{thermal-xx} -- \eqref{thermal-xp},  with additional off-diagonal couplings $\sigma_{x\varphi_s} $ between system and bath suddenly switched on at $t=0$ as explained in more detail in Appendix \ref{AppA-initial}.

One crucial point left to mention here, however, is that beyond the static approximation, the widths $\sigma_{xx},\sigma_{pp},\sigma_{xp}$ actually \emph{do} evolve non-trivially in time, even in the adiabatic approximation, when $\mathcal C(t)$ is assumed to vary slowly in time. This is because the relaxation time for the widths of the system particle to approach their stationary limits is given by  $1/\gamma $, and this relaxation time is in general not negligible compared to the characteristic time scale $\delta t$ of the variations $\delta\mathcal C(t)$. Assuming, in the adiabatic approximation, that the heat-bath dof's are fast compared to this characteristic time $\delta t$  is totally different from assuming that $1/\gamma $ is. In fact, for small damping $\gamma $ we expect to have $1/\gamma \gg \delta t \gg 2\pi/\omega_s $ for the relevant high frequencies that dominate the Ohmic bath.
We will further comment on this in Section \ref{gauss-extract-sf} below, 
after elaborating on the colored noise needed in either case.

\subsubsection{Colored Noise}
\label{gauss-colored-noise}

Step (iii) in our approach to modelling the heat bath in the GSA by the quantum mechanical expectation value $\xi(t) \equiv \langle\hat\xi(t)\rangle  $ of the stochastic quantum force  from Eq.~(\ref{xi-op-averaged})
requires specifying initial conditions for the thermal correlations of the bath oscillator expectation values  $\Phi_s = \langle\hat\varphi_s\rangle_\alpha $ and  $\Pi_s = \langle \hat\pi_s\rangle_\alpha $ in coherent states. In particular, the discussion leading to (\ref{rho_ho_gsa}) implies that their initial thermal variances are given by the classical variances \eqref{classical-thermal=widths} for each heat-bath oscillator, 
\begin{align}
 \langle\Phi_s(0) \Phi_{s'}(0)\rangle_\beta  &=  \sigma_{\varphi_s\varphi_{s'}}^c(0) = \delta_{ss'} \,  n_{\text B}(\omega_s)/\omega_s\, , \nonumber \\ 
  \langle\Pi_s(0) \Pi_{s'}(0)\rangle_\beta &=   \sigma_{\pi_s\pi_{s'}}^c(0) = \delta_{ss'}\, \omega_s\, n_{\text B}(\omega_s)\, , \nonumber\\
  \langle\Phi_s(0) \Pi_{s'}(0)\rangle_\beta  &=     \sigma_{\varphi_s\pi_{s'}}^c(0) =0\, .
\end{align}
Eq.~(\ref{xi-op-averaged}) then yields
\begin{align}
     \langle \xi(t) \xi(t') \rangle_\beta = \sum_s \frac{g_s^2}{\omega_s} \, n_{\text B}(\omega_s)  \, \cos\big( \omega_s (t-t')\big) \, .
\end{align}
With the definition of the spectral density of the heat bath in Eq.~\eqref{bath-spectral-density} 
we can thus finally represent the unequal-time correlations of $\xi(t)$ in the form,
\begin{align}
    \langle \xi(t) \xi(t') \rangle_\beta = 
\int_0^\infty \frac{\dif\omega}{\pi} J( \omega )\, n_{\text B}( \omega ) \cos(\omega(t-t'))
\end{align}
for an arbitrary spectral distribution $J(\omega)$ of oscillators in the bath.
For the Ohmic bath~(\ref{ohmic-spectral-density}) in the limit $ \Lambda \to \infty $ the integral can be solved analytically, yielding
\begin{align}
    \langle \xi(t) \xi(t') \rangle_\beta &=  \gamma T \, \bigg( -\frac{ \pi T }{   \sinh^2( \pi T (t-t') ) }  \label{time_correlation}\\
    &\hskip 3cm + \frac{1 }{ \pi T (t-t')^2 }\bigg) \nonumber
\end{align}
in the time domain, where the terms in brackets approach $2 \delta(t-t')$ for $T\to\infty $, i.e.~the classical-statistical limit with a noise term $\xi(t)$ in the equations of motion which is Gaussian and local in time~\cite{Schweitzer:2020noq}. For numerical purposes and completeness, in frequency domain the noise in \eqref{time_correlation} corresponds to
\begin{align}
    \langle | \xi(\omega) |^2 \rangle_\beta &= \langle \xi(-\omega) \xi(\omega) \rangle_\beta \\
    &= \gamma\omega \Big( \coth\Big(\frac{\omega}{2T}\Big) - \sign{\omega} \Big)  \nonumber \\
    &= 2\gamma \omega \, n_{\text B}(\omega) \, , \hspace{0.5cm} \text{for $\omega > 0$}\, ,\nonumber
\end{align}
where the white-noise limit is recovered from the classical Rayleigh-Jeans law with $ n_{\text B}(\omega) \to T/\omega $ at high temperatures.
Therefore, $\xi(t)$ represents a `colored' noise term with a Gaussian autocorrelation, but with the quantum contribution subtracted, in the sense that $ \xi(t) $ vanishes identically at $ T=0 $.
This is due to the fact that in our framework, zero-point fluctuations are already naturally taken into account by the Gaussian widths, $\sigma_{xx}^0, \sigma_{xp}^0, \sigma_{pp}^0$ and therefore do not contribute to the fluctuation of the mean coordinate and momentum.

\subsubsection{Extracting the Gaussian spectral function}
\label{gauss-extract-sf}
In order to extend the definition of the (anti-symmetrized) classical-statistical spectral function in Eq.~(\ref{eq:simple_spectr_func}) to the correct quantum spectral function that respects 
the fluctuation-dissipation relation with the colored-noise distribution of the heat bath in the GSA, first note that the FDR~\eqref{eq:FDT} must be replaced by
\begin{align}
    \im F(\omega)
    &= \frac{K(\omega)}{2\gamma \omega} \,  2\pi \im\, \rho(\omega),
\end{align}
for a general heat-bath kernel $ K(\omega)=\langle |\xi(\omega)|^2 \rangle_\beta $.
Now we use the definition \eqref{class-rho-FDR} of the classical-statistical spectral function to arrive at a balance-type equation
\begin{align}
    \rho(\omega) = \frac{2\gamma T}{K(\omega)} \, \rho_\text{c}(\omega)  = 
    \frac{T}{\omega n_{\text B}(\omega)} \,
    \rho_\text{c}(\omega),
    \label{gauss_sf}
\end{align}
which allows us to use the classical-statistical extraction scheme for the spectral function, i.e.~using Eq.~(\ref{eq:simple_spectr_func}) in the GSA as well, and just rescale the result to obtain the corresponding quantum spectral function.

Having defined the extraction scheme of the spectral function in the GSA, we can further elaborate on the problems that occur when we keep the full time evolution of the Gaussian widths \eqref{eom-xx-ad} -- \eqref{eom-pp-ad}.
Since the GSA is only an approximation to the infinite hierarchy of the time evolution of higher moments of the (in principle exact) Wigner quasi-probability distribution $w(x,p)$ in phase space, to quadratic order~\cite{Buividovich:2017kfk}, one can no-longer guarantee that classical and quantum dynamics are strictly divided into the evolution of the expectation values $X,P$, and the second order moments $\sigma_{xx},\sigma_{pp},\sigma_{xp}$, respectively.
Therefore some classical contributions to the spectral function are also contained in the time evolution of the widths.
Extracting the spectral function na\"ively as in \eqref{gauss_sf} is therefore not sufficient, if the quantum corrections are highly non-Gaussian by themselves, when the full time evolution of the widths is included.
In this case, one would actually need some improved procedure to correctly extract these non-Gaussian contributions contained in the second-order unequal-time correlators.
This undesirable effect is explicitly demonstrated for sufficiently large anharmonicity $\lambda$  in Section~\ref{results}.

In summary, a complete simulation eventually comprises the following steps:
\begin{enumerate}
    \item Generate a random realization of the stochastic force $\xi(t)$ distributed according to the colored noise correlations in (\ref{time_correlation}).
    \item Integrate the Gaussian equations of motion (\ref{gauss-eom-x}), \eqref{gauss-eom-p} and \eqref{eom-xx-ad} -- \eqref{eom-pp-ad} numerically.
    \item Calculate the spectral function from the particular time history of the expectation values
    via (\ref{gauss_sf}).
    \item Finally, repeat steps 1 through 3 and average the spectral function over all realizations of the stochastic force to obtain the thermal equilibrium spectral function.
\end{enumerate}
More details of steps (i)  and (ii)  are described in Appendices \ref{sec:colored_noise_synthesis} and \ref{sec:leapfrog_algorithm}, respectively.

\section{Real-time FRG}
\label{rt-frg}

As another possibility for real-time calculations we have also explored the functional renormalization group (FRG)~\cite{Wetterich:1992yh,Berges:2000ew,Gies:2006wv} on the closed time path~\cite{Berges:2012ty}.
In the FRG one aims to compute the \emph{effective action} $\Gamma$, which is obtained from the generating functional $Z$ of the theory~\cite{Wipf:2013vp}.
Assuming the so-called {\em effective average action} $\Gamma_\Lambda$ is known at some initial energy-scale $\Lambda$ in the ultraviolet (UV),
the essential idea is to construct the full $\Gamma$ step-by-step by `interpolating' $\Gamma_k$ from the microscopic UV-action $\Gamma_\Lambda$ to the macroscopic action $\Gamma$ in the infrared (IR). This is achieved by introducing a parameter $k$ corresponding to the energy scale down to which the theory is valid. Using an auxiliary device called the \emph{regulator} $R_k$, one suppresses both thermal and quantum fluctuations of modes with $\omega  < k$.
In particular, assuming that at sufficiently high energies (and momenta, where $S \gg \hbar$) the theory behaves classically, one may start the interpolation in the UV with $\Gamma_{\Lambda} = S$, the classical bare action,
see Eq. \eqref{bareKeldyshAction} below.
(This can be shown rigorously by a saddle-point approximation, where the regulator term acts as a $\delta$-functional in the limit $k \to \infty$ \cite{Huelsmann:2020xcy}.)
One then computes the \emph{flow} of the effective average action through `theory space'~\cite{Gies:2006wv} until it reaches the full macroscopic effective action $\Gamma_{k\to 0} = \Gamma$ of the theory.
Ideally, one would solve the flow from $k\to\infty$ to $k=0$, but in almost all practical applications it is sufficient to start with some large but finite UV cutoff $k=\Lambda$, and to stop at some sufficiently small finite value $k=k_\text{IR}$ in the IR.

\subsection{Flow Equation}

The FRG flow is determined by the equation that was derived by Wetterich~\cite{Wetterich:1992yh,Berges:2000ew} in the imaginary-time formalism, but here formulated on the closed time path (CTP)~\cite{Berges:2012ty},
\begin{align}
	\label{eqn:wetterich-equation}
	\partial_k \Gamma_k[\phi] = \frac{\im}{2} \mbox{Tr}\, \left\{ (\partial_k R_k) \circ \left( \Gamma_k^{(2)}[\phi] + R_k \right)^{-1} \right\} ,
\end{align}
for a real scalar field $\phi^T(x) = (\phi^c(x),\phi^q(x))$ in $D=d+1$ spacetime dimensions and Keldysh space, where we adopt the convention of Ref.~\cite{Tan:2021zid} and define the Keldysh rotation to be measure-preserving, i.e.
\begin{align}
    \phi^c = \frac{1}{\sqrt{2}} \left( \phi^+ + \phi^- \right),
    \quad
    \phi^q = \frac{1}{\sqrt{2}} \left( \phi^+ - \phi^- \right),
\end{align}
for the classical (or average) field $\phi^c$ and the quantum (or response) field $\phi^q$, and vice-versa
\begin{align}
    \phi^+ = \frac{1}{\sqrt{2}} \left( \phi^c + \phi^q \right),
    \quad
    \phi^- = \frac{1}{\sqrt{2}} \left( \phi^c - \phi^q \right)
\end{align}
for the inverse transformation.
By $\phi^\pm$ we denote the field components that live on the forward $(+)$ and backward $(-)$ parts of the closed time path.
This convention has the convenient property that for a symmetric potential $V(\phi)$ the potential term $ -V(\phi^+) + V(\phi^-) $ on the closed time path is invariant under the interchange $\phi^c \leftrightarrow \phi^q$, i.e. does not distinguish between classical and quantum field components.
For a detailed introduction to the formalism and a derivation of the flow equation see for example Refs.~\cite{Berges:2012ty} and \cite{Huelsmann:2020xcy}.

The flow in Eq.~\eqref{eqn:wetterich-equation} on the right hand side includes the regulator $R_k$ and $\Gamma_k^{(2)}[\phi]$, the Hessian of $\Gamma_k[\phi]$,
\begin{align}
    \Gamma_k^{(2)}[\phi] =
    \begin{pmatrix}
        \Gamma^{cc}_k[\phi] & \Gamma^{cq}_k[\phi] \\
        \Gamma^{qc}_k[\phi] & \Gamma^{qq}_k[\phi]
    \end{pmatrix},
\end{align}
where we have already use a notation defined below, in Eq.~\eqref{fncDiffNotation}, for brevity.
$R_k$ and $\Gamma^{(2)}_k$ both have the form of self-energies on the closed time path, i.e.\ $2 \times 2$ matrices in Keldysh $(c, q)$ space.
Therefore, $\circ$ denotes $2 \times 2$ matrix multiplication, and the trace also implies integration over adjacent coordinates.
We denote functional derivatives of the effective average action as
\begin{align}
    \Gamma_k^{\alpha_1 ... \alpha_n}[\phi^c, \phi^q](x_1, ..., x_n) = \frac{\delta^n \Gamma_k[\phi^c, \phi^q]}{\delta \phi^{\alpha_1}(x_1) ... \delta \phi^{\alpha_n}(x_n)} , \label{fncDiffNotation}
\end{align}
where Greek indices from the beginning of the alphabet denote CTP indices, $\alpha_1, ..., \alpha_n \in \{c,q\}$.
Correspondingly, $\Gamma_k^{(n)}[\phi]$ denotes the tensor of rank $n$ containing all functional derivatives w.r.t. the classical and quantum fields.
The full scale-dependent propagator $G_k$ in front of the background field expectation value $ \phi $ is given by
\begin{equation}
    - G_k^{-1}[ \phi ] =  R_k + \Gamma^{(2)}_k[ \phi ] \label{frgPropagatorDef}
\end{equation}
in compact matrix notation, or explicitly
\begin{subequations}
\begin{align}
    G_k^{\tilde{K}}[\phi_{0,k}] &= 0, \label{propAnom} \\
    G_k^R[\phi_{0,k}] &= -\left( \Gamma_k^{qc}[\phi_{0,k}] + R_k^R \right)^{-1} , \label{propRet} \\
    G_k^A[\phi_{0,k}] &= -\left( \Gamma_k^{cq}[\phi_{0,k}] + R_k^A \right)^{-1} \label{propAdv} , \\
    G_k^K[\phi_{0,k}] &= G_k^R \circ \left( \Gamma_k^{qq}[\phi_{0,k}] + R_k^K \right) \circ G_k^A \label{propKel}
\end{align}
\end{subequations}
at the scale-dependent minimum $\phi_{0,k}$, which satisfies the quantum equations of motion $\delta \Gamma[ \phi_{0,k} ] = 0$.
The superscripts ${A}$, ${R}$, ${K}$ and ${\tilde{K}}$ denote the \emph{advanced}, \emph{retarded}, \emph{Keldysh} and \emph{anomalous} components, respectively.
For the expressions in an arbitrary background field configuration $\phi(x)$ see for example Ref.~\cite{Berges:2012ty}.

Being the imaginary part of the retarded propagator, the spectral function $\rho_k(\omega)$ can be computed from the retarded 2-point function in the usual way,
\begin{align}
    \rho_k(\omega) = \frac{1}{\pi} \frac{\Im\, \Gamma_k^{qc}(\omega)}{\left( \Re\, \Gamma_k^{qc}(\omega) \right)^2 +\left( \Im\, \Gamma_k^{qc}(\omega) \right)^2}
    \, . \label{frgSpectralFnc}
\end{align}

\subsection{Causal regulators}

The need of respecting causality in the process of constructing regulators for the real-time FRG was already mentioned in Ref.~\cite{Duclut:2016jct}, where the advantages of using such a causal regulator were manifest in the results for dynamical critical exponents.
In this section, we construct such a causal regulator step-by-step in the $0+1$ dimensional case. The construction is based on considering the regulator term as an additional  self-energy with the causal matrix structure of the Keldysh action and can be generalized to field theories in higher dimensions. For our single real degree of freedom $\phi^T(t) = (\phi^c(t),\phi^q(t))$, we first add a term to the Keldysh action of the form, 
\begin{equation} 
  \Delta S_k[\phi] = \frac{1}{2} \int_{-\infty}^\infty \!\!\dif t \int_{-\infty}^\infty \!\!\dif t'\, \phi^T(t) R_k(t,t') \phi(t')
  \, ,\label{regulator1}
\end{equation}
with
\begin{align}
    R_k(t,t') =
    \begin{pmatrix}
    R_k^{\tilde{K}}(t,t')&R_k^A(t,t')\\
    R_k^R(t,t')&R_k^K(t,t')
    \end{pmatrix}
\end{align}
being a $2 \times 2$-matrix on the CTP.
We call a regulator \emph{causal} if it maintains the causal structure of the corresponding Keldysh action, as e.g.~defined in Chapter 2.7 of Ref.~\cite{Kamenev:2011}.
Most importantly, a causal regulator has to ensure that the retarded (advanced) propagator has to stay retarded (advanced) during the flow.

Here, we propose a new quite general construction of causal regulators which proceeds as follows:
We assume that the bilinear term $\Delta S_k[\phi]$ really is the result of a coupling of the field $\phi$ to an ensemble of Gaussian degrees of freedom which have been integrated out, e.g.\ see Chapter 3.2~of Ref.~\cite{Kamenev:2011}.  In fact, the other way round, any term quadratic in the fields can be linearized via Hubbard-Stratonovich transformation to replace it by a linear coupling to the Gaussian Hubbard fields representing the ensemble. Therefore, the assumption that  $\Delta S_k[\phi]$ is the result of integrating an ensemble of Gaussian degrees of freedom should still be fairly general. As such, after Fourier transform, the corresponding retarded/advanced components can readily be written as spectral integrals
\begin{equation}
R_{k}^{R/A}(\omega) = - \int_0^\infty  \frac{\dif\omega'}{2\pi} \frac{2\omega' J_k(\omega') }{(\omega\pm \im\varepsilon)^2 -\omega'^2}\, , \label{spectralRep}
\end{equation}
so that the spectral density $J_k(\omega)\ge 0$ for $\omega > 0$ of the fictitious Gaussian ensemble is given by the imaginary part of the retarded bath propagator $D_\mathrm{bath}^R(\omega) $ which is here represented, after the Gaussian integration of the bath, by $D_\mathrm{bath}^R(\omega) = - R_k^R(\omega)$, i.e.
\begin{equation}
    J_k(\omega) \equiv - 2 \, \mbox{Im}\, D_\mathrm{bath}^R(\omega) =  2\,  \mbox{Im}\,  R_k^R(\omega) \, ,
\end{equation}
as can explicitly be checked from (\ref{spectralRep}).
Note that for a single oscillator with frequency $\omega_k$ and coupling $g_k$ in the bath, it is here normalized to $\omega J_k(\omega)= \pi g_k^2  \, \big( \delta(\omega-\omega_k) + \delta(\omega+\omega_k) \big)$, and we generally have $J(-\omega) = - J(\omega)$.
This fixes the retarded/advanced components of the regulator.

Since it is furthermore desirable to keep a system in thermal equilibrium during the flow, if it was in equilibrium initially, one may also require the symmetry 
\begin{align}
\label{reg-thermal-equilib}
\Delta S_k[\mathcal{T}_\beta \phi^c, \mathcal{T}_\beta \phi^q] = \Delta S_k[\phi^c, \phi^q] \, ,
\end{align}
which is a sufficient condition for thermal equilibrium on the closed time path~\cite{Sieberer:2015hba}, where the transformation $\mathcal{T}_\beta$ is defined as
\begin{align*}
	\mathcal{T}_\beta 
	\begin{pmatrix} \phi^c(\omega) \\
	\phi^q(\omega) 
	\end{pmatrix}
	= \begin{pmatrix}
	    \cosh\left( \frac{\beta \omega}{2} \right) & 
	    - \sinh\left( \frac{\beta \omega}{2} \right)\\
	    -\sinh\left( \frac{\beta \omega}{2} \right) &
	    \cosh\left( \frac{\beta \omega}{2} \right)
	\end{pmatrix}  
	\begin{pmatrix} 
	\phi^c(-\omega) \\ \phi^q(-\omega) 
	\end{pmatrix} 
\end{align*}
for our real degree of freedom $\phi(t)$.
We can now  insert the general ansatz for the regulator term and check that this condition is satisfied, if 
\begin{align}
R_k^{\tilde K}(\omega) &\equiv 0\, , \;\; \mbox{and}\\ 
R_k^K(\omega) &= \coth(\beta\omega/2) \big(R_k^R(\omega)-R_k^A(\omega)\big) \nonumber\\
&=  \coth(\beta\omega/2) \, \im J_k(\omega) \, .    
\nonumber
\end{align}
This implies that our fictitious Gaussian ensemble should then represent a heat bath at the same temperature $T$ as that of the equilibrium system that is being regulated.

\begin{figure}[t]
\centering
\includegraphics[width=0.9\linewidth]{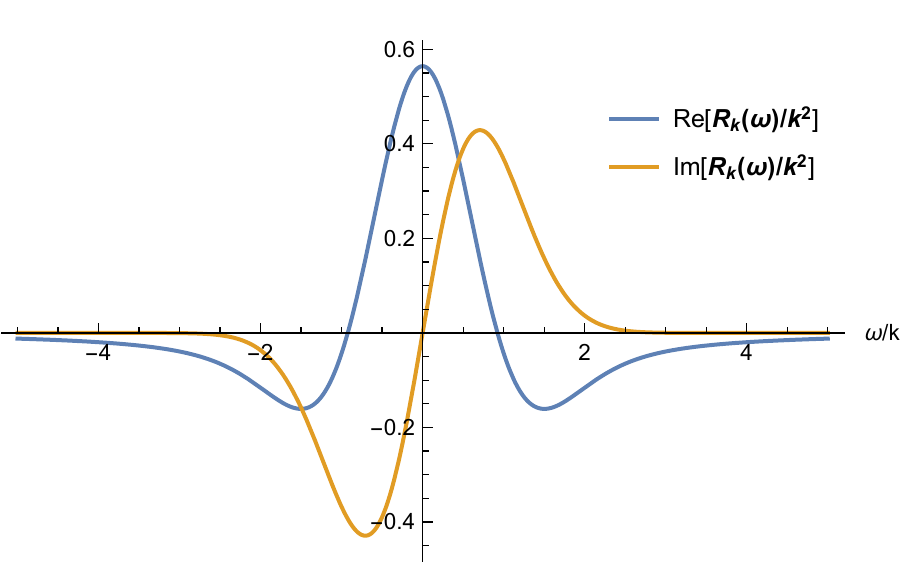}
\caption{Real and imaginary parts of $R_k^R(\omega)$ in units of $k^2$ for the bath with the analytic cutoff in Eq.~(\ref{AnaBath}). \label{plot4}}
\end{figure}

Our construction of causal regulators for thermal equilibrium systems  therefore starts at specifying suitable FRG scale $k$ dependent spectral densities $J_k(\omega) $ to represent some fictitious heat bath.
Here, we specifically use an analytic spectral density of the form
\begin{equation}
J_k(\omega) = k \omega \, \exp\big\{- \omega^2/k^2 \big\} \,. \label{AnaBath}
\end{equation}
which has the desired regulating property by giving rise to an Ohmic bath with damping constant $\gamma_k = k/2$ in the IR, while it rapidly goes to zero towards the UV,
\begin{equation}
    J_k(\omega) \to \begin{cases}
    k\omega & \text{for $\omega \ll k$} \\
    0 & \text{for $\omega \gg k$}
    \end{cases} .
\end{equation}
The suppression of low-frequency modes is thus realized by the coupling to the fictitious heat bath with an FRG-scale dependent damping constant which starts out large, of the order of $\Lambda$  in the UV, and vanishes with $k\to 0$ towards the IR.   
Inserting~(\ref{AnaBath}) into the spectral representation~(\ref{spectralRep}) explicitly yields for the corresponding causal heat-bath regulator in our example,
\begin{align}
	\label{eqn:hb-reg-analytic}
		R_{\text{HB},k}^\mathrm{R/A}(\omega)
		&= \\
&\hskip -1cm		\frac{1}{\sqrt{\pi}} \left( \frac{1}{2} k^2 - k \omega F\left( \frac{\omega}{k} \right) \right) \pm \frac{ \im k\omega}{2} \exp \left\{ -\frac{\omega^2}{k^2} \right\},\nonumber
\end{align}
with the Dawson function
\begin{align}
	F(x) \equiv \frac{2}{\sqrt{\pi}} \mathrm{e}^{-x^2} \mathrm{erfi}(x) = \mathrm{e}^{-x^2} \int_0^x \dif t\,\mathrm{e}^{t^2} .
\end{align}
Real and imaginary part of the retarded regulator are shown in Figure~\ref{plot4}.
Note, however, that its feature of representing a causal Green function unavoidably entails,
\begin{equation}
  \int_{-\infty}^\infty \frac{d\omega}{2\pi} \big( R_k^R(\omega) + R_k^A(\omega) \big)
  = \int_{0}^\infty \frac{d\omega}{2\pi} \, 4 \, \mbox{Re} \, R_k^R(\omega) = 0\, .
\end{equation}
In particular, whenever we construct a causal regulator in this way, the real part must have a zero crossing. It must start out negative for $\omega \gg k$ in the UV, and with a single zero crossing as here, it thus turns positive for $\omega \ll k$ in the IR.

Note that a \emph{positive} real part at $\omega = 0$  corresponds to a \emph{negative} mass-squared shift which can affect the propagator poles for sufficiently large $k$ in an uncontrollable and unphysical way, a feature that is at least unpleasant for an FRG regulator. 
The easiest way out here seems to add a frequency independent counter-term that has the form of a Callan-Symanzik regulator. Such a frequency independent term is certainly causal, and it can offset $R_{\text{HB},k}^{R/A}(\omega)$ by a $k$-dependent constant such that the real part of the resulting regulator stays strictly negative during the flow. We therefore introduce an additional positive parameter $\alpha $ to define
\begin{align}
    R_{k}^{R/A}(\omega)
    \equiv 
    R_{\text{HB},k}^{R/A}(\omega) - \alpha k^2 \label{csShift} . 
\end{align}
Since the absolute value of the real part of $R_{\text{HB},k}^{R/A}(\omega)$ is monotonically decreasing for $\omega > 0$, we restrict $\alpha$ from below by requiring $ \alpha > R_{\text{HB},k}^{R/A}(0) / k^2 $.

\begin{figure*}[t]
\includegraphics[width=\textwidth]{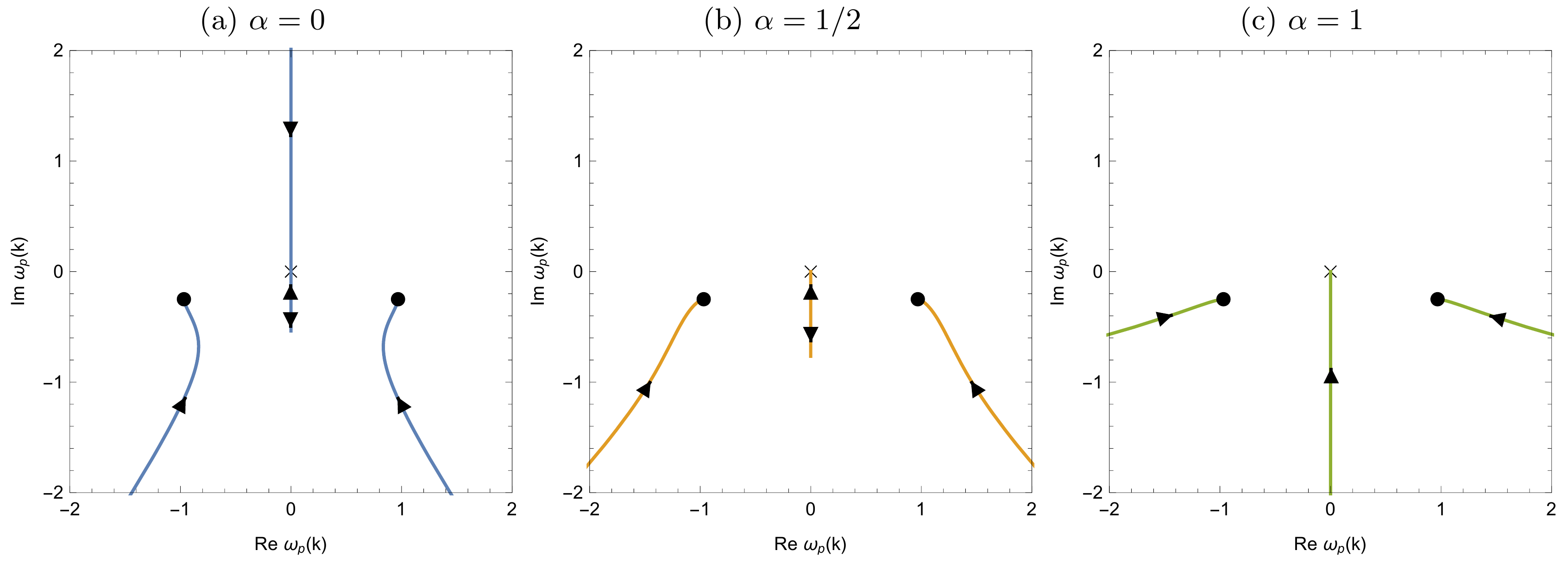}
    \caption{Trajectories of the poles $\omega_p(k)$ of the retarded propagator \eqref{examplePropagator} with the Drude regulator \eqref{drudeReg} in the complex plane, moving with the FRG scale $k$ from the UV towards the IR, with  $\m =1$, $\gamma=0.5$ and three different values of $\alpha$. The black dots mark the quasi-particle poles \eqref{quasiParticlePoles} of the propagator at $k=0$, where the regulator vanishes. 
    They move with $k$ in the lower half-plane as indicated by the arrows, but never cross the real axis.
    The crosses at the origin mark the points where the regulator-induced third poles disappear with $k\to 0 $ in the IR. Their flows are indicated by arrows as well. For $\alpha < 1/2$ this relaxational regulator pole violates causality: It starts in the upper half-plane and only 
    crosses at a finite value of the FRG scale $k$ during the flow  into the lower where it has a turning point (at the end of the line) before it moves up again and disappears in the origin. 
    } \label{regulated-propagator-poles}
\end{figure*}

The  effect of this can be understood in two equivalent ways: First, 
as every bath ensemble, cf.~Eq.~(\ref{bathMassShift}), the regulator heat-bath 
necessarily induces a negative shift of the system particle's mass or oscillator frequency squared, by  $- \Delta \omega_\text{HB}^2(k)$, which here is $k$-dependent as our regulator bath is and can explicitly be written in the form
\begin{equation}
    \Delta \omega_\text{HB}^2(k) = \int_0^\infty \frac{\dif\omega}{\pi} \, \frac{J_k(\omega)}{\omega}  = \, R_{\text{HB},k}^{R/A}(0)\,.
\end{equation}
With our analytic spectral density (\ref{AnaBath}), for example, we have $\Delta \omega_\text{HB}^2(k) =  k^2/{\sqrt{4\pi}}$, and 
$\alpha$ must be chosen large enough to compensate this mass shift, $\alpha k^2 \geq  \Delta \omega_\text{HB}^2(k)$. This is necessary for the theory to remain causal during the flow in the first place; see the discussion that follows below.
Secondly, in order to regularize all infrared modes we must have a negative real part of the regulator at $\omega = 0$, requiring the strict inequality $\alpha k^2 >  R_{\text{HB},k}^{R/A}(0) $.

To further illustrate how the analytic structure of the propagators is changed by the heat-bath regulator, and how a suitable Callan-Symanzik counter-term can solves the issue, it is constructive to consider a spectral density for the regulator bath based on the Drude model, cf.~Sec.~\ref{Caldeira-Leggett-Sec} and Ref.~\cite{Ingold_2002}, as a simpler alternative
which yields with $2\gamma = \omega_\text{D} = k$, 
\begin{align}
    J_k(\omega) = \frac{\omega k}{1 + (\omega/k)^2} .
\end{align}
Using Eq.~(\ref{spectralRep}) again, we then obtain explicitly, 
\begin{equation}
    R_k^{R/A}(\omega) = \frac{1}{2} \frac{k^2}{1 \mp \im \omega/k} - \alpha k^2, \label{drudeReg}
\end{equation}
where we already included the Callan-Symanzik counter-term for which we expect to require $\alpha > 1/2 = \Delta \omega_\text{HB}^2/k^2 $.
To see how the regulator affects the poles during the flow, we consider an exemplary retarded (advanced) propagator of the form
\begin{align}
    G_k^{R/A}(\omega) = -\frac{1}{\omega^2 \pm \im \gamma \omega - \m^2 + R_k^{R/A}(\omega)} . \label{examplePropagator}
\end{align}
The two are related by the symmetry $G_k^R(\omega) = {G_k^A{}}^*(-\omega)$.
The simplicity of the regulator~(\ref{drudeReg}) allows to derive analytic expressions for the poles $\omega_p = \omega_p(k)$ by solving the cubic equation
\begin{equation}
    \omega_p^2 \pm \im \gamma \omega_p - \m^2 + \frac{k^2}{2(1 \mp \im \omega_p/k)} - \alpha k^2 = 0 .
\end{equation}
To keep the analytic structure intact, when the regulator is switched on, the retarded (advanced) propagator must only have poles in the lower (upper) half plane.
For simplicity, we assume in our illustration here that $\m^2$ and $\gamma$ stay constant during the flow.
The resulting poles $\omega_p(k)$ of the retarded propagator are shown in Figure~\ref{regulated-propagator-poles} for different choices of $\alpha$.
For $k=0$ the physical poles of the retarded propagator are located at
\begin{equation}
    \omega_{p, \pm}(k=0) = -\im\frac{\gamma}{2} \pm \sqrt{\m^2 - \frac{\gamma^2}{4}} \label{quasiParticlePoles}
\end{equation}
in the complex plane, as represented by the black dots in Figure~\ref{regulated-propagator-poles}, corresponding to the expected quasi-particle excitations.
We see that these two poles $\omega_p(k)$ are located symmetrically around the real-part-zero axis. They move upwards with $k$ towards the IR, coming from \emph{lower} values of their imaginary parts, and thus always stay in the lower half plane and never cause problems with causality, even for vanishing Callan-Symanzik term with $\alpha = 0$.

\begin{figure}[t]
    \centering
    \includegraphics[width=0.9\linewidth]{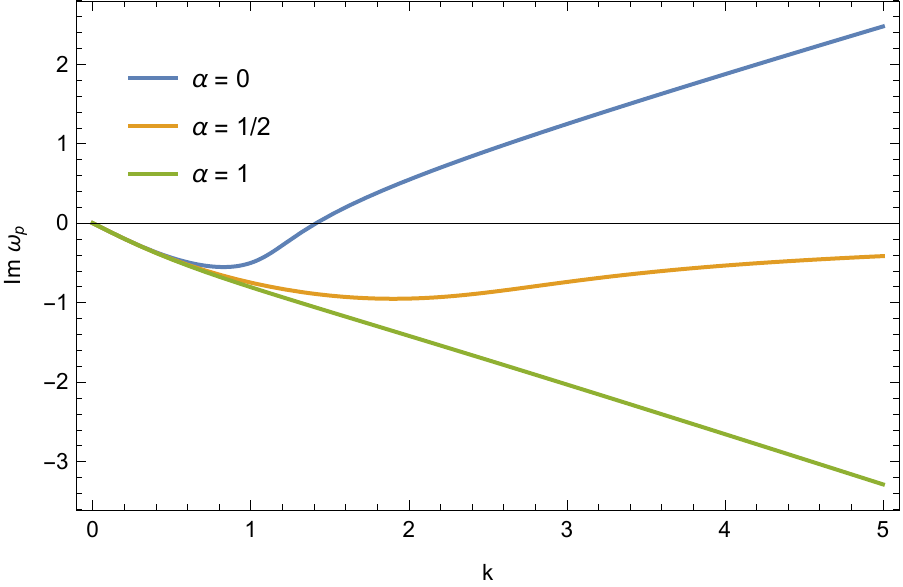}
    \caption{FRG scale $k$ dependence of the imaginary parts $\mathrm{Im}\,\omega_p(k)$ of the regulator-induced relaxational poles in the retarded propagators of Fig.~\ref{regulated-propagator-poles}. 
    Here, $\alpha = 1/2$ is the limiting case, i.e.~causality is always violated at large $k$ for $\alpha <1/2$.} \label{regulated-propagator-poles-im-part}
\end{figure}

For non-vanishing $k>0$ there is a third pole with vanishing real~part moving along the imaginary axis, however, which is entirely due to the regulator.
In the time domain, it represents the purely relaxational contribution \cite{Weiss_2012} here arising from our regulator heat bath.
The crosses in the origins in Fig.~\ref{regulated-propagator-poles} mark the point where it disappears for $k \to 0^+$ in the IR.
The corresponding FRG scale $k$ dependence of its imaginary parts $\mathrm{Im}\,\omega_p(k)$ are shown for the same three values of $\alpha$  in Figure~\ref{regulated-propagator-poles-im-part}.

Following these from the IR towards the UV we see that for $\alpha \leq 1/2$ the imaginary part of the relaxational regulator pole first moves to smaller values. Eventually, however, it turns around to increase again towards the UV, for $\alpha < 1/2$ without bound. In this case it thus always crosses the real axis and moves into the upper half-plane (where a retarded self-energy should be analytic) so that causality is violated by the regulator at finite FRG scale $k$.  
For $\alpha=1/2 $ it turns around as well, but approaches $0^-$ for $k\to\infty $ in the UV and never moves into the upper half-plane. This is the liming case where $\alpha $ is chosen precisely such that the total regulator in \eqref{drudeReg} has a root at $\omega = 0$. 
For larger values $\alpha  > 1/2 $ the imaginary part of the relaxational regulator pole remains strictly negative, and the regulator never violates  causality (here, for $\alpha =1 $ it decreases monotonically, in fact). Its real part has no zero-crossings anymore, and always leads to a \emph{positive} mass/frequency shift, because the Callan-Symanzik counter-term is large enough to over-compensate the negative shift in the squared mass/frequency caused by $\Delta \omega_\text{HB}^2 $ from the heat-bath regulator.

\subsection{Truncation of the Effective Average Action}

Before we discuss the details of our truncation, we first give the explicit expression for the effective average action at the starting point $k=\Lambda $, where it equals the bare Keldysh action, $\Gamma_\Lambda = S$.
For the anharmonic oscillator \eqref{hamiltonian-anh-osc} coupled to an external Ohmic heat bath with damping constant~$\gamma$ the Keldysh action is given by~\cite{Kamenev:2011}
\begin{widetext}
\begin{align}
    \label{bareKeldyshAction}
    S[\phi] &=
    \frac{1}{2} \int_{-\infty}^\infty \frac{d\omega}{2\pi}
    \phi^T(-\omega)
    \begin{pmatrix}
        0 & \omega^2 -\im \gamma \omega - \m^2 \\
        \omega^2 +\im \gamma \omega - \m^2 & 2\im \gamma \omega \coth\left( \frac{\omega}{2T} \right)
    \end{pmatrix}
    \phi(\omega) \\ \nonumber
    &\hspace{2cm} - \frac{2\lambda}{4!} \int_{-\infty}^\infty \hspace{-1em} dt\,\left( \phi^c(t) \phi^c(t) \phi^c(t) \phi^q(t) + \phi^c(t) \phi^q(t) \phi^q(t) \phi^q(t) \right)
    \, ,
\end{align}
\end{widetext}
where we already used the Fourier transform of the quadratic part in the action for convenience. With this general structure of the bare action in mind, one possibility to truncate the Wetterich equation for the full effective average action is a functional Taylor expansion in terms of the 1-PI $n$-point vertex functions~\cite{Berges:2000ew}. 
Using the origin in field space, $\phi_0 = \left( \phi^c_0, \phi^q_0 \right) = 0$, as the expansion point, and the abbreviation
\begin{align*}
    \Gamma_k^{(n)}(x_1,\dots,x_n) &\equiv \Gamma_k^{(n)}[\phi^c = 0, \phi^q = 0](x_1,\dots,x_n) 
\end{align*}
to denote the $n$-point vertex $\Gamma_k^{(n)}$, it reads
\begin{align}
    \Gamma_k[\phi] &=
    \sum_{n=1}^Q \frac{1}{n!} \int d^D x_1 \, \dots \, d^D x_n \, \times \label{generalVertexExp} \\ \nonumber
    &\hspace{2em} \Gamma_k^{\alpha_1 \dots \alpha_n}(x_1, \dots, x_n) \, \phi^{\alpha_1}(x_1) \dots \phi^{\alpha_n}(x_n) , 
\end{align}
such that the $Q$-point vertex is the highest one that is taken into account.

In this work, we use this vertex expansion \eqref{generalVertexExp} up to the order $Q = 6$ in combination with a loop expansion of the corresponding right hand sides of the flow equations.\footnote{In the symmetric phase of $\phi^4$-theories, without spontaneous symmetry breaking or tunneling in quantum mechanics for $\omega_0^2>0$, all odd $n$-point functions vanish identically, and the minimum of the effective average action is fixed at $\phi_0 = (\phi_0^c, \phi_0^q) = 0$, independent of~$k$.}
For the latter we adopt an ordering scheme tied to the vertex expansion in such a way that the highest $n$-point function (with $n=Q$) is assumed to be given by a frequency (and momentum) independent (but scale $k$ dependent) vertex, while successively higher loops are included for the lower $n$-point functions. Specifically, we include $(Q-n)/2$-loop structures
for the  $n$-point functions with $n=2$, \dots $Q$. With $Q=6 $ here, this amounts to taking into account the 2-loop structure of the 2-point functions, the 1-loop structure of the 4-point functions, and the scale-dependent constant 6-point vertex without substructure (corresponding to the order zero in the loop expansion).

Note that our combined vertex and loop-structure expansion at the order $Q=4$ would essentially only yield a mass resp.\ frequency shift of the main peak in the spectral function, corresponding to the $0 \leftrightarrow 1$, $1\leftrightarrow 2$  and higher one-step transitions. In order to describe effects such as collisional broadening or higher resonance excitation frequencies in the spectral function, as e.g.\ corresponding to $ 0 \leftrightarrow 3 $ or $1\leftrightarrow 4 $ transitions, one needs non-local (here meaning frequency dependent) corrections of one-loop form in the 4-point function \cite{Huelsmann:2020xcy}. In the combined scheme we adopt here, with $Q=6$ this implies self energies of two-loop structure, and it then automatically also includes the local but $k$-dependent 6-point vertex which leads to a further quantitative improvement.

To explain the truncation in more detail, we start with the formal expression for the effective average action,
\begin{widetext}
\begin{align}
	\Gamma_k[\phi] &=
	\frac{1}{2} \int_{xx'} \, \phi^T(x)
	\begin{pmatrix}
		0 & \Gamma_k^{(2),A}(x,x') \\
		\Gamma_k^{(2),R}(x,x') & \Gamma_k^{(2),K}(x,x')
	\end{pmatrix}
	 \phi(x')	
	+ \frac{3}{4!} \int_{xx'} \, \phi^\alpha(x)  \phi^\beta(x) \Gamma_k^{\alpha\beta;\beta'\alpha'}(x,x') \phi^{\beta'}(x') \phi^{\alpha'}(x')
	\nonumber \\
	&\qquad - \frac{1}{6!} \int_{x} \left( \frac{3}{2}\mu_k (\phi^c(x))^5 \phi^q(x) + 5 \mu_k (\phi^c(x))^3 (\phi^q(x))^3 + \frac{3}{2} \mu_k \phi^c(x) (\phi^q(x))^5 \right)
	+ O(\phi^8) ,
	 \label{vertexExpansion}
\end{align}
\end{widetext}
where we denote spacetime integrations over~$x$ in short by
\begin{align}
    \int_x \dots &\equiv \int \dif^{\,D}\hspace{-0.7ex}x \, \dots \; .
\end{align}
The first line in \eqref{vertexExpansion} corresponds to the 2-loop exact 2-point function, the second line to the 1-loop exact 4-point function, and the third line to the `0-loop' exact 6-point function.
Their detailed structures are explained in reversed order, starting from the 6-point function, in Subsections~\ref{sct:eff-pot}, \ref{sct:4-point-function}, and  \ref{sct:2-point-function}, respectively.

At this point it is convenient to follow Ref.~\cite{Huelsmann:2020xcy} and to introduce the shorthand notations  $B_k^R, B_k^A, B_k^K$ as follows,
\begin{subequations}
\begin{align}
    B_k^R &= G_k^R \circ \partial_k R_k^R \circ G_k^R, \label{bubbleRet} \\
    B_k^A &= G_k^A \circ \partial_k R_k^A \circ G_k^A, \label{bubbleAdv} \\
    \begin{split}
    B_k^K &= G_k^R \circ \partial_k R_k^K \circ G_k^A + G_k^R \circ \partial_k R_k^R \circ G_k^K \\&\qquad + G_k^K \circ \partial_k R_k^A \circ G_k^A, \label{bubbleKel}
    \end{split}
\end{align}
\end{subequations}
for convolutions of propagators and regulator insertions  with fixed outgoing legs $(c,q)$, $(q,c)$ and $(c,c)$, respectively. These are counterparts of the retarded, advanced and Keldysh propagators $ G_k^R$, $G_k^A$ and $G_k^K $ with all possible ways of inserting one regulator term $\partial_k R_k$ in between.

\subsubsection{6-Point Function and Effective Potential}
\label{sct:eff-pot}

Working out general flow equations for $n$-point couplings by the diagrammatic method can rather cumbersome, especially in the case of the 6-point coupling that we are interested in.
In fact, it is much more convenient to first consider the flow equation for the scale dependent force from the effective potential $V_k(\varphi)$, here defined as a function of the rescaled classical field $\varphi \equiv \phi^c/\sqrt{2}$ by
\begin{align}
	- V_k'(\varphi) \equiv \frac{1}{\sqrt{2}} \, \frac{\delta \Gamma_k[\phi]}{\delta \phi^q(x)} \bigg\rvert_{\substack{\phi^c = \sqrt{2}\,\varphi = \text{const.}\,\\ \phi^q = \,0 \phantom{\sqrt{2}\varphi = \text{const}}}},
	\label{effPotDef}
\end{align}
and generally valid with any ansatz for the effective average action  $\Gamma_k[\phi]$, where the prime denotes ordinary differentiation w.r.t.~the constant classical field variable $\varphi$. 
In a general non-equilibrium situation, the potential might be 
spacetime $x=(x^0, \mathbf{x})$ dependent. This is not the case, however, for a spatially homogeneous system in thermal equilibrium.

The definition in \eqref{effPotDef} is motivated by the form of the potential term $S_V[\phi^c, \phi^q]$ in the bare Keldysh action on the closed time path, given by 
\begin{align}
    &S_V[\phi^c, \phi^q]= \label{keldyshActionPot} \\ \nonumber
    &\quad \int_x \left[ -V\left(\frac{\phi^c + \phi^q}{\sqrt{2}}\right) + V\left(\frac{\phi^c - \phi^q}{\sqrt{2}}\right) \right],
\end{align}
for a general potential $V(\varphi)$ in the Lagrangian of the theory, where $-V'(\varphi)$ is the force in the classical field equations obtained from the $\phi^q\to 0$ limit.

For the spacetime independent vertices we are interested in the flow equations for the Taylor coefficients $V_k^{(n)}(0)$ of the scale dependent effective potential $V_k(\varphi)$ used in \eqref{effPotDef}, when expanded around a possibly likewise scale dependent minimum $\varphi_{0,k}$. In our case, $\varphi_{0,k} \equiv 0$ for all $k$, and we are interested, in particular, in the sixth order Taylor coefficient which is precisely our 6-point coupling constant, $\mu_k \equiv V^{(6)}_k(0)$.

Using Eq.~\eqref{effPotDef} to define the scale dependent effective potential (up to a constant), we can then furthermore relate the desired Taylor coefficients to the spacetime integrals of the corresponding $n$-point functions, via
\begin{align}
    V_k^{(n)}(0)
    = -2^{n/2-1} \int_{x_2 ... x_n} \Gamma^{qc \dots c}_k(x,x_2,\dots,x_n) ,
    \label{effPotNPointCorrespondence}
\end{align}
where we have also used the exchange symmetries of the $n$-point functions,
\begin{align}
    \Gamma_k^{\dots \alpha\beta \dots}(\dots,x,y,\dots) =
    \Gamma_k^{\dots \beta\alpha \dots}(\dots,y,x,\dots) , 
\end{align}
which are  valid specifically for a real scalar field theory~\cite{Rivers:1987hi}, in order to combine equivalent terms in \eqref{effPotNPointCorrespondence}.

To obtain the flow equation for derivative of the effective potential, we thus have to project the Wetterich equation (\ref{eqn:wetterich-equation}) on constant classical field configurations accordingly.
To achieve this, we first take the functional derivative with respect to the \emph{quantum} field $\phi^q(x)$ on both sides of the Wetterich equation, and then set  $\phi^q = 0$, $\phi^c = \text{const.}$ which diagrammatically corresponds to the equation, see e.g.~\cite{Tan:2021zid},
\begin{align}
	\partial_k \frac{\delta \Gamma_k[\phi]}{\delta \phi^q(x)}
	=
	-\frac{\im}{2}
	\;\;
	\begin{tikzpicture}[baseline=-0.5ex]
	 	\centerarc[pgreen](0,0)(-90:90:0.5)
	 	\centerarc[pgreen](0,0)(90:270:0.5)
		\draw[pred] (0,-0.5) -- (0,-0.5-0.42) node[anchor=south west,black] {$x$};
	 	\fill[black] (0,-0.5) circle (0.1);
	 	\draw[black] (-0.1,0.5-0.1) -- (+0.1,0.5+0.1);
	 	\draw[black] (-0.1,0.5+0.1) -- (+0.1,0.5-0.1);
	\end{tikzpicture}
	\;\; .
	\label{effPotFlowDiagram}
\end{align}
Due to the functional derivative, the flow of the zero-point energy $V_k(0)$ is lost, of course. This in generally true on the closed-time path, however,
where the  Keldysh action contains no information on the zero point energy either, because the contributions to a constant  offset  in $V(\varphi)$ 
from the forward and backward branches exactly cancel,  cf.~Eq.~\eqref{keldyshActionPot}.

For the flow of the higher Taylor coefficients of the effective potential, we can set $\phi^q= 0$ in \eqref{effPotFlowDiagram},
but we need to maintain the dependence on the constant classical field $\varphi $. This implies that one would need a partially field-dependent full 3-point vertex function
\[ \Gamma^{qcc}_{\varphi,k}(x,x_2,x_3) \equiv \Gamma^{qcc}_k[\phi^c=\sqrt 2\varphi,\phi^q = 0](x,x_2,x_3) \]
in the loop diagram on the r.h.s.~of \eqref{effPotFlowDiagram} which obeys its own flow equation involving successively higher $n$-point functions as usual. 
At this point we employ a \emph{local-vertex approximation} in the sense that we  neglect possible spacetime dependent substructures but maintain the required field dependence in the local part. Consistency with Eq.~\eqref{effPotNPointCorrespondence} then requires us to use,
\begin{align}
    \Gamma_{\varphi,k}^{qcc}(x,x_2,x_3) =  -\frac{1}{\sqrt{2}}V_k'''(\varphi) \, \delta(x-x_2)\delta(x-x_3) ,  \label{local-vert-app}
\end{align}
where the dependence on the constant classical field $\varphi $ will be needed for the higher-order derivatives later on. Using this local-vertex approximation and the definition in \eqref{bubbleKel}, the flow equation for the effective potential from \eqref{effPotDef} and \eqref{effPotFlowDiagram} becomes,
\begin{align}
	&\partial_k V_k'(\varphi)= \label{effPotFlowEq} -\frac{\im}{4} V_k''' (\varphi) \int \frac{d^D p}{(2\pi)^D} \, B_{\varphi,k}^K(p) .
\end{align}
The notation $B_{\varphi,k}^K$ on the right indicates that analogously field dependent propagators $G_{\varphi,k} \equiv G_k[\sqrt 2\varphi, 0 ]$
are being used inside the loop. To further illustrate the technique needed for further derivatives w.r.t.~the classical field expectation value $\varphi$, first consider for example the fully field-dependent retarded propagator $G_k^R[\phi^c, \phi^q]$.
For the successive Taylor coefficients of the effective potential we can again set the classical field $\phi^c(x) = \sqrt{2}\varphi$ to its constant expectation value and the quantum field to zero, $\phi^q(x) = 0$.
For the partially field dependent $G_{\varphi,k}^R \equiv G_k^R[\sqrt 2\varphi, 0 ]$ this implies that we can relate the ordinary partial derivative w.r.t.~the constant field expectation value $\varphi$ to its functional derivative w.r.t.~the classical field $\phi^c(x) $,
\begin{align}
    \frac{\partial}{\partial \varphi} G_{\varphi,k}^R = \sqrt{2}\int d^D x\, \frac{\delta G_k^R[\phi^c, \phi^q]}{\delta \phi^c(x)} \bigg\rvert_{\substack{\phi^c(x) = \sqrt{2}\,\varphi \\ \phi^q(x) =\, 0 \phantom{~~\varphi} }} . \label{gRderivVarphi}
\end{align}
To evaluate the functional derivative inside the integral, we make use of \eqref{propRet} which, after applying the functional chain and product rules, directly tells us that 
\begin{align}
    \frac{\delta G_k^R[\phi^c, 0]}{\delta \phi^c(x)} = G_k^R \circ \Gamma_k^{qcc}(\cdot,x,\cdot) \circ G_k^R . \label{GRfuncDeriv}
\end{align}
The dots in the arguments of the intermediate 3-point function indicate that the middle argument is fixed by the external differentiation point $x$, whereas the first and the third argument are convoluted with those of the propagators at the outgoing and incoming legs. After this functional derivative we can now insert our local-vertex approximation from Eq.~\eqref{local-vert-app} which then from \eqref{gRderivVarphi} yields,
\begin{align}
    \frac{\partial}{\partial \varphi} G_{\varphi,k}^R = -V_k'''(\varphi) \, G_{\varphi,k}^R \circ G_{\varphi,k}^R . \label{GRderivRel}
\end{align}
This is the final derivative relation for the retarded propagator that we intended to illustrate here.
Together with the FDR, we can now construct arbitrarily high $\varphi$-derivatives of the retarded, advanced and Keldysh propagators in the local-vertex approximation,
\begin{widetext}
\begin{subequations}
\begin{align}
	\partial_\varphi G^{R/A}_{\varphi,k}(\omega) &= - V_k'''(\varphi) \left( G^{R/A}_{\varphi,k}(\omega) \right)^2, \label{eqn:lpa-effpot-flow-helper-1} \\
	\partial_\varphi G^{K}_{\varphi,k}(\omega) &= - V_k'''(\varphi) G^{K}_{\varphi,k}(\omega) \left( G^{R}_{\varphi,k}(\omega) + G^{A}_{\varphi,k}(\omega) \right), \label{eqn:lpa-effpot-flow-helper-2} \\
	\partial_\varphi B^{R/A}_{\varphi,k}(\omega) &= -2 V_k'''(\varphi) B^{R/A}_{\varphi,k}(\omega) G^{R/A}_{\varphi,k}(\omega), \; \text{and} \label{eqn:lpa-effpot-flow-helper-3} \\
	\partial_\varphi B^{K}_{\varphi,k}(\omega) &= - V_k'''(\varphi) \left\{ \left( B^{R}_{\varphi,k}(\omega) + B^{A}_{\varphi,k}(\omega) \right) G^{K}_{\varphi,k}(\omega) + B^{K}_{\varphi,k}(\omega) \left( G^{R}_{\varphi,k}(\omega) + G^{A}_{\varphi,k}(\omega) \right) \right\} \label{eqn:lpa-effpot-flow-helper-4} .
\end{align}
\end{subequations}
\end{widetext}
These relations form the basis for a set of recurrence relations, because 
we can now obtain flow equations for the higher derivatives of the scale dependent effective potential from 
\eqref{effPotFlowEq} by iterating these relations \eqref{eqn:lpa-effpot-flow-helper-1} -- \eqref{eqn:lpa-effpot-flow-helper-4}. Setting the classical field variable $\varphi $ to the expansion point (here at  $\varphi = 0$) afterwards, then finally results in corresponding flow equations for the Taylor coefficients $V^{(n)}_{k}(0) $ which are related to the $n$-point coupling constants via \eqref{effPotNPointCorrespondence}.
The resulting flow equation for the sixth order Taylor coefficient $\mu_k =  V^{(6)}_k(0)$ is derived explicitly in Appendix \ref{sct:appendix-flow-equations}.

\subsubsection{4-Point Function}
\label{sct:4-point-function}

In a real-time $\phi^4$ theory there are three different types of 4-point vertices at 1-loop level, namely (a) the classical $\phi^c \phi^c \phi^c \phi^q$ vertex, (b) the quantum $\phi^c \phi^q \phi^q \phi^q$ vertex and (c) the `anomalous' $\phi^c \phi^q \phi^c \phi^q$ vertex~\cite{Huelsmann:2020xcy}.
The former vertices (a), (b) already exist at tree level in the bare Keldysh action~\eqref{bareKeldyshAction}, and acquire (non-local) corrections during the FRG flow.
In contrast, the anomalous vertex (c) does not exist at tree level and is first generated at 1-loop order.
Since in our truncation scheme we want the flow of the 4-point function to be 1-loop exact, we have to consider all three vertices (a), (b) and (c).

We start with a few general remarks on how to truncate the flow equation for a 4-point function $\Gamma^{\alpha\beta\beta'\alpha'}_k$ consistently within the framework of our truncation scheme, with the upper indices $\alpha\beta\beta'\alpha'$ corresponding to
\begin{enumerate}
    \item[(a)] the classical $\alpha\beta\beta'\alpha' = cccq$ vertex,
    \item[(b)] the quantum $\alpha\beta\beta'\alpha' = cqqq$ vertex, and
    \item[(c)] the anomalous $\alpha\beta\beta'\alpha' = cqcq$ vertex.
\end{enumerate}
The flow equation for each of these 4-point functions is obtained from a corresponding   forth-order functional derivative of the Wetterich equation (\ref{eqn:wetterich-equation}).
This is straightforward but tedious, so it is not explicitly repeated here. The result, evaluated at the origin in field space (see e.g.~\cite{Blaizot:2005wd}), can be compactly summarized~\cite{Berges:2012ty} as follows, 
\newcommand{\LegFontSize}{\scriptsize}

\begin{widetext}
\begin{align}
	\label{flow-4pt-graph}
	\partial_k \Gamma_k^{\alpha\beta\beta'\alpha'}(x,y,y',x') = -\im
	\left\{
	\begin{tikzpicture}[baseline=-0.5ex]
	 	\centerarc[pgreen](0,0)(-90:90:0.5)
	 	\centerarc[pgreen](0,0)(90:270:0.5)
	 	\draw[black] (-0.5,0) -- (-0.5-0.3,0.3) node[anchor=south] {\LegFontSize $(y,\beta)$};
	 	\draw[black] (-0.5,0) -- (-0.5-0.3,-0.3) node[anchor=north] {\LegFontSize $(x,\alpha)$};
	 	\draw[black] (0.5,0) -- (0.5+0.3,0.3) node[anchor=south] {\LegFontSize $(y',\beta')$};
	 	\draw[black] (0.5,0) -- (0.5+0.3,-0.3) node[anchor=north] {\LegFontSize $(x',\alpha')$};
	 	\fill[black] (-0.5,0) circle (0.1);
	 	\fill[black] (+0.5,0) circle (0.1);
	 	\draw[black] (-0.1,0.5-0.1) -- (0.1,0.5+0.1);
	 	\draw[black] (-0.1,0.5+0.1) -- (+0.1,0.5-0.1);
	\end{tikzpicture}
	+
	\begin{tikzpicture}[baseline=-0.5ex]
	 	\centerarc[pgreen](0,0)(-90:90:0.5)
	 	\centerarc[pgreen](0,0)(90:270:0.5)
	 	\draw[black] (0,0.5) -- (-0.3,0.5+0.3) node[anchor=east] {\LegFontSize $(y,\beta)$};
	 	\draw[black] (0,0.5) -- (+0.3,0.5+0.3) node[anchor=west] {\LegFontSize $(y',\beta')$};
	 	\draw[black] (0,-0.5) -- (-0.3,-0.5-0.3) node[anchor=east] {\LegFontSize $(x,\alpha)$};
	 	\draw[black] (0,-0.5) -- (+0.3,-0.5-0.3) node[anchor=west] {\LegFontSize $(x',\alpha')$};
	 	\fill[black] (0,+0.5) circle (0.1);
	 	\fill[black] (0,-0.5) circle (0.1);
	 	\draw[black] (-0.5-0.1,-0.1) -- (-0.5+0.1,+0.1);
	 	\draw[black] (-0.5-0.1,+0.1) -- (-0.5+0.1,-0.1);
	\end{tikzpicture}
	+
	\begin{tikzpicture}[baseline=-0.5ex]
	 	\centerarc[pgreen](0,0)(-90:90:0.5)
	 	\centerarc[pgreen](0,0)(90:270:0.5)
	 	\draw[black] (0,0.5) -- (-0.3,0.5+0.3) node[anchor=east] {\LegFontSize $(y,\beta)$};
	 	\draw[black] (0,-0.5) -- (-0.3,-0.5-0.3) node[anchor=east] {\LegFontSize $(x,\alpha)$};
	 	\fill[black] (0,+0.5) circle (0.1);
	 	\fill[black] (0,-0.5) circle (0.1);
	 	\draw[black] (-0.5-0.1,-0.1) -- (-0.5+0.1,+0.1);
	 	\draw[black] (-0.5-0.1,+0.1) -- (-0.5+0.1,-0.1);
	 	\centerarc[black](0,-0.5)(90:0:1)
	 	\node at (1,-0.8) {\LegFontSize $(x',\alpha')$};
	 	\centerarc[black](0,+0.5)(-90:-35:1)
	 	\centerarc[black](0,+0.5)(-25:0:1)
	 	\node at (1,+0.8) {\LegFontSize $(y',\beta')$};
	\end{tikzpicture}
	\right\}
	-\frac{\im}{2}
	\begin{tikzpicture}[baseline=-0.5ex]
	 	\centerarc[pgreen](0,0)(-90:90:0.5)
	 	\centerarc[pgreen](0,0)(90:270:0.5)
	 	%
	 	\draw[black] (0,-0.5) -- (-0.15,-0.5-0.3);
	 	\draw[black] (0,-0.5) -- (+0.15,-0.5-0.3);
	 	\draw[black] (0,-0.5) -- (-0.45,-0.5-0.3);
	 	\draw[black] (0,-0.5) -- (+0.45,-0.5-0.3);
	 	\fill[black] (0,-0.5) circle (0.1);
	 	\draw[black] (-0.1,0.5-0.1) -- (0.1,0.5+0.1);
	 	\draw[black] (-0.1,0.5+0.1) -- (+0.1,0.5-0.1);
	\end{tikzpicture} \, .
\end{align}
\end{widetext}
At this level, when using scale-dependent local vertices on the right-hand side of the flow, there is a natural separation into $s$, $t$ and $u$-channel contributions, corresponding to the first three diagrams in \eqref{flow-4pt-graph}.
We furthermore  omitted the explicit labelling of the legs in the right-most diagram containing the 6-point function which is local anyway, at the order $Q=6$ of our truncation scheme, and the order of the external legs hence irrelevant.

With scale-dependent local vertices inside the flow, we can now apply our loop expansion by splitting the 4-point function $\Gamma_k^{\alpha\beta\beta'\alpha'}$ into a sum of $s$, $t$ and $u$-channel contributions each of which are local in two of the three relative coordinates $x-x'$, $x-y$ and $x'-y'$,
\begin{align}	
	\label{eqn:4pt-stu}
	\Gamma^{\alpha\beta\beta'\alpha'}_k(x,y,y',x') &=
	 \\ \nonumber
	&\hskip -1.6cm\phantom{+} \delta(x-y)\delta(x'-y') \Gamma^{\alpha\beta;\beta'\alpha'}_k(x,x') \quad \text{($s$-channel)}
	\\ \nonumber
	&\hskip -1.6cm +\delta(x-x')\delta(y-y') \Gamma^{\alpha'\alpha;\beta\beta'}_k(x,y) \quad \text{($t$-channel)}
	\\ \nonumber
	&\hskip -1.6cm +\delta(x-y')\delta(y-x') \Gamma^{\alpha\beta';\beta\alpha'}_k(x,x') \;\; \text{($u$-channel)}
\end{align}
with the symmetries $\Gamma^{\alpha\beta;\beta'\alpha'}_k(x,x') = \Gamma^{\beta'\alpha';\alpha\beta}_k(x',x)$, $\Gamma^{\alpha\beta;\beta'\alpha'}_k = \Gamma^{\beta\alpha;\beta'\alpha'}_k$ and $\Gamma^{\alpha\beta;\beta'\alpha'}_k = \Gamma^{\alpha\beta;\alpha'\beta'}_k$.
Note that such an  $s$, $t$ and $u$-channel splitting is not generally possible beyond one-loop: Although \eqref{flow-4pt-graph} formally has a one-loop structure, 
with full 4 and 6-point vertices on the right-hand side of \eqref{flow-4pt-graph},
it would contain structures of arbitrarily high loop order. 
In general, i.e.~with non-local vertex functions inside the loops, Equation \eqref{eqn:4pt-stu} would therefore represent an additional approximation. 


Here we stick with scale-dependent local vertices inside the loops and decompose Eq.~\eqref{flow-4pt-graph} into the three channels \eqref{eqn:4pt-stu}
which are all determined by essentially the same two-point correlation
$\Gamma^{\alpha \beta; \beta' \alpha'}_k(x,x')$ with suitably permuted indices and arguments.
We are therefore free to arbitrarily select one out of the three equivalent channels.
For the $s$-channel the flow equation reads,
\begin{widetext}
\begin{align}
    \label{vertex-flow-s}
	&\partial_k \Gamma^{\alpha\beta;\beta'\alpha'}_k
	(x,x') = 
	-\im
	\int\displaylimits_{\substack{x-y,\\x'-y'}}
	\hspace{-1ex}
	\begin{tikzpicture}[baseline=-0.5ex]
	 	\centerarc[pgreen](0,0)(-90:90:0.5)
	 	\centerarc[pgreen](0,0)(90:270:0.5)
	 	\draw[black] (-0.5,0) -- (-0.5-0.3,0.3) node[anchor=south] {\LegFontSize $(y,\beta)$};
	 	\draw[black] (-0.5,0) -- (-0.5-0.3,-0.3) node[anchor=north] {\LegFontSize $(x,\alpha)$};
	 	\draw[black] (0.5,0) -- (0.5+0.3,0.3) node[anchor=south] {\LegFontSize $(y',\beta')$};
	 	\draw[black] (0.5,0) -- (0.5+0.3,-0.3) node[anchor=north] {\LegFontSize $(x',\alpha')$};
	 	\fill[black] (-0.5,0) circle (0.1);
	 	\fill[black] (+0.5,0) circle (0.1);
	 	\draw[black] (-0.1,0.5-0.1) -- (0.1,0.5+0.1);
	 	\draw[black] (-0.1,0.5+0.1) -- (+0.1,0.5-0.1);
	\end{tikzpicture}
	-\frac{\im}{6}
	\int\displaylimits_{\substack{x-y,\\x'-y'}}
	\begin{tikzpicture}[baseline=-0.5ex]
	 	\centerarc[pgreen](0,0)(-90:90:0.5)
	 	\centerarc[pgreen](0,0)(90:270:0.5)
	 	%
	 	\draw[black] (0,-0.5) -- (-0.15,-0.5-0.3);
	 	\draw[black] (0,-0.5) -- (+0.15,-0.5-0.3);
	 	\draw[black] (0,-0.5) -- (-0.45,-0.5-0.3);
	 	\draw[black] (0,-0.5) -- (+0.45,-0.5-0.3);
	 	\fill[black] (0,-0.5) circle (0.1);
	 	\draw[black] (-0.1,0.5-0.1) -- (0.1,0.5+0.1);
	 	\draw[black] (-0.1,0.5+0.1) -- (+0.1,0.5-0.1);
	\end{tikzpicture}
	\, ,
\end{align}
\end{widetext}
where an extra factor of $1/3$ in front of the 6-point diagram occurs because we evenly distribute the local 6-point contributions to the $s$, $t$ and $u$-channels, cf.\ Eq.~\eqref{eqn:4pt-stu}. 
As full 4 and 6-point functions on the r.h.s.~of the flow equation in \eqref{vertex-flow-s} would invalidate this $s$, $t$ and $u$-channel splitting, they are therefore beyond the truncation scheme used here.
At the present order $Q=6$ this scheme requires the flow of the 4-point function only to be one-loop exact, and we may therefore replace the full 4-point and 6-point vertex functions
by effective local coupling constants~$\nu_k$ and $\mu_k$, respectively.
For the  4-point functions on the r.h.s.~of \eqref{vertex-flow-s} this amounts to inserting,
\begin{align}
	\label{4-point-local}
	&\Gamma^{\alpha\beta\beta'\alpha'}_k(x,y,y',x') \to
	\\ \nonumber
	&\hspace{3em}
	-\frac{1}{2} \nu_k^{\alpha\beta\beta'\alpha'} \delta(x-y)\delta(x-x')\delta(x'-y')
\end{align}
with effective local coupling constants $\nu_k^{\alpha\beta\beta'\alpha'}$ (see Eq.~\eqref{frgext-eff-coupling} below.)
Moreover, at our truncation order $Q=6$ the 6-point function is determined  by the scale-dependent local (classical) vertex $V_k^{(6)}(0) = \mu_k$ already, cf.\ Eq.~\eqref{effPotNPointCorrespondence}.
Inserting \eqref{4-point-local} into \eqref{vertex-flow-s}, the r.h.s.~of the flow equation~\eqref{vertex-flow-s} therefore becomes,
\begin{align}
\label{eqn:flow-4pt-eff-coupling}
	&\partial_k \Gamma^{\alpha\beta;\beta'\alpha'}_k(x,x') =
	\\ \nonumber
	&\hspace{3em}
	-\frac{\im}{4}\,
	\nu_k^{\alpha \beta \gamma \delta}
	\nu_k^{\delta' \gamma' \beta' \alpha'}
	B^{\gamma \gamma'}_k(x,x') G^{\delta \delta'}_k(x,x')
	\\ \nonumber
	&\hspace{3em}
	+ \frac{\im}{24} \,  \mu_k\,\delta(x-x') B_k^{K}(x,x') ,
\end{align}
where summation over repeated indices is implied again.
The effective local coupling constants $\nu_k$ are given by summing up all the (non-local) contributions of the 4-point function,
\begin{align}
	-\frac{1}{2} \nu_k^{\alpha \beta \beta' \alpha'} = \int_{x-y} \int_{x-x'} \int_{x'-y'} \Gamma^{\alpha \beta \beta' \alpha'}_k (x,y,y',x') .
	\label{frgext-eff-coupling}
\end{align}
We can then insert the ansatz \eqref{eqn:4pt-stu} into \eqref{frgext-eff-coupling} and do a Fourier transform,
\begin{align}
    \label{frgext-eff-coupling-p}
	&-\frac{1}{2} \nu_k^{\alpha \beta \beta' \alpha'} =
	\left( \Gamma^{\alpha\beta;\beta'\alpha'}_k + \Gamma^{\alpha'\alpha;\beta\beta'}_k + \Gamma^{\alpha\beta';\beta\alpha'}_k \right) \bigg\rvert_{p = 0},
\end{align}
where one can readily see that the $\nu$'s are just the sum over the the three channels of \eqref{eqn:4pt-stu} evaluated at zero momentum,
consistent with the 1-loop expansion of the flow equation.
The relation \eqref{frgext-eff-coupling-p} together with the flow equation \eqref{eqn:flow-4pt-eff-coupling} constitute our final system of one-loop exact flow equations for the full 4-point function $\Gamma^{\alpha\beta\beta'\alpha'}_k$ in which the scale-dependent constant vertices are calculated self-consistently.

Having the general flow equations for the 4-point functions of type (a), (b) and (c) at hand, we need to briefly discuss one minor additional subtlety.
Recall that the anomalous vertex (c) is first generated at one-loop order. 
It hence has a structure that is highly non-local in spacetime.
The local approximation \eqref{frgext-eff-coupling} is therefore not suited  in this case, the anomalous vertex has no constant contribution at tree level.
It is therefore in fact more accurate to set the effective coupling constant for this anomalous vertex (c) to zero, i.e.\ $ \nu_k^{cqcq} = 0 $, on the r.h.s.~of Eq.~\eqref{eqn:flow-4pt-eff-coupling} at our truncation order,
and only employ \eqref{frgext-eff-coupling} for the classical and quantum vertices (a) and (b).
We emphasize that we nevertheless  still solve \eqref{eqn:flow-4pt-eff-coupling} for \emph{all} of the three vertex functions (a), (b) and (c).
Setting $ \nu_k^{cqcq} = 0 $ to zero only effectively removes all diagrams from the r.h.s.~of the flow equations \eqref{eqn:flow-4pt-eff-coupling} that would represent a constant contribution from the anomalous vertex.
As a result, the anomalous vertex is not fed back at all into the flow equations for the 4-point functions.
One would have to go beyond the one-loop expansion on the r.h.s.~of the flow equations \eqref{eqn:flow-4pt-eff-coupling} in order to do this in a consistent way, see~\cite{Huelsmann:2020xcy}. In our combined vertex and loop-expansion scheme that would imply to increase the truncation order to at least $Q=8$. 
However, the calculated anomalous 4-point vertex function \emph{will} enter the flow of the two-point function, and can thus not be neglected entirely, although it does not re-enter the flow equations for the 4-point functions themselves.

We conclude this Subsection by introducing a few further notations that will be convenient in the following Subsections.
For the two-point vertex functions we define the shorthand notations
\begin{subequations}
\begin{align*}
	V^{cl}_k(x,x') &\equiv \Gamma^{cc;cq}_k(x,x') \, , \\
	V^{qu}_k(x,x') &\equiv \Gamma^{cq;qq}_k(x,x') \, , \; \text{and} \\
	V^{an}_k(x,x') &\equiv \Gamma^{cq;cq}_k(x,x') \, ,
\end{align*}
\end{subequations}
to emphasize the distinction between the classical (a), quantum (b) and anomalous (c) vertex functions.
We furthermore introduce the diagrammatic notation
\begin{align}
    \Gamma^{\alpha\beta;\beta'\alpha'}_k(x,x') =
    \begin{tikzpicture}[baseline=-1ex]
		%
	 	\draw[black] (-0.354,0) -- (-0.3-0.354,-0.3)  node[anchor=east] {\LegFontSize $(x,\alpha)$};
	 	\draw[black] (0.354,0) -- (+0.3+0.354,-0.3)  node[anchor=west] {\LegFontSize $(x',\alpha')$};
	 	\draw[black] (-0.354,0) -- (-0.3-0.354,+0.3)  node[anchor=east] {\LegFontSize $(x,\beta)$};
	 	\draw[black] (0.354,0) -- (+0.3+0.354,+0.3)  node[anchor=west] {\LegFontSize $(x',\beta')$};
	 	\draw[fill=black] (0,0) ellipse (0.354 and 0.354/4);
	\end{tikzpicture} \, , \label{2-pt-corr-4}
\end{align}
to denote the two-point correlation $\Gamma^{\alpha\beta;\beta'\alpha'}_k(x,x') $ in each channel of  the full 4-point vertex function, to emphasize its one-loop structure.
We will use this notation in the next Subsection to represent the two-loop order in the flow equation of the two-point function $\Gamma_k^{\alpha\alpha'}(x,x')$.

\subsubsection{2-Point-Function}
\label{sct:2-point-function}

The exact flow equation of the 2-point function at the minimum $\phi_0 = 0$ has the formal structure of a tadpole diagram, except that it contains the full 4-point vertex function~\cite{Huelsmann:2020xcy,Berges:2012ty},
\begin{align}
	\partial_k \Gamma_k^{\alpha\alpha'}(x,x') =
	-\frac{\im}{2}\;\;
	\hspace{-0.5cm}
	\begin{tikzpicture}[baseline=-0.5ex]
	 	\centerarc[pgreen](0,0)(0:360:0.5)
	 	\draw[black] (-0.5,-0.5) node[anchor=north] {\LegFontSize $(x,\alpha)$} -- (0.5,-0.5) node[anchor=north] {\LegFontSize $(x',\alpha')$};
	 	\fill[black] (0,-0.5) circle (0.1);
	 	\draw[black] (-0.1,0.5-0.1) -- (0.1,0.5+0.1);
	 	\draw[black] (-0.1,0.5+0.1) -- (+0.1,0.5-0.1);
	\end{tikzpicture}
	\, . \label{gamma2-flow}
\end{align}
According to our truncation scheme at order $Q=6$, we need to solve the flow equation for the two-point function in a two-loop exact way, which we  achieve by inserting our one-loop exact 4-point functions \eqref{eqn:4pt-stu} from the previous section without further approximation into the r.h.s.~of the flow equation \eqref{gamma2-flow}.  With the notation just introduced in \eqref{2-pt-corr-4} above, this yields 
\begin{widetext}
\begin{align}
	\partial_k \Gamma_k^{\alpha\alpha'}(x,x') =
	-\frac{\im}{2} \left\{
	\begin{tikzpicture}[baseline=-2ex]
	 	\centerarc[pgreen](0,0)(0:360:0.5)
	 	\draw[black] (0,-0.8) -- (-0.5,-0.8) node[anchor=north] {\LegFontSize $(x,\alpha)$};
	 	\draw[black] (0,-0.8) -- (+0.5,-0.8) node[anchor=north] {\LegFontSize $(x',\alpha')$};
		\draw[fill=black] (0,-0.65) ellipse (0.15/4 and 0.15);
	 	\draw[black] (-0.1,0.5-0.1) -- (0.1,0.5+0.1);
	 	\draw[black] (-0.1,0.5+0.1) -- (+0.1,0.5-0.1);
	\end{tikzpicture}
	+
	\begin{tikzpicture}[baseline=-1ex]
	 	\centerarc[pgreen](0,0)(-45:90:0.5)
	 	\centerarc[pgreen](0,0)(90:225:0.5)
	 	\draw[black] (-0.354,-0.354) -- (-0.3-0.354,-0.354-0.3) node[anchor=north] {\LegFontSize $(x,\alpha)$};
	 	\draw[black] (0.354,-0.354) -- (+0.3+0.354,-0.354-0.3) node[anchor=north] {\LegFontSize $(x',\alpha')$};
		\draw[fill=black] (0,-0.354) ellipse (0.354 and 0.354/4);
	 	\draw[black] (-0.1,0.5-0.1) -- (0.1,0.5+0.1);
	 	\draw[black] (-0.1,0.5+0.1) -- (+0.1,0.5-0.1);
	\end{tikzpicture}
	+
	\begin{tikzpicture}[baseline=-1ex]
	 	\centerarc[pgreen](0,0)(-45:90:0.5)
	 	\centerarc[pgreen](0,0)(90:225:0.5)
	 	\centerarc[black](-0.354+0.1,-0.354)(0:-90:0.608)
	 	\centerarc[black](+0.354-0.1,-0.354)(180:240:0.608)
	 	\centerarc[black](+0.354-0.1,-0.354)(250:270:0.608)
	 	\node at (-0.708,-0.354-0.608) {\LegFontSize $(x,\alpha)$};
	 	\node at (+0.708,-0.354-0.608) {\LegFontSize $(x',\alpha')$};
		\draw[fill=black] (0,-0.354) ellipse (0.354 and 0.354/4);
	 	\draw[black] (-0.1,0.5-0.1) -- (0.1,0.5+0.1);
	 	\draw[black] (-0.1,0.5+0.1) -- (+0.1,0.5-0.1);
	\end{tikzpicture}
	\right\}
	. \label{gamma2-flow-stu}
\end{align}
\end{widetext}
The $s$ and $u$-channel contributions (the second and the third diagram in \eqref{gamma2-flow-stu}) are responsible for generating a dynamic frequency dependence in the two-point function, and therefore in the spectral function, likewise.
This is why we had to go to at least the order $Q=6$ in the first place, to see effects in the spectral function that are beyond a constant mass/frequency shift.
These contributions generate a dynamic frequency dependence since they explicitly depend on the external frequency (and momentum in higher dimensions) that flows through the diagram.
In contrast, the $t$-channel contribution (first diagram in \eqref{gamma2-flow-stu}) is proportional to $\sim \delta(x-x')$ and thus only contributes a constant shift to the bare mass/frequency~$\m$.
Translating the diagrams into formal expressions then finally leads to
\begin{widetext}
\begin{align}
	\partial_k \Gamma^{cq}_{k}(p) &=
	-\frac{\im}{2} V_k^{cl}(p = 0) \int \hspace{-0.3em} \frac{\dif^{\,D}\hspace{-0.7ex}q}{(2\pi)^D} B_k^{K}(q) - \im \int \hspace{-0.3em} \frac{\dif^{\,D}\hspace{-0.7ex}q}{(2\pi)^D} \left( B_k^{K}(q) V_k^{cl}(p-q)+ B_k^{A}(q) V_k^{an}(p-q) \right) , \\
	\partial_k \Gamma^{qq}_{k}(p) &= -\im \int \hspace{-0.3em} \frac{\dif^{\,D}\hspace{-0.7ex}q}{(2\pi)^D} \left(B_{k}^{K}(q) V^{an}_{k}(p-q)+B_{k}^{R}(q) V^{qu}_{k}(p-q)+B_{k}^{A}(q) V^{qu}_{k}(q-p)\right) ,
\end{align}
\end{widetext}
here for the advanced $A$ and Keldysh $K$ components of the 2-point function.
Together with the flow equations for the 4-point vertex functions (see \eqref{frgext-eff-coupling-p} and \eqref{eqn:flow-4pt-eff-coupling}), and the scale-dependent 6-point vertex (see Appendix~\ref{sct:appendix-flow-equations}) they now represent a closed system of flow equations.
We solve this coupled system of flow equations numerically with a method outlined in Appendix~\ref{sct:numerical-impl}, and then calculate the spectral function in the IR ($k \to 0$) via \eqref{frgSpectralFnc}.
As we have explained in the beginning, this system of flow equations is then fully consistent with our expansion order $Q=6$ in the sense that the flows of the two-point functions are two-loop exact, that of the 4-point functions are one-loop exact with self-consistent tree-level contributions, and that of  the 6-point function, being a scale-dependent constant, is self-consistent at tree level. 

At this point we note for completeness that the classical-statistical limit is readily implemented in the real-time FRG flows by simply deleting the quantum $\phi^c \phi^q \phi^q \phi^q$ vertex from the microscopic action \eqref{bareKeldyshAction} together with replacing the quantum distribution function by the classical Rayleigh-Jeans distribution, which amounts to replacing  $\coth(\beta \omega/2) \to 2T/\omega$ for $\omega \ll T$, e.g.~see \cite{Kamenev:2011}.

\subsubsection{UV Initial Conditions}

In the UV (at $k=\Lambda$) the effective average action is given by the bare action \eqref{bareKeldyshAction}, from which we can therefore read off the initial conditions for two, four and 6-point functions,
\begin{subequations}
\begin{align}
	\label{eqn:frgext-uv-gamma2-cc}
	\Gamma_\Lambda^{(2),\tilde K}(\omega) &= 0, \\
	\label{eqn:frgext-uv-gamma2-cq-qc}
	\Gamma_\Lambda^{(2),R/A}(\omega) &= \omega^2 \pm \im \gamma_\Lambda \omega - {\m}_\Lambda^2, \\
	\label{eqn:frgext-uv-gamma2-qq}
	\Gamma_\Lambda^{(2),K}(\omega) &= 2 \im \gamma_\Lambda \omega \coth\left( \beta\omega/2 \right),\\
	V_\Lambda^{cl}(\omega) &= V_\Lambda^{qu}(\omega) = -\frac{\lambda_\Lambda}{6} \, , V_\Lambda^{an}(\omega) = 0 \, , \label{eqn:frgext-uv-vert} \\ \nonumber
	\text{and } \mu_\Lambda &= 0 \, ,
\end{align}
\end{subequations}
with the bare damping $\gamma_\Lambda$, the inverse temperature $\beta = 1/T$ (entering through the FDR), the bare coupling constant $\lambda_\Lambda$, and the bare frequency ${\m}_\Lambda$.

\section{Results}
\label{results}

For the discussion of the results, we set the bare mass/frequency $\m = 1$ to unity, which corresponds to implicitly expressing energies in units of $\m$ and all dimensionful parameters in appropriate powers thereof (e.g.~the coupling $\lambda$ is here  measured in units of $\m^3$). We furthermore use a rather small damping of $\gamma=0.06$, in order to warrant the reliability of our benchmark solution from the exact diagonalization.

\begin{figure*}[t]
    \centering
    \begin{minipage}{.49\textwidth}
        \centering
        {(a) $\lambda=1/32$}
        \includegraphics[width=\textwidth]{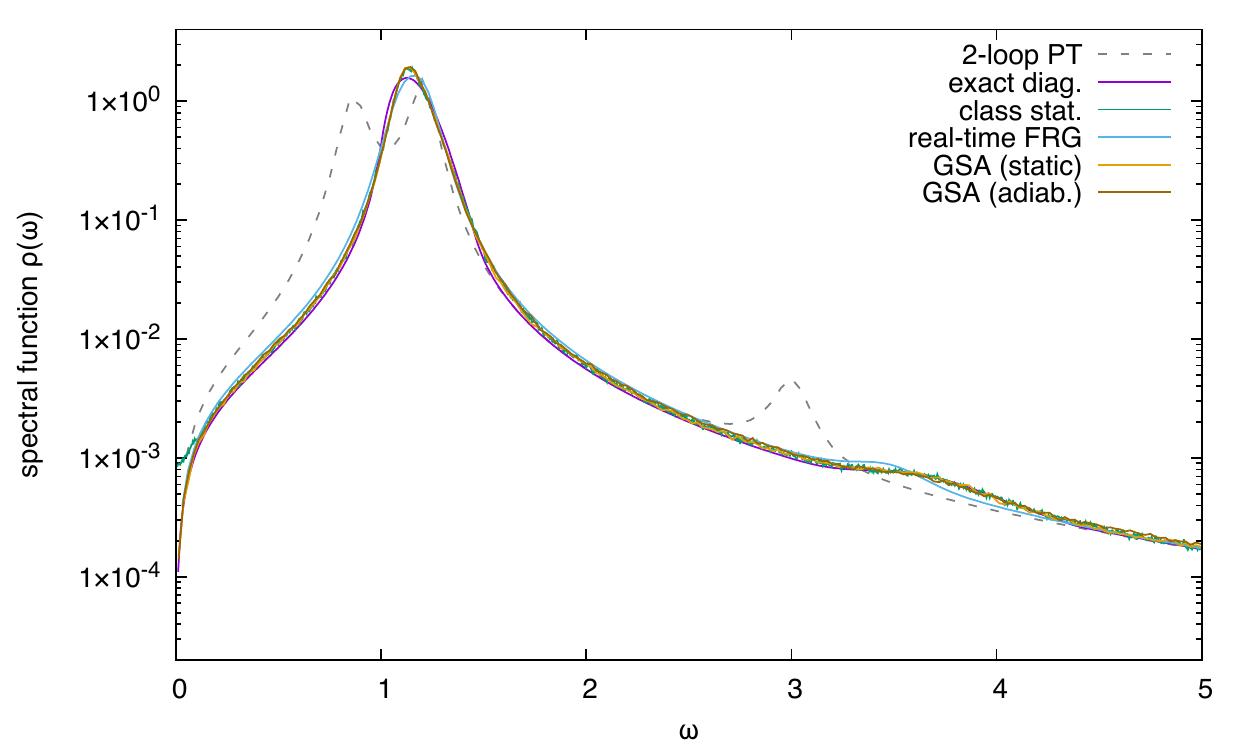}
    \end{minipage}
    \begin{minipage}{.49\textwidth}
        \centering
        {(b) $\lambda=4$}
        \includegraphics[width=\textwidth]{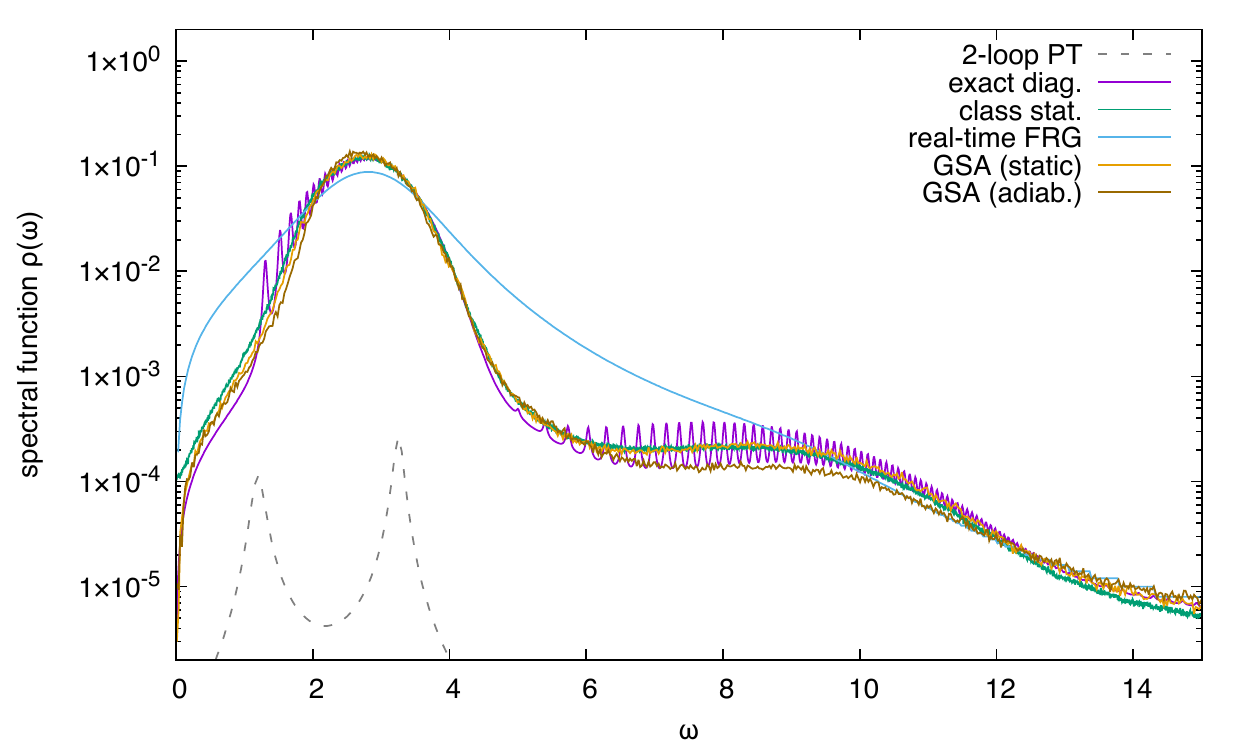}
    \end{minipage}
    \caption{Comparison of high-temperature spectral functions, at $T=32$ with weak damping $\gamma=0.06$, over frequency (all in units of $\m $) from the various methods. The panel in (a) on the left shows results at a weak coupling of $\lambda=1/32$, and that in (b) on the right the corresponding ones at a rather strong coupling of $\lambda=4$. The sharp individual peaks from the quantized transition energies, cf.~Fig.~\ref{QanOscSF}, gradually build up the broad continuum distributions observed at high temperatures in (a). These represent the classical limit in which the classical-statistical spectral functions agree with the GSA results (static/adiabatic), and all of them coincide with the solution from the exact diagonalization. Increasing the coupling at fixed temperature increases the splitting between the transition energies so that the individual peaks reemerge in (b).
    The main peak represents the ensemble of one-step transitions $\ket n \leftrightarrow \ket{n+1}$, the second one that of the three-step transitions $\ket n \leftrightarrow \ket{n+3}$ at higher excitation energies.
    The time-dependent second moments $\sigma_{xx}(t)$, $\sigma_{xp}(t)$, and $\sigma_{pp}(t)$ beyond the static approximation in the GSA produce contributions which are dismissed when extracting the spectral function from the quasi-classical method described in Sec.~\ref{gauss-extract-sf}. These contributions, which can be neglected in the nearly harmonic system (a), are mainly responsible for the differences between static and adiabatic GSA results in (b). The classical limit also serves to assess the truncation used in the real-time FRG calculations (performed on a  frequency grid  with $512$ points in the interval $\omega \in [0,15] $).  Whether or not explicitly employing the classical limit in the real-time FRG flow equations makes no noticeable differences here. The corresponding NLO 2-loop perturbative results are shown as dashed lines for comparison, and agree with those of Ref.~\cite{Huelsmann:2020xcy}.
    \label{sf_lam0.03125-and-4_T32}
    }
\end{figure*}

We start to discuss our results at a rather high temperature of $T=32$ (in units of $\m$) in Figure~\ref{sf_lam0.03125-and-4_T32}.
The high-temperature spectral functions in principle contain all possible energy differences $E_n - E_m$ in the spectrum which are allowed by selection rules, e.g.~as due to the conserved parity in our case,
weighted by the appropriate factor $e^{-\beta E_n} |\langle n |\hat{x}| m \rangle|^2$ that quantifies the probability for the transition in the thermal mixed state of the canonical ensemble, cf.~Eq.~\eqref{sfgamma-in-eigenstates}.
At a small value of $\lambda=1/32$ for the coupling in the quartic anharmonicity in panel (a) on the left, the system behaves nearly harmonically, and all approaches more or less coincide. The classical approximation for $T\gg\omega $ is well justified.  The main peak has a Breit-Wigner-like shape  which arises from the thermal ensemble of one-step transitions $\ket n\to\ket{n+1}$, see Fig.~\ref{QanOscSF}, where the $\ket n$ are distributed according to the Boltzmann weight $\e^{-\beta E_n}$, cf.~Eq.~\eqref{sfgamma-in-eigenstates}.
Because the coupling is small, the transition energies for the different $n$ that contribute are all close to that of the ground-state transition which itself is only a little larger than $\m = 1$ in the harmonic case. The central frequency $\omega_c$ of this rather narrow main peak at  $\omega_c \gtrsim 1$ is therefore also close to $\m$. 
All non-perturbative methods describe this main peak in perfect agreement with the exact solution.
In contrast, the results from NLO 2-loop perturbation theory (taken from Ref.~\cite{Huelsmann:2020xcy}) show spurious double peak structures. Most importantly, such a splitting of the main peak does not occur in the classical-statistical limit.
Although the quartic coupling $\lambda = 1/32$ is rather small in Fig.~\ref{sf_lam0.03125-and-4_T32} (a), this is not surprising, however, because  the relevant \emph{thermal coupling} $\lambda T = 1$ is comparatively large: 
In fact, one can rescale variables in the classical-statistical theory to trade the explicit dependence on the quartic coupling $\lambda$ for a dependence only on the combination $\lambda T$
(which is not possible in the full quantum theory).
Hence, the effective expansion parameter of the classical-statistical theory in the perturbative series is not $\lambda$, but the thermal coupling $\lambda T$.
Since $\lambda T=1$ in Figure~\ref{sf_lam0.03125-and-4_T32} (a) and even $\lambda T = 128$ in Figure~\ref{sf_lam0.03125-and-4_T32} (b), we must not expect the perturbative expansion to be valid here.
This once again emphasizes the need of non-perturbative real-time methods here, and it might help to appreciate the huge qualitative improvements brought about by the FRG. This is particularly reassuring for field-theory applications beyond the classical-statistical limit where we neither have exact solutions nor ab-initio results from real-time simulations.

The second peak represents the corresponding thermal ensemble of three-step transitions,
$\ket n\to\ket{n+3}$. It would be absent in the harmonic case and is therefore small because $\lambda$ is. In the NLO 2-loop perturbative calculation it occurs at $\omega = 3 \omega_0 $ corresponding to the unperturbed energy difference in all these transitions. Diagrammatically it originates from non-local `sunset' diagrams  at 2-loop level which contain three \emph{bare} propagators.
Cutkosky's cutting rule then implies that the spectral function, i.e.~the imaginary part of the retarded propagator, peaks at $\omega \approx 3 \omega_0$. It is therefore 
neither at the correct frequency nor of the correct shape. The perturbative calculation overestimates its height and underestimates its width. This is a manifestation of the fact that the perturbative expansion is not valid here, because the effective expansion parameter of the classical-statistical theory, $\lambda T=1$, is not small.

Beyond perturbation theory, the spread of the individual transition energies generally grows with $\omega$ and the second peak therefore tends to be wider than the main peak as well. The resonance frequency of the second peak is somewhat larger but close to $ 3\omega_c$ (which is still well below $T=32$ here). In fact, in our present truncation to the real-time FRG flows, the second peak has to be at  $\omega \approx 3\omega_c$, as discussed below. The fact that its central frequency is thus somewhat underpredicted by the real-time FRG  in  Fig.~\ref{sf_lam0.03125-and-4_T32} (a) is therefore a first hint at the systematic errors due to the truncation. All other methods (except perturbation theory, of course) are in perfect agreement with the exact diagonalization result around this second peak as well.
In particular, we observe no noticeable differences between the static and adiabatic GSA results. This also implies that extracting the Gaussian spectral functions quasi-classically, as described in Section~\ref{gauss-extract-sf}, is justified because the spectral function of the nearly harmonic system depends almost solely on the evolution of the expectation values $X$ and $P$.

Results at the same temperature $T=32$ but a considerably larger  
coupling of $\lambda = 4$ are shown in Figure~\ref{sf_lam0.03125-and-4_T32}~(b).
This fairly strong coupling increases the splitting between the individual transition lines, whose widths are due to the heat-bath coupling with the same small $\gamma=0.06$ as before, in each of the two ensembles so that the two corresponding peaks are  broadened significantly and their substructure becomes clearly visible in the exact-diagonalization solution. For the same reason, these peaks have moved up in energy to about $3 \m$ and $ 9 \m$, respectively, which also implies that the classical limit $T\gg\omega$ is less well satisfied here, especially in second peak. 
As compared to Figure~\ref{sf_lam0.03125-and-4_T32}~(a) this second peak is more pronounced, on the other hand, because the matrix elements for the three-step transitions are larger at larger coupling.

The real-time FRG, at this order $Q=6$ in the truncation outlined in Section~\ref{rt-frg}, has notable problems reproducing the classical limit for large coupling as also seen in Figure~\ref{sf_lam0.03125-and-4_T32} (b).
Its main peak is so broad that it almost swamps the second peak. 
To improve the truncation, on one hand, one has to go to higher orders $Q\ge 8$ in our combined  vertex and loop  expansion. On the other hand, self-consistent solutions might also be important, as shown for example in Ref.~\cite{Huelsmann:2020xcy}, where the 4-point function was flowed self-consistently. Quite expectedly, this turned out to increase the analytical as well as the numerical effort tremendously. Unfortunately, it also led to a decreased numerical stability of the flow at the same time, however.
As shown in Ref.~\cite{Huelsmann:2020xcy} the relevant parameter in the loop expansion is the thermal quartic coupling $\lambda T$. It was observed there already that the FRG spectral functions start to deviate from the classical-statistical result
for rather small but certainly non-perturbative values of $\lambda T \approx 4$ when the 4-point function of 1-loop structure is employed.
This should be compared to the comparatively huge value $\lambda T = 128$ in Figure~\ref{sf_lam0.03125-and-4_T32} (b), which exceeds this proposed range of validity by nearly two orders of magnitude.
As visible in Figure~\ref{sf_lam0.03125-and-4_T32} (a) and also below in Figure~\ref{sf_lam4_T0.5-to-4} (d) for comparatively small but already non-perturbative values of $\lambda T \sim 1-2$ the real-time FRG is still capable of accurately describing the shape of the  spectral function.
Hence there seems to be a limiting value $\lambda T \sim 4$ for our combined loop and vertex expansion at the present order as well. 
At this value e.g.~the splitting of the 4-point function in $s$, $t$ and $u$ channels is no-longer sufficient, and higher loop structures have to be taken into account. Compared to the perturbative results this is a tremendous improvement, however, as already discussed in relation with Figure~\ref{sf_lam0.03125-and-4_T32}. The NLO 2-loop result is not able to describe the spectral functions reasonably well for any of the parameters that we have considered.
Despite its difficulty with the necessarily very large thermal couplings $\lambda T$ of the classical limit, the real-time FRG is thus nevertheless very well suited to describe non-perturbative phenomena at lower temperatures where the classical-statistical calculations have to break down, eventually.

Due to the sufficiently high temperature, as in Figure~\ref{sf_lam0.03125-and-4_T32}~(a),  the classical and the static GSA spectral functions still coincide, with minute differences due to the higher relevant frequencies here only in the second peak. 
They both interpolate the substructure of the exact solution and agree with the average  shape very well.
However, the GSA in the adiabatic approximation, with the non-trivial time evolution of the Gaussian  widths included, now shows visible deviations  from those averages  in the second peak of the three-step ensemble.
This is because for a coupling as strong as $\lambda=4$ used here, our quasi-classical approach of extracting the spectral function described in Section \ref{gauss-extract-sf} becomes inconsistent, which we can infer from the fact that the classical spectral function better matches the exact one than the adiabatic GSA result does. Beyond the static approximation, the time-dependence of the Gaussian widths $\sigma_{xx}(t)$, $\sigma_{xp}(t)$, and $\sigma_{pp}(t)$ produces  contributions to the spectral functions that are not contained in the quasi-classical extraction scheme based on the classical unequal-time correlators of $X(t)$ and $P(t)$ alone.  
This explicitly demonstrates the relevance of the discussion at the end of Section \ref{gauss-extract-sf}.

\begin{figure*}[t]
    \centering
    \begin{minipage}{.49\textwidth}
        \centering
    	{(a) $\lambda=4,T=4$}
    	\includegraphics[width=\textwidth]{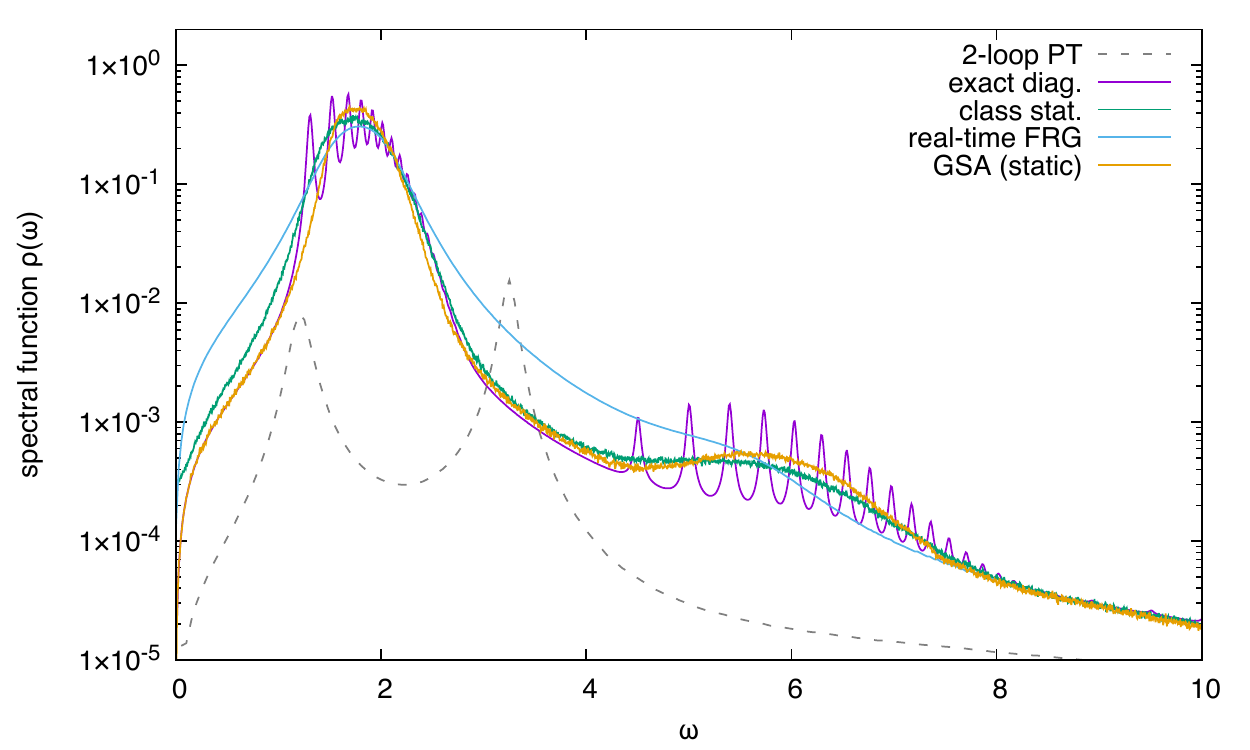}
    	{(c) $\lambda=4,T=1$}
    	\includegraphics[width=\textwidth]{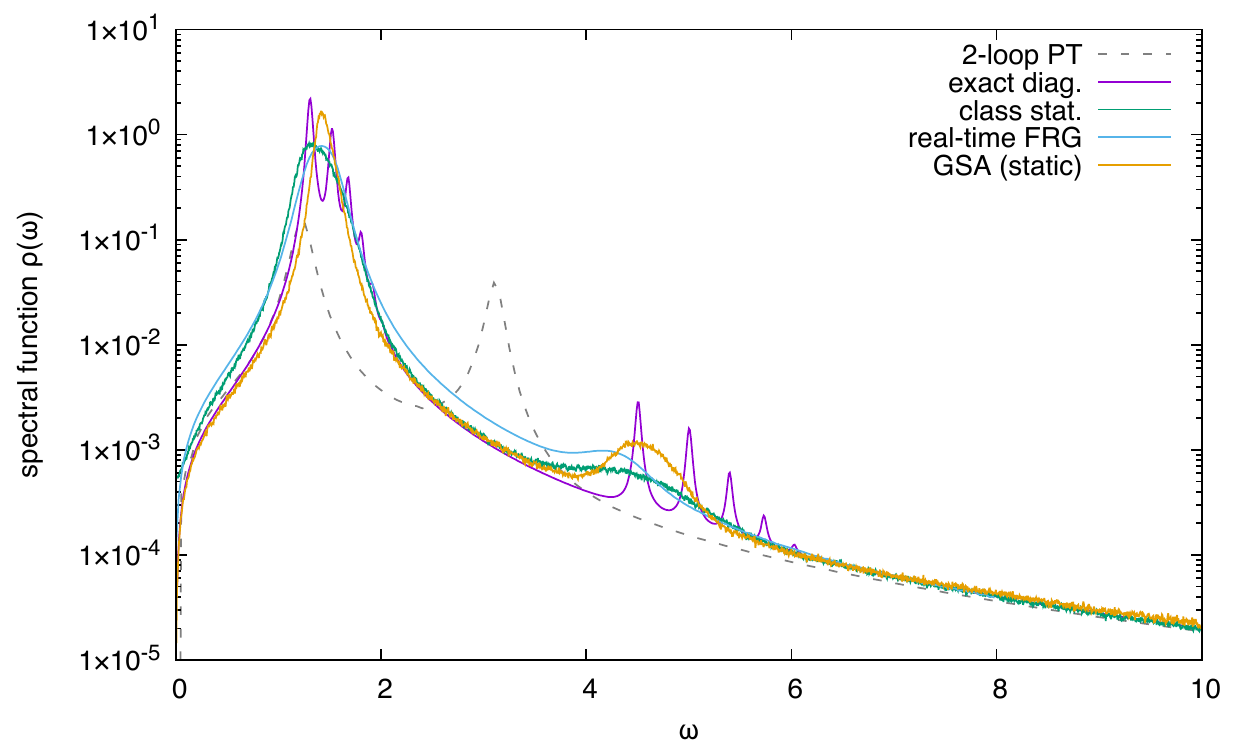}
    \end{minipage}
    \begin{minipage}{.49\textwidth}
        \centering
    	{(b) $\lambda=4,T=2$}
    	\includegraphics[width=\textwidth]{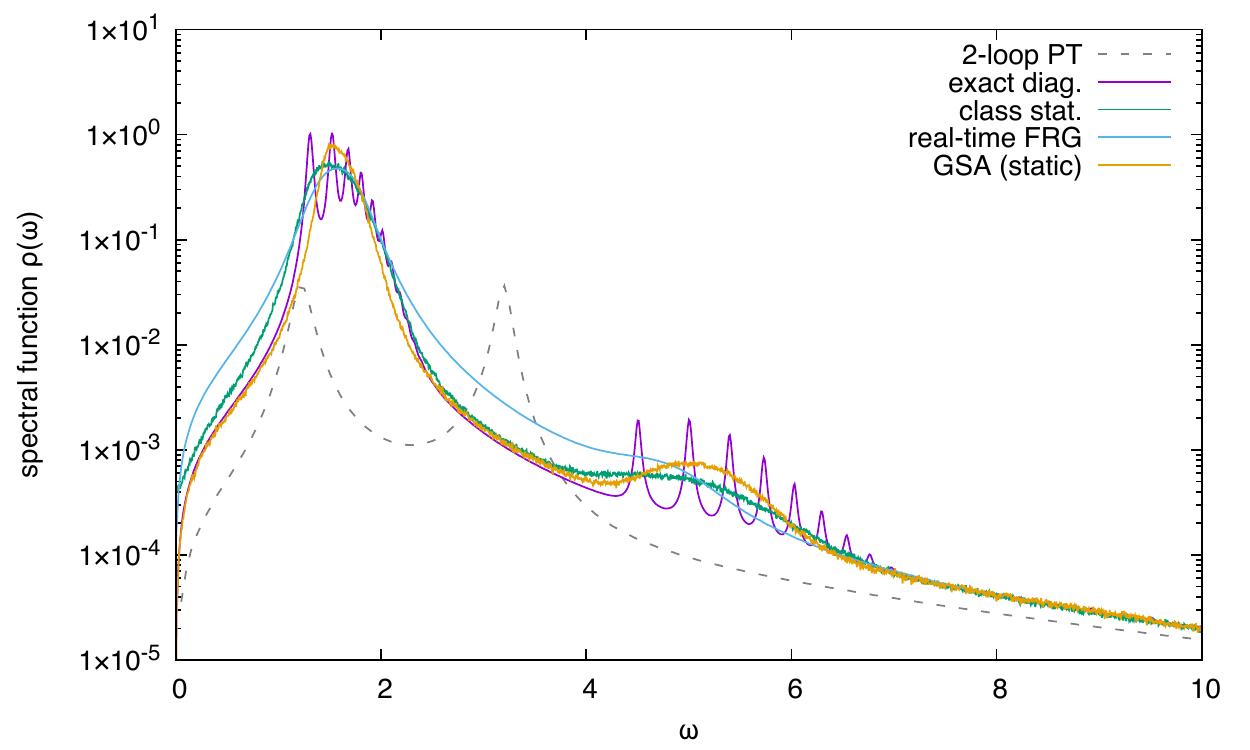}
    	{(d) $\lambda=4,T=0.5$}
    	\includegraphics[width=\textwidth]{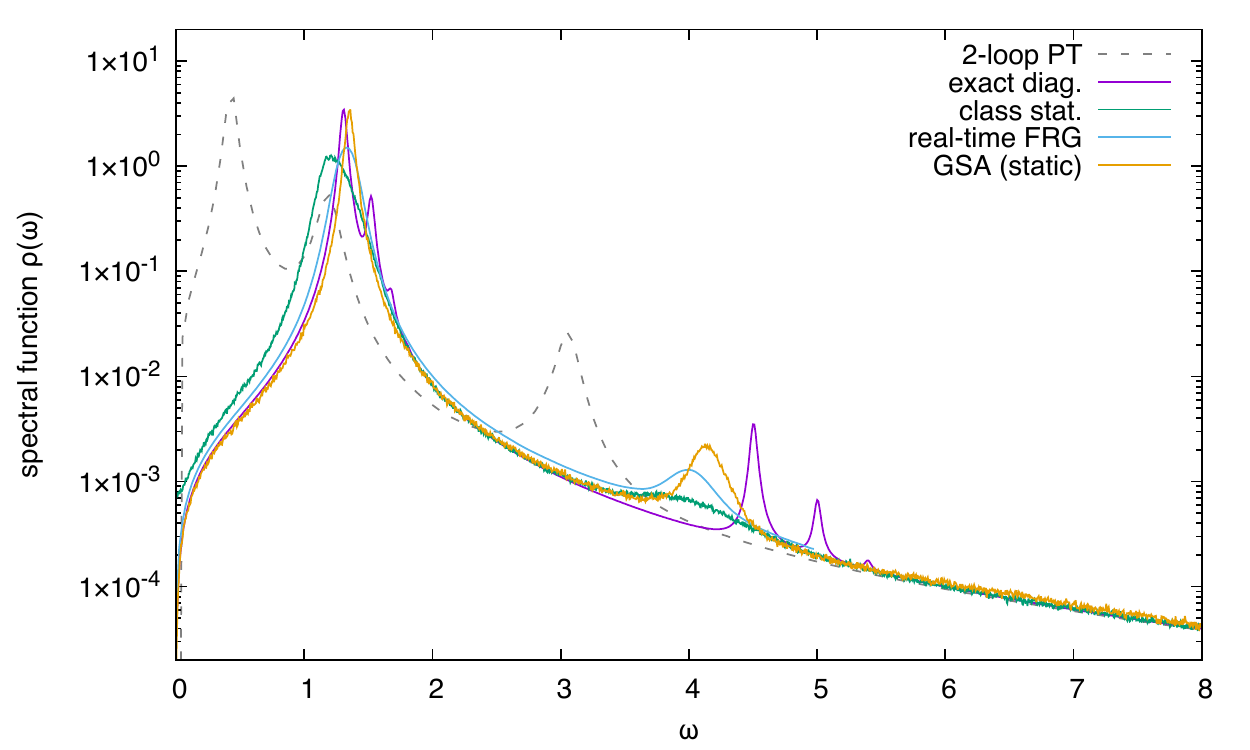}
    \end{minipage}
    \caption{Strong-coupling spectral functions with 
    $\lambda=4$ (and  $\gamma=0.06$) at successively lower temperatures starting with  $T=4$ in (a) down to $T=0.5 $ in (d).  With decreasing temperature the contributions from higher states in the one-step and three-step transitions  get exponentially suppressed more and more so that the corresponding ensembles of main and second peak in the spectral functions get compressed  towards their lowest transition frequencies until only a few individual transition lines remain. The (static) GSA spectral function follows the ensemble averages of main and second peak more closely than the classical one as temperature is lowered beyond the range of validity of the classical approximation ($ T\gg \omega$). In our present truncation of the real-time FRG, the second peak for $\omega \approx 3\omega_c$ stays put at about three times the central frequency $\omega_c$ of the main peak (a frequency grid with $320$ points on $\omega\in [0,8] $
    was used for $T=1$, $2$, $4$, and one with $200$ points on $\omega\in [0,5]$ for $T=0.5$).  As in Fig.~\ref{sf_lam0.03125-and-4_T32}, the corresponding NLO 2-loop perturbative results are included with dashed lines for comparison.
    \label{sf_lam4_T0.5-to-4}}
\end{figure*}

\begin{figure}[t]
    \centering
    \begin{minipage}{.49\textwidth}
        \centering
        \includegraphics[width=\textwidth]{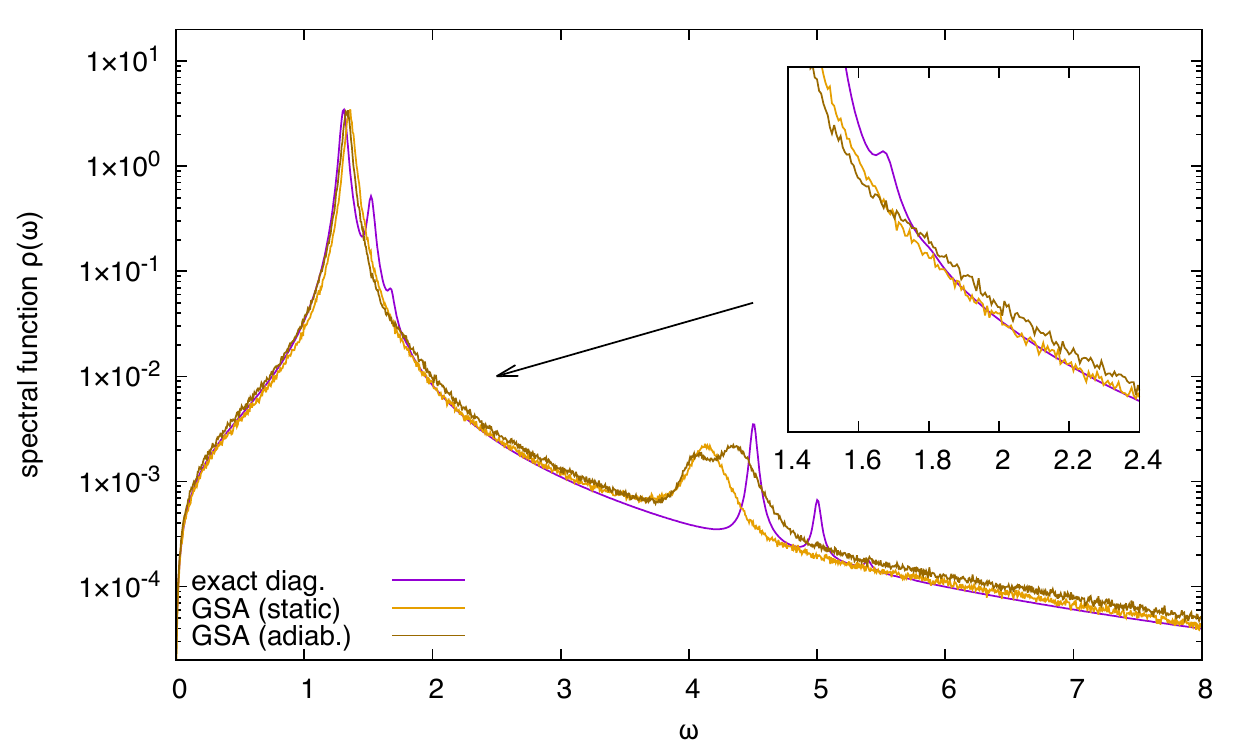}
    \end{minipage}
    \caption{Comparison of \emph{static} and  \emph{adiabatic} approximation in the GSA at strong coupling ($\lambda=4$ with $\gamma=0.06$ as before) and  low temperature ($T=0.5$). With the adiabatic corrections included in the time evolution, the central frequency of the main peak matches the exact one slightly better than in the static approximation. An additional `flank' on its right (shown in the insert) resembles the multi-peak substructure of the individual transitions from the exact diagonalization. Moreover, we observe a splitting of the second peak in the adiabatic approximation, which at least qualitatively resembles the first two distinct three-step transitions in the exact diagonalization.} \label{sf_lam4_T0.5-td-gauss-gauss}
\end{figure}

Our results for the same strong coupling of $\lambda=4$ (and $\gamma =0.06$) but now at successively lower temperatures between $T=4$ and $T=0.5$ are summarized in Figure~\ref{sf_lam4_T0.5-to-4}. We leave out the GSA results from the adiabatic approximation with the time-dependent widths in this summary because they do not represent improvements over the static GSA results. A dedicated comparison of the two approximation schemes for the GSA is deferred to Figure~\ref{sf_lam4_T0.5-td-gauss-gauss} and discussed below.  

The two ensembles of one-step and three-step transition lines in the exact spectral functions of Figure~\ref{sf_lam4_T0.5-to-4} become sparser with decreasing temperature because the transitions with larger $n$ get sequentially more suppressed. Because that removes strength form the higher-frequency side of each of the two, the corresponding peaks become narrower and their central frequencies shift to smaller values with decreasing temperature. 
In contradistinction to the high-temperature limit and the previous Figure~\ref{sf_lam0.03125-and-4_T32}, for very low temperatures these ensembles can not be represented by broad peaks in the first place anymore, the quantization in terms of the individual transition lines can eventually no-longer be neglected.

While all methods reproduce the general trend of the overall infrared shift and reduction of width of the main peak in the spectral function, there are quantitative differences worth discussing:
At the starting temperature of $T=4$ in Figure~\ref{sf_lam4_T0.5-to-4} (a), classical and GSA spectral functions are still very similar and both agree well (on average) with the exact result, while the FRG solution shows problems analogous to those discussed in relation with Figure~\ref{sf_lam0.03125-and-4_T32} (b) before.
Reducing the temperature to $T=2$, we see in Figure~\ref{sf_lam4_T0.5-to-4} (b) that the classical and GSA spectral functions start to separate from each other.  
The classical result tends to underestimate the central frequency of the main peak, and the GSA more closely follows its shape on average. In particular, the GSA tends to reproduce better its 
rather abrupt start due to the relatively sharp and well isolated quantum mechanical ground-state transition line on the low-frequency side.

When it comes to the second peak representing the three-step transition lines, the GSA result is clearly able to describe their ensemble average significantly better than the classical spectral function. This trend continues towards lower temperatures where the GSA results are able to follow these ensemble averages more closely than the classical spectral functions, and show an enhanced strength in the second peak as compared to the classical-statistical result.
Consequently, all these effects become more pronounced at $T=1$ in Figure~\ref{sf_lam4_T0.5-to-4} (c).
At the lowest temperature $T=0.5$ of this comparison in Figure~\ref{sf_lam4_T0.5-to-4} (d) it becomes obvious that the GSA in our static approximation
eventually also tends to underestimate the central frequency of the second peak in the Caldeira-Leggett model.

The classical spectral function is bound to approach its mean-field value in the limit $T \to 0$, which is here simply given by the Breit-Wigner form with width $\gamma$ around the unperturbed $\omega_0$. It should therefore only be considered valid for high temperatures.
In contrast, the exact vacuum  spectral function (for $T \to 0$) in the interacting theory still contains all possible ground state transitions $|0 \rangle \leftrightarrow |1 \rangle, |3\rangle, |5\rangle, \ldots $ which are inherently quantum mechanical.
Because even the lowest energy difference, between ground and first excited state, increases in presence of the quantum-self-interactions which are not included in the classical limit, the latter thus necessarily fails to describe the low-temperature mass shift correctly. We also notice that it produces a main peak which is systematically too broad in comparison with the exact solution.
In the language of the closed time path and the Martin-Siggia-Rose (MSR) path integral formulation of classical-statistical mechanics, it is missing a quantum $\phi^c \phi^q \phi^q \phi^q$ vertex. 
The real-time FRG, which includes such a vertex, therefore becomes better for smaller temperatures, and the location of its main peak, with central frequency $\omega_c$, fits the quantum-mechanical solution quantitatively quite well.
The strength from the higher excitations in the second peak is best reproduced by the real-time FRG as well, although its central frequency stays closer to the one in the classical spectral function, as most prominently seen in the $T=0.5$ plot in Figure \ref{sf_lam4_T0.5-to-4} (d).
The reason for this can be understood from the truncation described in Section~\ref{rt-frg}. The effective mass-shift of the main peak is indeed generated by the tadpole diagram in~\eqref{gamma2-flow-stu}. Responsible for the second peak, however, are non-local `sunset' diagrams, e.g.~see Ref.~\cite{Huelsmann:2020xcy}. They contain three Green functions which all include the correct effective mass-shift as explained above. The imaginary part of this diagram from Cutkosky's cutting rule then fixes the location of the second peak in the spectral function to $\omega\approx 3\omega_c$.
This could best be improved upon with self-consistency and by including  higher-order vertex corrections.

To further assess the influence of the static approximation in the GSA, we also ran the adiabatic GSA simulations at our lowest temperature $T=0.5$. The results are shown in  Figure~\ref{sf_lam4_T0.5-td-gauss-gauss}.
We observe that the adiabatic approximation of the GSA quantitatively yields a slightly better estimate of the central frequency in the main peak. An additional structure on its right side, as shown in the inserts, might even be interpreted as a remnant of the individual transition-line substructure. 
Such a process is not contained in the classical-statistical approach.

The three-step transitions at higher frequencies are captured by both of the GSA results. In the adiabatic version of the GSA a double-peak structure thereby emerges resembling the corresponding substructure in the exact diagonalization result which is a purely quantum-mechanical effect (and which is never observed in the classical spectral function, cf.~Figure \ref{sf_lam4_T0.5-to-4} (d)). Quantitatively, however, both GSA results underestimate the exact frequencies, and appear to too broad as compared to the exact diagonalization solution. 
The continuous high-frequency fall-off of the spectral function from there on is in fact described best by the static GSA result which includes the quasi-classical correction factor from the colored quantum noise, cf.~Section~\ref{gauss-colored-noise}. In contrast, the large-$\omega$ behavior of the adiabatic GSA result does not match that of the exact solution so precisely anymore. 
This is another indication that the quasi-classical method to measure the spectral functions as described in Section~\ref{gauss-extract-sf} eventually becomes inconsistent with the full time evolution of the Gaussian widths in the adiabatic GSA at very low temperatures (or very high frequencies), in agreement with the discussion of the classical limit of the GSA in  Figure~\ref{sf_lam0.03125-and-4_T32} above.

\section{Conclusion \& outlook}
\label{conclusion}
In Sections~\ref{quart-anh-osc} to \ref{rt-frg} we have established four different real-time methods for calculating spectral functions. These are based on exact diagonalization, classical-statistical field theory, the Gaussian-state approximation (GSA), and the functional renormalization group (FRG) formulated on the Keldysh closed-time path.  
We have compared the results from these various methods in Section~\ref{results} for the spectral function of the quartic anharmonic oscillator coupled to an Ohmic heat bath at finite temperature as in the Caldeira-Leggett model.

Having established the underlying quantum mechanical structure of the spectral function  for the anharmonic Caldeira-Leggett oscillator from exact diagonalization in Section~\ref{quart-anh-osc}, we were able to demonstrate explicitly how the quantum mechanical system with its discrete transition lines gradually turns into a classical one when temperature is increased, as illustrated here by the changing behavior of the spectral function.
The quartic anharmonicity with coupling strength $\lambda $ of the system is responsible for the emergence of distinct thermal ensembles, each of which consists of individual transition lines whose separation increases with $\lambda $.  
With increasing temperature $T$, more and more transitions between states higher up in the spectrum contribute to each ensemble. It then is the ensemble averages that tend to the continuous broad peaks of the spectral function in the classical limit $T\gg\omega $.
Although computationally expensive, such an approach could be used to calculate spectral functions for field theories or at least small many-body systems exactly, which is an interesting topic for future work. Here it provides our benchmark solution.

We have summarized the most important concepts needed for classical-statistical simulations in Section~\ref{class-sf}. These can be used for ab-initio calculations of spectral functions in the classical limit of large bosonic occupation numbers. This limit is realized at high temperatures or low frequencies, i.e.~for $T\gg\omega $, and it can therefore also be used to study the static and dynamic universal behavior near a thermal phase transition at $T_c$  of the critical infrared modes with $\omega\ll T_c$ \cite{Berges:2009jz,Schlichting:2019tbr,Schweitzer:2020noq,Schweitzer:2021iqk}. Here it served us as the limiting case, to verify the validity of the other approaches in the high-temperature limit. We have observed that in this limit the classical result describes the exact spectral function perfectly well as a coarse-grained version of the quantum mechanical one. However, the classical approach is not able to resolve the quantum-mechanical substructure from the individual transition lines, and it also underestimates the frequencies of main and second peak in the spectral function when it reaches its limitations at smaller temperatures.

In Section~\ref{gauss-state-approx} we have shown how to construct a consistent description of an external heat bath in the Gaussian state approximation, describing the effects of thermalization, dissipation, and quantum-mechanical decoherence.
There, we have employed two distinct approximations, named `static' and `adiabatic', and discussed their effects on the spectral functions in Section~\ref{results}.
We have verified that the GSA is perfectly consistent with the high-temperature limit, where it reproduces the classical results as expected. In the static approximation the GSA is able to follow the ensemble averages of the exact solution towards lower temperatures than the classical spectral functions, thus extending their range of applicability by some margin.   
 While the GSA can serve as a useful qualitative tool to study the existence and rough structure of quantum effects as (small) corrections to classical-statistical simulations towards lower temperatures, it leaves the discrete substructure in the ensemble peaks unresolved, likewise.
An interesting opportunity for future studies would be to investigate alternatives to Gaussian distributions of pure states, e.g.~based on the logistic function as suggested in \cite{Adamian:1997qna}, for the system particle at large anharmonicities where already the leading corrections to the classical limit are expected to be non-Gaussian.

While the GSA in the static approximation presented here might not be so well-suited for high-precision calculations at strong coupling $\lambda$, it does incorporate the \emph{exact} classical-statistical field theory limit near a thermal second order phase transition at a critical temperature $T_c$, where the dynamics are dominated by critical infrared modes.
However, at a finite `distance' to the critical point or in non-equilibrium phase transitions along trajectories in the phase diagram that get close to it, quantum mechanical effects might well become important. In such a situation,  as relevant e.g.~in heavy-ion collision experiments searching for the QCD critical point, the GSA could serve a useful indicator for that to happen.

For our real-time FRG calculations  of spectral functions in Section~\ref{rt-frg} we have introduced the novel concept of heat-bath regulators. These are constructed from coupling the system to a fictitious external heat bath, which is introduced in the spirit of the Caldeira-Leggett model as an ensemble of harmonic oscillators,
controlled by an FRG scale $k$ dependent spectral density $J_k(\omega)$. This provides a rather intuitive picture of suppressing infrared modes by overdamping, and it is particularly well-suited for near-equilibrium real-time calculations. The construction includes regulating real \emph{and} imaginary parts of self-energies, while the causal structure of the Keldysh action is built-in. At the same time, this  causal structure of such a regulator added to the Keldysh action necessarily also requires adding a Callan-Symanzik counter-term as well, in order to avoid acausal regulator singularities and to suppress long-wavelength infrared modes. With these counter terms included, however, our heat-bath regulators can straightforwardly be generalized to field theories in $d$ spatial dimensions, which is work ongoing. An analogous construction scheme might also be feasible, for example by imagining the bath to be an ensemble of Klein-Gordon fields, which brings along further subtleties such as e.g.~Lorentz covariance, that are not present in our 0+1 dimensional example here.

After establishing our causal regulators, we have adopted a truncation scheme for real-time FRG calculations, which is a modification of that used in Ref.~\cite{Huelsmann:2020xcy} in that it combines vertex and loop expansions. 
At our present truncation order it includes a self-consistent vertex expansion of the self-energies up to two-loop order. We have demonstrated in Section \ref{results} that such a truncation yields robust and quantitatively very reasonable results for the main peak in the spectral function at low temperatures. Noticeable problems occurred only  in reproducing the classical limit at strong coupling $\lambda$, which indicates that higher-order non-local structures in the vertex functions might have to be taken account  to describe this regime in parameter space.

With regulators and truncation scheme for real-time FRG calculations in place, the potential of this new direction of field-theory applications for the future includes studying  dynamic critical phenomena and non-equilibrium phase transitions; work we have not touched upon here but which is also under way already. 

First FRG calculations of dynamical critical exponents~$z$ for various dynamical models~\cite{Mesterhazy:2015uja,Duclut:2016jct,Canet:2006xu} or critical spectral functions~\cite{Tan:2021zid} have only relatively recently become available and will be systematically expanded.
Moreover, the approach of building causal regulators might be analogously possible also for fermions, leading to the perspective of studying different dynamics in renormalizable  chiral effective theories such as the Quark-Meson Model, as an alternative and extension to the ana\-lytically continued Euclidean FRG flows \cite{Kamikado:2013sia,Tripolt:2013jra,Tripolt:2014wra,Jung:2021ipc}, or the chiral Parity-Doublet Model for nuclear and chirally symmetric hadronic matter \cite{Weyrich:2015hha,Tripolt:2021jtp} within our real-time FRG framework.

For non-equilibrium  phase transitions, both the real-time FRG and classical-statistical simulations are well-suited to study off-equilibrium phenomena such as finite-time trajectories in the vicinity of a critical point in the phase diagram  \cite{tauber_2014}.
It is believed that the theoretical understanding of such systems might yield protocols to locate the QCD critical point through the data obtained from heavy-ion collisions (see e.g.~\cite{Berdnikov:1999ph,Rajagopal:2019xwg,Stephanov:1998dy,Stephanov:1999zu}). Both the classical-statistical simulations, possibly extended by the static GSA, as well as the real-time FRG provide promising computational frameworks for further investigations of such non-equilibrium systems in the future.

\section*{Acknowledgements}
We thank Pavel Buividovich, Sören Schlichting and Philipp Scior for many inspiring and helpful discussions. We are also grateful to G.G.~Adamian and N.V.~Antonenko for 
valuable comments on our manuscript.      
This work was supported by the Deutsche Forschungsgemeinschaft (DFG) through the grant CRC-TR 211 ``Strong-interaction matter under extreme conditions.'' 

\appendix

\renewcommand{\theequation}{\thesection.\arabic{equation}}

\section{GSA Details}
\subsection{Equations of Motion for Gaussian Widths}
\label{gauss-heat-bath-eoms-derivation}

Based on an adiabatic approximation, in this appendix we derive expressions for the stochastic forces $K_{xp}(t)$ and $K_{pp}(t)$ on the variances $\sigma_{xp}(t) $ and $\sigma_{pp}(t)$ in Eqs.~\eqref{sigma-xp-fluc} and \eqref{sigma-pp-fluc} which are given by the irreducible correlators 
\begin{align}
    K_{xp}(t) \equiv \llangle \hat x(t)\hat \eta(t) \rrangle , \;\;\mbox{and}\;\;
    K_{pp}(t) \equiv \llangle \hat p(t)\hat \eta(t) \rrangle . \label{flucF}
\end{align}
With the explicit expression for the fluctuating force in Eq.~(\ref{xi-op}), without the transient initial shift, these are
\begin{align}
    K_{xp}(t) &=
    \label{Kxp-exp} \\ \nonumber
    &\quad \sum_s g_s \left( G_{x \varphi_s}(t) \cos( \omega_s t ) + \frac{ 1 }{ \omega_s } G_{x \pi_s}(t) \sin(\omega_s t) \right),
    \\
    K_{pp}(t) &=
    \label{Kpp-exp} \\ \nonumber
    &\quad \sum_s g_s \left( G_{p \varphi_s}(t) \cos( \omega_s t ) + \frac{ 1 }{ \omega_s } G_{p \pi_s}(t) \sin(\omega_s t) \right),
\end{align}
where
\begin{subequations}
\begin{align}
    G_{x \varphi_s}(t) &\equiv \llangle \hat{x}(t) \hat{\varphi}_s(0) \rrangle , \;
    G_{x \pi_s}(t) \equiv \llangle \hat{x}(t) \hat{\pi}_s(0) \rrangle ,
    \label{Gx-phipi} \\
    G_{p \varphi_s}(t) &\equiv \llangle \hat{p}(t) \hat{\varphi}_s(0) \rrangle , \;
    G_{p \pi_s}(t) \equiv \llangle \hat{p}(t) \hat{\pi}_s(0) \rrangle ,
    \label{Gp-phipi}
\end{align}
\end{subequations}
describe the irreducible correlations between the heat bath oscillators and the system particle.
To calculate these correlators we first consider their respective equations of motion obtained straightforwardly from their
time derivatives together with the HLEs \eqref{qeom-x} and \eqref{qeom-p} for $\hat x(t)$ and $\hat p(t)$, 
\begin{subequations}
\begin{align}
    \dod{}{t} G_{x \varphi_s}(t) &= G_{p \varphi_s}(t)
    \label{Gx-phi-eom0}
    \\
    \dod{}{t} G_{x \pi_s}(t) &= G_{p \pi_s}(t)
    \label{Gx-Pi-eom0}
    \\
    \dod{}{t} G_{p \varphi_s}(t) &= - \int_0^t \dif t'\, \gamma(t-t') \, G_{p \varphi_s}(t')     \label{Gp-phi-eom0}
    \\
    & \hskip .4cm - \llangle V'(\hat{x}(t)) \hat{\varphi}_s(0) \rrangle +
     \llangle \hat{\xi}(t) \hat{\varphi}_s(0) \rrangle \nonumber\\
    \dod{}{t} G_{p \pi_s}(t) &= -  \int_0^t \dif t'\, \gamma(t-t') \, G_{p \pi_s}(t') 
    \label{Gp-Pi-eom0} \\
    & \hskip .4cm -  \llangle V'(\hat{x}(t)) \hat{\pi}_s(0) \rrangle 
    + \llangle \hat{\xi}(t) \hat{\pi}_s(0) \rrangle \nonumber
\end{align}
\end{subequations}
To evaluate the last cumulants on the right of Eqs.~\eqref{Gp-phi-eom0} and \eqref{Gp-Pi-eom0}, we first note that 
the bath oscillators have (equally distributed and independent) minimum uncertainty at the beginning,
\begin{align}
    \llangle \hat{\varphi}_s(0) \hat{\varphi}_{s'}(0) \rrangle &= \Vhbar \delta_{ss'} \, 1/(2\omega_s) , \\
    \llangle \hat{\pi}_s(0) \hat{\pi}_{s'}(0) \rrangle &= \Vhbar \delta_{ss'} \, 
    \omega_s/2 .
\end{align}
Therefore, Eq.~\eqref{xi-op} immediately entails that
\begin{subequations}
\begin{align}
    \llangle \hat{\xi}(t) \hat{\varphi}_s(0) \rrangle &= \frac{ \Vhbar g_s}{2\omega_s} \cos(\omega_s t) ,
    \\
    \llangle \hat{\xi}(t) \hat{\pi}_s(0) \rrangle &= \Vhbar \frac{g_s}{2}  \, \sin(\omega_s t).
\end{align}
\end{subequations}
Having in mind the limit $\Lambda\to\infty $ for later, the contribution from the  transient initial shift $ \gamma(t) \hat{x}(0) \to 2\gamma \delta(t) \hat x(0) $ is dismissed here again. 

More difficult to compute are the cumulants with the conservative force terms on the right of Eqs.~\eqref{Gp-phi-eom0} and \eqref{Gp-Pi-eom0}. The application of Wick's theorem is a priori only justified for equal-time correlators from the Gaussian state \eqref{general-gauss-state}. If we use it also for the unequal-time correlators here, we obtain 
\begin{align}
    \llangle V'(\hat{x}(t)) \hat{\varphi}_s(0) \rrangle
    = \mathcal{C}( X(t), \sigma_{xx}(t)) \, G_{x \varphi_s}(t), \label{unequalTimeWick}
\end{align}
with the curvature of the potential given by 
\begin{equation} 
\mathcal{C}(X, \sigma_{xx}) = \m^2 + \frac{\lambda}{2} (X^2 + \sigma_{xx}), \label{tdcurv}
\end{equation} 
and an analogous expression for the corresponding force term in \eqref{Gp-Pi-eom0}.
As in the main text, we abbreviate the time-dependent curvature in the following simply by
\[ 
\mathcal{C}(t) \equiv \mathcal{C} \left( X(t), \sigma_{xx}(t) \right),
\]
and its inital value by $\mathcal C_0= \mathcal C(0)$.

Putting everything together, and combining (\ref{Gx-phi-eom0}), (\ref{Gx-Pi-eom0}) with (\ref{Gp-phi-eom0}), (\ref{Gp-Pi-eom0}) then yields two decoupled second-order differential equations for the irreducible correlations of $\hat x(t)$  with  the initial heat-bath coordinates and momenta. These differential equations
describe driven harmonic motion with damping (including memory effects) and, most importantly, with an in general time-dependent frequency given by the square root of the curvature $\mathcal{C}(t)$,
\begin{subequations}
\begin{align}
    \dod[2]{}{t} G_{x \varphi_s}(t) &+ \int_0^t \dif t'\, \gamma(t-t') \, \dod{}{t'}G_{x \varphi_s}(t')   \label{Gphi-eom}\\
    &\hskip .8cm + \mathcal{C}(t) \, G_{x \varphi_s}(t) = \frac{\Vhbar g_s}{2\omega_s} \cos(\omega_s t),\nonumber
    \\
    \dod[2]{}{t} G_{x \pi_s}(t) &+ \int_0^t \dif t'\, \gamma(t-t') \, \dod{}{t'} G_{x \pi_s}(t')     \label{GPi-eom} \\
    &\hskip .8cm + \mathcal{C}(t) \, G_{x \pi_s}(t) = \Vhbar \frac{g_s}{2} \sin(\omega_s t). \nonumber
\end{align}
\end{subequations}
For low frequencies $ \omega_s \ll \Lambda $, on the time-scales relevant for the driving force, the memory integrals over the damping kernel reduce to ordinary (local in time) damping terms $ \gamma \tod{}{t} G_{x \varphi_s}(t)$, $\gamma \tod{}{t} G_{x \pi_s}(t) $, cf.~Eq.~\eqref{gamma-sharp}.

An exact analytic solution to the general oscillator problem with time-dependent restoring force $\mathcal C(t) $ is unfortunately not known, at least to us. Therefore, we have to resort to an additional \textit{adiabatic} approximation,
assuming that the curvature of the potential fluctuates slowly about a temperature-dependent equilibrium value $\mathcal{C}_0(T) $ obtained from a mean-field prescription,
\begin{equation}
    \mathcal{C}_0(T) \equiv \langle \mathcal{C}(X,\sigma_{xx}) \rangle_\beta =  \m^2 + \frac{\lambda}{2} \langle\hat{x}^2\rangle_\beta ,
\end{equation}
To go beyond this approximation,
we furthermore split the time-dependent value of the curvature into this constant equilibrium value plus a small perturbation,
\begin{align}
    \mathcal{C}(t) = \mathcal{C}_0(T) + \delta \mathcal{C}(t) ,
\end{align}
and treat $ \delta \mathcal{C}(t) $  as a correction to the exactly solvable Eqs.~(\ref{Gphi-eom}), (\ref{GPi-eom}) for the constant $ \mathcal{C}_0 \equiv \mathcal C_0(T) $ (dropping the temperature dependence in the following) in time-dependent  perturbation theory with inital condition $\delta C(0) = 0$ so that the equilibrium value $\mathcal C_0 $ is our initial value for $\mathcal C(t)$ at the same time.

\subsubsection{Static Solution}
\label{AppA-static}

The zeroth-order static solution is then the textbook problem of the driven harmonic oscillator. After the transient initial time, when all contributions from solutions to the homogeneous equations have died out due to the damping, the solutions are given by
\begin{subequations}
\begin{align}
    G_{x \varphi_s}^{0}(t) &= \frac{g_s}{2\omega_s} \, A(\omega_s) \, \cos(\omega_s t - \theta(\omega_s)),
    \label{Gphi-sol-0}
    \\
    G_{x \pi_s}^{0}(t) &= \frac{g_s}{2} \, A(\omega_s) \, \sin(\omega_s t - \theta(\omega_s)),
    \label{Gpi-sol-0}
\end{align}
\end{subequations}
with amplitude 
\begin{align}
    A(\omega) = \frac{1 }{ \sqrt{ (\mathcal{C}_0 - \omega^2 )^2 + \gamma^2 \omega^2 }  }
\end{align}
and phase shift $ \theta(\omega) $,
\begin{align}
    \tan \theta(\omega) = \frac{\gamma \omega}{\mathcal{C}_0 -\omega^2} . \label{phase-shift}
\end{align}
Inserting these static solutions \eqref{Gphi-sol-0} and \eqref{Gpi-sol-0} into the fluctuating force of
Eq.~(\ref{Kxp-exp}), for example, we obtain 
\begin{align}
    K_{xp}(t)
    &= \sum_s g_s \left( G_{x \varphi_s}^0(t) \cos( \omega_s t ) + \frac{ 1 }{ \omega_s } G_{x \pi_s}^0(t) \sin(\omega_s t) \right)
    \nonumber \\
    &= \Vhbar \sum_s \frac{ g_s^2 }{ 2\omega_s } \frac{  \mathcal{C}_0 - \omega_s^2 }{  (\mathcal{C}_0 - \omega_s^2 )^2 + \gamma^2 \omega_s^2  } ,
\end{align}
where trigonometric relations were used to simplify the result in the last line.
The analogous calculation for the force in \eqref{Kpp-exp} yields
\begin{align}
   K_{pp}(t) =
     \sum_s \frac{g_s^2}{2}  \frac{ \Vhbar \gamma  \omega_s
    }{ (\mathcal{C}_0 - \omega_s^2 )^2 + \gamma^2 \omega_s^2  } .
\end{align}
Using the definition~(\ref{bath-spectral-density}) of the spectral function $J(\omega)$ of the bath modes, the fluctuating forces on the variances defined in \eqref{flucF} finally become,
\begin{align}
    K_{xp}(t)
    &=  \int_0^\infty \frac{\dif\omega}{2 \pi} \frac{J(\omega) \left(  \mathcal{C}_0 - \omega^2 \right) }{  (\mathcal{C}_0 - \omega^2 )^2 + \gamma^2 \omega^2  } , \label{Kxp-integral}
\end{align}
and
\begin{align}
    K_{pp}(t) &=
    \gamma \int_0^\infty \frac{\dif\omega}{2 \pi} \frac{ J(\omega) \, \omega^2
    }{ (\mathcal{C}_0 - \omega^2 )^2 + \gamma^2 \omega^2  } . \label{Kpp-integral}
\end{align}
Note that in the static approximation with constant $\mathcal C_0$, the stationary solutions \eqref{Gphi-sol-0}, \eqref{Gphi-sol-0} lead to stochastic forces $K_{xp}$ and $K_{pp}$ that are in fact time-independent as well.

If we now introduce the Ohmic heat bath $J_\Lambda(\omega) $, we observe that these constant stochastic forces are logarithmically UV divergent in the limit $\Lambda \to \infty$, however. This is a known artifact of the Ohmic bath without cutoff as mentioned in the main text \cite{Hakim:1985zz,Weiss_2012}.
To remove this constant contribution, which is irrelevant for the time-evolution of $\sigma_{xx} $ in \eqref{gauss-eom-p}, we first introduce 
\begin{align}
    d(\Lambda) \equiv \int_0^\infty \frac{\dif\omega}{2 \pi} \frac{ J_\Lambda(\omega) \, \omega^2
    }{ (\mathcal{C}_0 - \omega^2 )^2 + \gamma^2 \omega^2  } , 
\end{align}
which for the Ohmic bath in Eq.~\eqref{ohmic-spectral-density} diverges with $\Lambda\to \infty $ as
\begin{align}
    d(\Lambda) \sim  \frac{\gamma}{2\pi} \, \ln\frac{\Lambda^2}{\mathcal C_0} \,.
\end{align}
With this constant contribution to the fluctuating forces we then define a subtracted $\sigma_{pp}^r$ via
\begin{align}
    \sigma_{pp}^r = \sigma_{pp} - d(\Lambda) . \label{sigmapp-subtract}
\end{align}
This removes the UV divergence in the stochastic forces from both equations, \eqref{sigma-xp-fluc} and \eqref{sigma-pp-fluc} at the same time. The constant stochastic force $K_{pp}$ on $\sigma_{pp} $ in Eq.~\eqref{gauss-eom-p} is then absorbed completely in the static approximation, and the subtracted force in  \eqref{sigma-xp-fluc} reads
\begin{align}
    K_{xp} +  d(\Lambda)  &=  \int_0^\infty \frac{\dif\omega}{2 \pi} \frac{J_\Lambda(\omega)   \mathcal{C}_0 }{  (\mathcal{C}_0 - \omega^2 )^2 + \gamma^2 \omega^2  }  \label{Kxpr-integral} \\
    &\to \frac{ \mathcal C_0}{4\pi \omega_\mathcal C } \bigg(\pi+ 2   \arctan\Big(\frac{\omega_\mathcal C^2 -\gamma^2/4}{\gamma\omega_\mathcal C}\Big)\bigg) , \nonumber
\end{align}
for $\Lambda\to\infty $, where we have introduced the shifted oscillator frequency, 
\begin{equation}
\omega_\mathcal C \equiv \sqrt{\mathcal C_0 -\gamma^2/4} > 0 ,
\end{equation}
assuming weak damping. The complete equations of motion \eqref{sigma-xx-fluc} -- \eqref{sigma-pp-fluc} for the relevant Gaussian widths, without the transient initial terms and  with the subtraction \eqref{sigmapp-subtract}, in the static approximation thus read,  
\begin{subequations}
\begin{align}
    \dod{}{t} \sigma_{xx} &= 2 \sigma_{xp}, \label{eom-xx-appendix} \\
    \label{eom-xp-appendix}
    \dod{}{t} \sigma_{xp} &= \sigma_{pp}^r -  \mathcal{C}_0\sigma_{xx}  - \gamma \sigma_{xp}
    + \mathcal C_0 \, F(\mathcal C_0), \\
    \label{eom-pp-appendix}
    \dod{}{t} \sigma_{pp}^r &= - 2 \mathcal{C}_0 \sigma_{xp}  - 2 \gamma \sigma_{pp}^r ,
\end{align}
\end{subequations}
where 
\begin{equation}
F(\mathcal C_0) =  \frac{1}{2\omega_\mathcal C } \bigg( \frac{1}{2} + \frac{1}{\pi}   \arctan\Big(\frac{\omega_\mathcal C^2 -\gamma^2/4}{\gamma\omega_\mathcal C}\Big)\bigg) .
\end{equation}
This is a rather simple set of inhomogeneous first-order differential equations with the particular solution, for weak damping $\gamma < 2 \sqrt{\mathcal C_0} $, when all homogeneous contributions have died out for $t\to\infty$,
\begin{align}
    \sigma_{xx}(t) &\to F(\mathcal C_0), \nonumber \\
    \sigma_{xp}(t) &\to 0, \label{statsol-app} \\
    \sigma_{pp}(t) &= \sigma_{pp}(0) + \sigma_{pp}^r(t) \to \sigma_{pp}(0). \nonumber
\end{align}

\subsubsection{Inital Conditions}
\label{AppA-initial}

The vanishing constant force on the subtracted $\sigma_{pp}^r$  in the last line of this static approximation is consistent with the initial condition $\sigma_{p\varphi_s}(0) = 0 $ , cf.~\eqref{Kpp-exp}  for $t=0$. Because of the constant subtracted force \eqref{Kxpr-integral} on $\sigma_{xp} $, however, we have implicitly assumed non-trivial  initial conditions for the $\sigma_{x\varphi_s}$. Suitable initial conditions that are consistent with the constant (subtracted) $K_{xp}$ can be read off from Eq.~\eqref{Kxp-exp} for $t=0$,
\begin{equation}
    \sigma_{x\varphi_s}(0) = \frac{g_s}{2\omega_s} \frac{ \mathcal C_0}{(\mathcal C_0 - \omega_s^2)^2+\gamma^2\omega_s^2}.
\end{equation}
With  $\sigma_{p\varphi_s}(0) = 0 $ and  $\sigma_{p\pi_s}(0) = 0$  the time derivative of Eq.~\eqref{Kxp-exp} for $t=0$ 
then implies that  $\sigma_{x\pi_s}(0) = 0 $ also.

In summary, the presence of the constant force on $\sigma_{xp} $ in \eqref{eom-xp-appendix} requires us to slightly modify  the initial covariance matrix in \eqref{initial-sigma}. As we switch on the coupling between the heat bath and the particle at $t=0$ we have implicitly assumed here that this happens via switching on
the off-diagonal $\sigma_{x\varphi_s} $ via the term 
\begin{align}
     \sigma_{x\varphi_s}(t) &= \Theta(t) \sigma_{x\varphi_s}(0^+) + \, \cdots  , \;\; \mbox{with} \nonumber\\ 
     & \hskip 1.4cm \sigma_{x\varphi_s}(0^+) = \frac{g_s}{2\omega_s} \mathcal C_0 A^2(\omega_s) . 
\end{align}
This is analogous to the discontinuities that arise in $\sigma_{xp}(t)$  and  $\sigma_{pp}(t) $ at $t=0$ as well, from the initial shifts in Eqs.~\eqref{sigma-xp-fluc} and \eqref{sigma-pp-fluc}.
Therefore the initial covariance matrix in \eqref{initial-sigma} strictly speaking refers to $\Sigma(0^-)$, whereas we start the integration of the equations of motion in the GSA at $t=0^+$ with
\begin{align}
\Sigma(0^+) &=     \label{initial-sigma-app}\\[6pt]
    &\hskip -.4cm \left(\begin{array}{cc|cccc}
        \sigma_{xx}(0)   & \sigma_{xp}(0^+)        & ...    & \sigma_{x\varphi_s}(0^+) & 0 & ... \\
        \sigma_{xp}(0^+) & \sigma_{pp}(0^+)        & ...    & 0 & 0 & ... \\ \hline
        \vdots           & \vdots                  & \ddots & \\
        \sigma_{x\varphi_s}(0^+)  & 0 &            &\sigma_{\varphi_s\varphi_s}(0) & 0          \\
        0 & 0 &          & 0          & \sigma_{\pi_s\pi_s}(0)  \\
        \vdots           & \vdots               &        &                              &                          & \ddots
   \end{array}\right) .\nonumber
\end{align}

\subsubsection{Adiabatic Approximation}
\label{AppA-adiabatic}

Assuming that the curvature of the potential $\mathcal C (t) $ varies slowly in time compared to the relevant bath oscillator frequencies  we can elevate the static approximation to an adiabatic one: Using the driven oscillator solutions in Eqs.~\eqref{Gphi-sol-0} and \eqref{Gpi-sol-0} for the fast bath oscillator degrees of freedom, we may then simply replace $\mathcal C_0$ by $\mathcal C(t)$ in the final expressions for the equations of motion. The stochastic forces then become time dependent as well, but they remain ultraviolet finite also for the Ohmic bath $J_\Lambda(\omega) $ with $\Lambda\to \infty $, after the time-independent subtraction of the static approximation. The adiabatic approximation then yields for the Gaussian widths,
\begin{subequations}
\begin{align}
    \dod{}{t} \sigma_{xx} &= 2 \sigma_{xp}, \label{eom-xx-ad-appendix} \\
    \label{eom-xp-ad-appendix}
    \dod{}{t} \sigma_{xp} &= \sigma_{pp}^r -  \mathcal{C}(t)\sigma_{xx}  - \gamma \sigma_{xp}   \\ 
    &\hskip 1.8cm + \mathcal C(t) F\big(\mathcal C(t)\big)  -\Delta_K\big(\mathcal C(t)\big) ,  \nonumber \\
    \dod{}{t} \sigma_{pp}^r &= - 2 \mathcal{C}(t) \sigma_{xp}  - 2 \gamma \sigma_{pp}^r  +2\gamma \Delta_K\big(\mathcal C(t)\big) , \label{eom-pp-ad-appendix} 
\end{align}
\end{subequations}
where $K_{xp}(t)$ is obtained from \eqref{Kxpr-integral} or \eqref{eom-xp-appendix} with $\mathcal C_0 \to \mathcal C(t)$, i.e.
\begin{equation}
     F\big(\mathcal C(t)\big)  =  \int_0^\infty \frac{\dif\omega}{2 \pi} \frac{J_\Lambda(\omega) }{  (\mathcal{C}(t) - \omega^2 )^2 + \gamma^2 \omega^2  } ,
\end{equation}
and an additional time-dependent force 
\begin{align}
    \Delta_K(t) =&   \int_0^\infty \frac{\dif\omega}{2 \pi} \, J_\Lambda(\omega)\, \times \\
&\hskip .2cm  \frac{2\omega^4 (\mathcal C(t) -\mathcal C_0) - \omega^2 (\mathcal C^2(t) - \mathcal C_0^2) }{ \big((\mathcal{C}(t) - \omega^2 )^2 + \gamma^2 \omega^2\big)\big( (\mathcal{C}_0 - \omega^2 )^2 + \gamma^2 \omega^2\big)} \nonumber
\end{align}
arises on both $\sigma_{xp}$ and $\sigma_{pp}$.

Finally, to include time-dependent (post-adiabatic) corrections to this adiabatic approximation, one might furthermore use the retarded Green function of the unperturbed oscillator, 
\begin{align}
     &g_R(t) = \frac{1}{ \omega_\mathcal{C} } \, \Theta(t) \, e^{ -\gamma t/2} \sin( \omega_\mathcal{C} t),  \\
     & \mbox{with}\quad  \Big( \dod[2]{}{t} + \gamma \dod{}{t} + \mathcal{C}_0 \Big) g_R(t) =   \delta(t).\nonumber
\end{align}
This allows to reformulate Eqs.~(\ref{Gphi-eom}), (\ref{GPi-eom}) as a self-consistency problem of time-dependent perturbation theory (with the initial condition that $\mathcal C(t=0) = \mathcal C_0$),
\begin{align}
    G_{x \varphi_s}(t) &= G_{x \varphi_s}^0(t) - \int_0^\infty \dif t' \, g_R(t-t') \delta \mathcal{C}(t') G_{x \varphi_s}(t') ,
    \label{Gphi-eom-sc}
    \\
    G_{x \pi_s}(t) &= G_{x \pi_s}^0(t) - \int_0^\infty \dif t' \, g_R(t-t') \delta \mathcal{C}(t') G_{x \pi_s}(t') ,
    \label{Gpi-eom-sc}
\end{align}
where the leading-order corrections are obtained upon inserting $G^0_{x \varphi_s} $ and $ G_{x \pi_s}^0$ for $ G_{x \varphi_s}$ and $ G_{x \pi_s}$ on the right-hand sides of these equations. To exploit the corresponding perturbative corrections or even self-consistent solutions might be an interesting opportunity for further studies.

\subsection{Colored noise synthesis}
\label{sec:colored_noise_synthesis}
As already discussed in Section~\ref{section:equations of motion} the colored noise will lead to our equations of motion \eqref{gauss-eom-x}, \eqref{gauss-eom-p} and (\ref{eom-xx-ad}) -- (\ref{eom-pp-ad}) being non-Markovian, i.e.\ non-local in time \cite{Ford:1988zz}. Therefore the stochastic noise $\xi(t)$ occurring in (\ref{gauss-eom-p}) can no-longer be generated `on the fly' as done in classical statistical simulations. However,  different realizations of colored noise and their numerical synthesis have already been extensively studied in the context of effective open quantum system formalisms, especially in quasi-classical Heisenberg-Langevin and Schrödinger-Langevin approaches, see e.g.~\cite{Eckern_1990, Hu_1992, Barrat_2011, Biele_2014, Katz_2016} and references therein.
We employ a commonly used approach profiting from fast Fourier transform (FFT) algorithms \cite{FFTW05}. The idea is to sample a realization of the stochastic forces in frequency space according to 
\begin{equation}
    \langle|\xi(\omega)|^2\rangle = K(\omega)
\end{equation}
with the noise kernel
\begin{equation}
    K(\omega) = 2\gamma \omega \, n_{\text B}(\omega)\, , \;\;\text{for $\omega > 0$}.
\end{equation}
Performing a discrete Fourier transform (DFT) of the $\xi(\omega)$ then yields a realization of the stochastic force in the time domain which (in the continuum limit) is described by the desired autocorrelation
\begin{align}
    \langle\xi(t)\xi(t+t')\rangle &= \frac{\gamma}{\pi}\int_0^\infty\frac{2\omega}{\exp(\omega\beta)-1}\cos(\omega t')\dif\omega  \\
    &= -\frac{\pi\gamma}{\beta^2 \sinh^2(\pi t' / \beta)} + \frac{\gamma}{\pi {t'}^2}.
\end{align}

To this end, the discrete and finite set of time points $t_i \in \{0,h,2h,\dots,t_\text{max}-h,t_\text{max}\}$ at which expectation values are calculated  is translated into the set of relevant frequencies $\omega_i \in \{-\frac{\pi}{h},-\frac{\pi}{h} + \frac{2\pi}{t_\text{max}},-\frac{\pi}{h} + \frac{4\pi}{t_\text{max}},\dots,\frac{\pi}{h}- \frac{2\pi}{t_\text{max}},\frac{\pi}{h}\}$.
For all frequencies, the stochastic force $\xi(\omega_i)$ is then sampled from a Gaussian distribution with variance $K(\omega_i)$.
Finally a DFT of $\xi(\omega_i)$ is performed yielding the desired $\xi(t_i)$.

\subsection{Leapfrog algorithm}
\label{sec:leapfrog_algorithm}
In the traditional leapfrog integration scheme, once before the first step, $P$ and $\sigma_{xp}$ are staggered backwards half a step according to
\begin{align}
    P(t_0-h/2) &= P(t_0) - \frac{h}{2} \dot{P}(t_0),\\
    \sigma_{xp}(t_0-h/2) &= \sigma_{xp}(t_0) - \frac{h}{2} \dot{\sigma}_{xp}(t_0).
\end{align}

After that, each step follows the same procedure.
First evolve $P$ and $\sigma_{xp}$ a full step forward in time
\begin{align}
    P(t+h/2) &= P(t-h/2) + h\dot{P}(t) + \sqrt{h}\,\xi(t) ,\\
    \sigma_{xp}(t+h/2) &= \sigma_{xp}(t-h/2) + h\dot{\sigma}_{xp}(t),
\end{align}
then evolve $X$, $\sigma_{xx}$ and $\sigma_{pp}$ by a full step using the new values of $P$ and $\sigma_{xp}$
\begin{align}
    X(t+h) &= X(t) + h \dot{X}(t+h/2),\\
    \sigma_{xx}(t+h) &= \sigma_{xx}(t) + h \dot{\sigma}_{xx}(t+h/2),\\
    \sigma_{pp}(t+h) &= \sigma_{pp}(t) + h \dot{\sigma}_{pp}(t+h/2).
\end{align} 
Note that $\dot{\sigma}_{pp}(t+h/2)$ would require the knowledge of $\mathcal{C}(X,\sigma_{xx})|_{t+h/2}$ which is approximated by
\begin{equation}
    \mathcal{C}(t+h/2) = \frac{\mathcal{C}(t) + \mathcal{C}(t+h)}{2}.
\end{equation}

To obtain all expectation values at the same point in time, $P$ and $\sigma_{xp}$ can be evolved forward by half a step
\begin{align}
    P(t) &= P(t-h/2) - \frac{h}{2} \dot{P}(t),\\
    \sigma_{xp}(t) &= \sigma_{xp}(t-h/2) - \frac{h}{2} \dot{\sigma}_{xp}(t).
\end{align}

\section{Details on the Real-Time FRG}

\subsection{Numerical Implementation}
\label{sct:numerical-impl}

The numerical implementation starts by incorporating the 2-point function on an $\omega$-grid using $2N-1$ grid points on an interval $[-L,L]$.
The number of grid points $2N-1$ is chosen to be odd to ensure that $\omega = 0$ is one of the grid points.
Therefore there are the $2N-1$ possible values $\omega = -L, -L+\delta, \dots, 0, \dots, L-\delta, L$ with $\delta = L/(N-1)$.

The advanced 2-point function $\Gamma^{(2),A}_k(\omega)$ and the vertex functions are therefore represented by a discrete data set
\begin{align}
	\Gamma^{(2),A}_{k,j} = \Gamma^{(2),A}_k(\omega_j),
	V^{cl}_{k,j} = V^{cl}_k(\omega_j) \text{ and }
	V^{an}_{k,j} = V^{an}_k(\omega_j)
\end{align}
with $\omega_j = -L + j\delta$ for $j = 0,\dots,2(N-1)$.
The numerical implementation also exploits the symmetry relation $V^{qu}_k(\omega) = V^{cl}_k(-\omega)$ to only calculate and store the classical and anomalous vertices.
Using these approximations the coupled set of flow equations consisting of partial and ordinary integro-differential equations is reduced to a finite set of ordinary integro-differential equations for the variables
\begin{align}
	\Gamma^{(2),A}_{k,j}, V^{cl}_{k,j}, V^{an}_{k,j} \text{ for } j = 0,\dots,2(N-1), \text{ and } \mu_k.
\end{align}
For simplicity, the resulting differential equations are numerically solved using an explicit Euler method.
To increase the efficiency, the equations are reformulated using the RG `time' parameter $t = \log(k/\Lambda)$.
In general, convolution integrals of the type
\begin{align}
	\int_{-\infty}^\infty \dif\omega\,f(\omega)g(\Omega-\omega),
\end{align}
with $\Omega$ fixed, have to be solved numerically, which are performed using a trapezoidal rule for integration.
Doing this for $N$ grid points has a numerical complexity of $O(N^2)$.
This may be optimized by performing a Fast Fourier transform (FFT) to real space, where the convolution integrals are simply multiplications, and then translating the result back into momentum space using an inverse FFT, as it is done in Ref. \cite{Huelsmann:2020xcy}.
Further, the convolution requires $g$ to be evaluated outside the region where the discrete data is available.
Therefore, the function value $g(\omega)$ must  be extrapolated from the discrete values $g_j = g(\omega_j)$.
Here it is important to roughly know the behaviour of $g(\omega)$ for $|\omega| \to \infty$.

If $g(\omega) \to \pm \infty$ for $|\omega| \to \infty$ (i.e. for $g = \Gamma_k^{(2),A}$), a second order Taylor expansion at the boundaries is employed, where the derivatives are evaluated using finite differences with a single-sided 3-point form for the first derivative and a single-sided 4-point form for the second derivative.

If $g(\omega) \to \text{const}$ for $|\omega| \to \infty$ (i.e. for $g = V^{cl}_k, V^{an}_k$), the function value $g(\omega)$ is just set to the limit $\lim_{\omega' \to \pm\infty} g(\omega')$.
The drawback is that this constant must be  known a priori, which is in generally not the case, but can be motivated from the UV limit~(\ref{eqn:frgext-uv-vert}), replacing $\lambda_\Lambda$ by the corresponding effective coupling following Eq.~(\ref{frgext-eff-coupling}).
This method works best if $L$ is chosen large enough that $g(\omega) \simeq \lim_{\omega' \to \pm\infty} g(\omega')$ for $|\omega| > L$ is a sensible approximation.

For all numerical calculations the flow parameters are characterized by $\Lambda = 20$, $k_\text{IR} = 0.01$.
In every case the $\omega$-grid is chosen large enough to cover the full range of all relevant excitations and fine enough to properly resolve individual peaks of width~$\gamma$.

\subsection{Regulator Dependence}
\label{sct:appendix-reg-dep}

An important test for a given truncation is how sensitive it is to the specific choice of the regulator.
In the optimal case, the effective average action $\lim_{k\to 0} \Gamma_{k}$ in the IR should not depend on the regulator.
But for any finite truncation a spurious dependence on the regulator will certainly be introduced.
To analyze this dependence, we test two possibilities of causal regulators, namely
\begin{enumerate}
	\item a linear combination $R_{\text{HB},k} - \alpha k^2$ of a heat bath regulator defined as \eqref{spectralRep} and the Callan-Symanzik regulator (in the following referred to simply as `heat bath regulator' or in short `HB+CS') and
	\item a Callan-Symanzik-like regulator where the damping is uniformly increased by $2k$ for all frequencies, i.e. $R^{R/A}_k(\omega) = -k^2 \pm 2\im k\omega$ (in the following referred to as `CS+Damping regulator'.)
\end{enumerate}
For convenience, the heat bath regulator is reparameterized according to $\alpha = \alpha_0 + \alpha'$, where the $\alpha_0$-term eliminates the negative mass shift quadratic in $k$ from the heat bath regulator and the $\alpha'$-term then adjusts the regulating mass.
To be explicit, we have $\alpha_0 = 1/\sqrt{4\pi}$ for the regulator in equation~(\ref{eqn:hb-reg-analytic}).

\begin{figure*}[t]
	\centering
	\begin{minipage}{0.49\textwidth}
	    \centering
	    {(a) $T=4$}
	    \includegraphics[width=\textwidth]{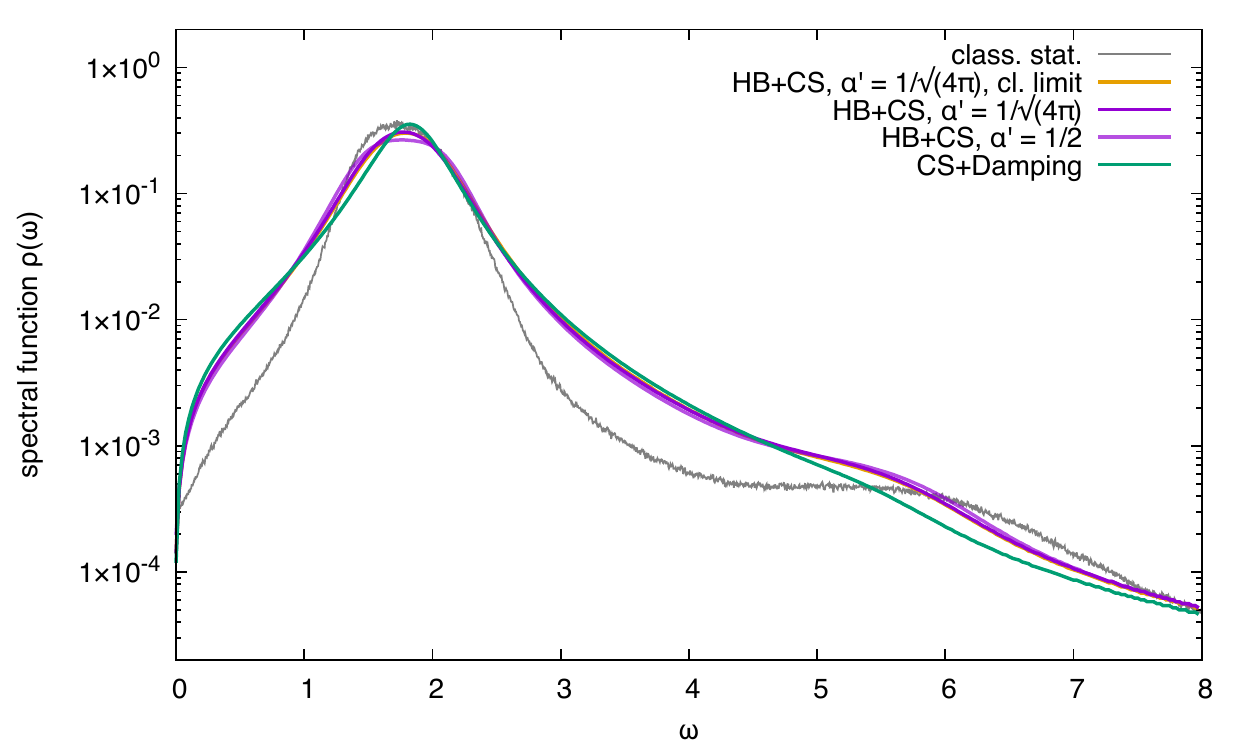}
	\end{minipage}
	\begin{minipage}{0.49\textwidth}
	    \centering
	    {(b) $T=0.5$}
	    \includegraphics[width=\textwidth]{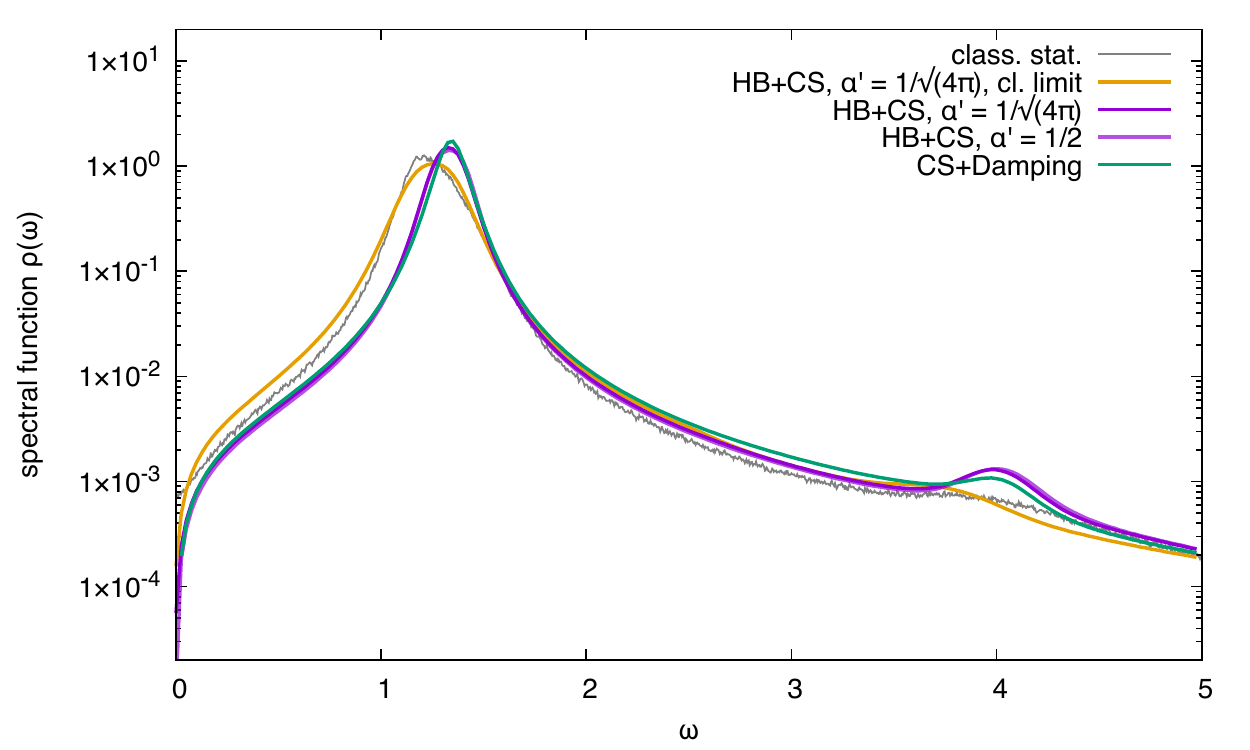}
	\end{minipage}
	\caption{Comparison between different choices for the Callan-Symanzik counter-term coefficient $\alpha$ at the UV parameters $ \lambda = 4, \gamma = 0.06$ at two temperaturs $T = 4$ (a) and $T = 0.5$ (b) for the heat bath regulator (`HB+CS'), the CS+Damping regulator, and results from the classical-statistical simulation. We also included the result for the classical limit in the real-time FRG, which arises by deleting the quantum $\phi^c \phi^q \phi^q \phi^q $ vertex and replacing the quantum distribution function with its Rayleigh-Jeans counterpart \cite{Kamenev:2011}. It agrees for $T=4$ very well with the full quantum result from the FRG, indicating that there we already are close to the classical limit. For $T=0.5$ it differs from the quantum result and matches the classical-statistical spectral function better, as expected. (For the FRG we use an $N=320, L=8$ grid for $T=4$, and an $N=200, L=5$ grid for $T=0.5$.)}
	\label{fig:reg-cmp}
\end{figure*}

For the parameters $\lambda = 4$ and $\gamma = 0.06$ and two different temperatures $T=0.5,4$ a comparison between the resulting spectral functions is shown in Fig. \ref{fig:reg-cmp}.
We also included the comparison with the FRG result when applying the classical limit, i.e. deleting the quantum $\phi^c \phi^q \phi^q \phi^q $ vertex and replacing the quantum distribution function by the Rayleigh-Jeans distribution, $\coth(\omega \beta/2) \to 2T/\omega$ \cite{Kamenev:2011}.
It becomes clear that the results indeed depend on the specific choice for the regulator.

We first focus on the $T=4$ case, visualized in Fig.~\ref{fig:reg-cmp} (a), where we know from the discussion in Section~\ref{results} that the main peak is quantitatively well described by the classical result.
Taking this classical-statistical spectral function as a benchmark for the FRG there, we see that the qualitative structure of the main peak is better described by a lower value for $\alpha'$.
In contrast, the second bump in the spectral function at around $\omega \approx 6$ is better described for a \emph{higher} value of $\alpha'$.
Therefore $\alpha'$ should be chosen such that both the main peak \emph{and} the second peak are both described sufficiently well, resulting in an optimization problem for $\alpha'$.\footnote{It is worth mentioning here that for higher values of $\alpha'$ than the ones listed in Fig.~\ref{fig:reg-cmp} the stability of the flow decreased significantly.}

For the CS+Damping regulator the main peak is slightly more sharpened, reducing the agreement with the classical-statistical spectral function in comparison with the heat bath regulator.
The CS+Damping regulator also shows a second bump, but which is much less pronounced than in the case of a heat bath regulator.
This 0+1 dimensional example already indicates the importance of using a heat bath regulator in practical calculations.
For completeness, we note that the classical limit of the FRG agrees very well with the full quantum result, which indicates that $T=4$ is indeed a temperature where the system behaves classically.

We now turn to the low temperature ($T=0.5$) case, shown in Fig. \ref{fig:reg-cmp} (b), where we observe qualitatively the same regulator-dependence as in the $T=4$ case in Fig. \ref{fig:reg-cmp} (a).
The resulting spectral function obtained from the FRG shows a very weak dependence on the Callan-Symanzik counter-term parameter $\alpha'$.
It also shows a slight broadening of the main peak and a minor enhancement of the second bump for increasing $\alpha'$.
The CS+Damping regulator produces a sharper main peak like in the $T=4$ case, and further a decrease in the height of the second bump. This suggests that the effects of the regulator noted above are stable under the variation of the temperature, such that the same conclusion in favor of the heat bath regulator applies.
Turning to the classical limit of the FRG, we see that the resulting spectral function lies significantly closer to the classical-statistical result, most visible through the shape of the main peak, as expected.
The shape of the second bump of the classical limit in the FRG also agrees with the classical-statistical spectral function. It is however located at lower frequencies, which is an effect of the structure of the non-local sunset diagrams in the flow equation of the 2-point function (cf. the discussion in Section \ref{results} in the context of Fig.  \ref{sf_lam4_T0.5-to-4}), and therefore a purely truncational issue.

We conclude from this discussion that a heat bath regulator  produces more accurate results than simpler options like the CS+Damping regulator, and it therefore should be the preferred choice for real-time calculations.

\subsection{Full flow equations}
\label{sct:appendix-flow-equations}

For the 4-point vertex, we choose to neglect the contributions arising from anomalous vertex on the r.h.s. of equation (\ref{eqn:flow-4pt-eff-coupling}) entirely, and to approximate both the classical and the quantum vertex by combining all (possibly non-local interaction terms) into one effective local interaction.
In terms of the effective coupling constants, this choice reads
\begin{subequations}
\begin{align}
	\label{eqn:eff-coupling-cccq}
	-\tfrac{1}{2}\nu_k^{cccq} &= 3\,\Gamma_k^{cc;cq}(p=0)
	,\\
	\label{eqn:eff-coupling-cqqq}
	-\tfrac{1}{2}\nu_k^{cqqq} &= 3\,\Gamma_k^{cq;qq}(p=0)
	,\\
	\label{eqn:eff-coupling-cqcq}
	-\tfrac{1}{2}\nu_k^{cqcq} &= 0
	\text{ and all others} = 0
\end{align}
\end{subequations}
where the effective coupling constants are the same for all permutations of the CTP indices.
Equipped with this choice of the effective coupling constants, it is straightforward to draw the diagrams in equation (\ref{eqn:flow-4pt-eff-coupling}) to arrive at the flow equations listed in equations~(\ref{eqn:vertex-flow-cl} -- \ref{eqn:vertex-flow-an}).

The flow equation for the 6-point function is gained in principle by functionally differentiating the flow equation~(\ref{eqn:wetterich-equation}) six times.
Since this is rather cumbersome in general, as mentioned above, we instead use the method of Taylor expanding the flow of the effective potential to sixth order as described in Subsection~\ref{sct:eff-pot}.
As a shorthand notation we define
\begin{align}
	\lambda_k^{cl} \equiv \nu^{cccq}_k \text{ and } \lambda_k^{qu} \equiv \nu^{cqqq}_k
\end{align}
for the effective coupling constants.
\begin{widetext}
\begin{subequations}
\begin{align}
    \label{eqn:vertex-flow-cl}
	\partial_k V^{cl}_k(x,x') &= -\frac{\im}{4} \lambda_k^{cl} \lambda_k^{cl} \left( B_k^{K}(x,x') G_k^{A}(x,x') + B_k^{A}(x,x') G_k^{K}(x,x') \right)+\frac{\im}{24} \mu_k \delta(x-x') B_k^{K}(x-x'=0)
	, \\
	\label{eqn:vertex-flow-qu}
	\partial_k V^{qu}_k(x,x') &= -\frac{\im}{4} \lambda_k^{cl} \lambda_k^{qu} \left( B_k^{K}(x,x') G_k^{R}(x,x') + B_k^{R}(x,x') G_k^{K}(x,x') \right)+\frac{\im}{24} \mu_k \delta(x-x') B_k^{K}(x-x'=0)
	, \\
	\label{eqn:vertex-flow-an}
	\partial_k V^{an}_k(x,x') &= -\frac{\im}{4} \lambda_k^{cl} \left( \lambda_k^{cl} B_k^{K}(x,x') G_k^{K}(x,x') + \lambda_k^{qu} B_k^{A}(x,x') G_k^{A}(x,x') +  \lambda_k^{qu} B_k^{R}(x,x') G_k^{R}(x,x') \right)
	, \\
	\begin{split}
	    \label{eqn:vertex-flow-6pt}
		\partial_k \mu_k &= -\frac{\im}{2} \int \frac{\dif^{\,D}\hspace{-0.7ex}p}{(2\pi)^D} \bigg\{ 15 \lambda _k^3 \bigg[B_k^A(p) G_k^K(p) \left(2 G_k^A(p)+G_k^R(p)\right)+B_k^R(p) G_k^K(p) \left(G_k^A(p)+2 G_k^R(p)\right)\\&\qquad+B_k^K(p) \left(G_k^A(p) G_k^R(p)+\left(G_k^A(p)\right){}^2+\left(G_k^R(p)\right){}^2\right)\bigg]\\&-\frac{15}{2} \mu_k \lambda _k \left[ G_k^K(p) \left(B_k^A(p)+B_k^R(p)\right)+B_k^K(p) \left(G_k^A(p)+G_k^R(p)\right)\right] \bigg\}
	\end{split}
\end{align}
\end{subequations}
\end{widetext}

\bibliographystyle{JHEP}
\bibliography{RealtimeMethodsRefs}

\end{document}